\documentclass[aps,prx,twocolumn,floatfix,superscriptaddress]{revtex4-2}
\usepackage[utf8]{inputenc}
\usepackage{multirow}
\usepackage{graphicx}
\usepackage{hyperref}
\usepackage[english]{babel}
\usepackage{braket}
\usepackage{amsmath}
\usepackage{siunitx}
\usepackage{nicefrac}
\usepackage[rgb]{xcolor} 
\usepackage{upgreek}
\usepackage{longtable}
\usepackage{cleveref}
\crefname{table}{}{}

\usepackage{xfrac}
\DeclareSIUnit\gauss{G}

\usepackage{xr}

\begin{document}
\title{Spectroscopy and modeling of $^{171}$Yb Rydberg states for high-fidelity two-qubit gates}

\author{Michael Peper}
\affiliation{Department of Electrical and Computer Engineering, Princeton University, Princeton, NJ 08544, USA}
\author{Yiyi Li}
\affiliation{Department of Electrical and Computer Engineering, Princeton University, Princeton, NJ 08544, USA}
\author{Daniel Y. Knapp}
\altaffiliation{Current address: Department of Physics and Astronomy, LaserLaB, Vrije Universiteit Amsterdam, de Boelelaan 1081, 1081 HV Amsterdam, The Netherlands}
\affiliation{Department of Electrical and Computer Engineering, Princeton University, Princeton, NJ 08544, USA}
\affiliation{Department of Physics, Princeton University, Princeton, NJ 08544, USA}
\author{Mila Bileska}
\affiliation{Department of Electrical and Computer Engineering, Princeton University, Princeton, NJ 08544, USA}
\affiliation{Department of Physics, Princeton University, Princeton, NJ 08544, USA}
\author{Shuo Ma}
\affiliation{Department of Electrical and Computer Engineering, Princeton University, Princeton, NJ 08544, USA}
\affiliation{Department of Physics, Princeton University, Princeton, NJ 08544, USA}
\author{Genyue Liu}
\affiliation{Department of Electrical and Computer Engineering, Princeton University, Princeton, NJ 08544, USA}
\author{Pai Peng}
\affiliation{Department of Electrical and Computer Engineering, Princeton University, Princeton, NJ 08544, USA}
\author{Bichen Zhang}
\affiliation{Department of Electrical and Computer Engineering, Princeton University, Princeton, NJ 08544, USA}
\author{Sebastian P. Horvath}
\affiliation{Department of Electrical and Computer Engineering, Princeton University, Princeton, NJ 08544, USA}
\author{Alex P. Burgers}
\altaffiliation{Current address: Department of Electrical and Computer Engineering, College of Engineering, University of Michigan, Ann Arbor, MI 48109, USA.}
\affiliation{Department of Electrical and Computer Engineering, Princeton University, Princeton, NJ 08544, USA}
\author{Jeff D. Thompson}
\email{jdthompson@princeton.edu}
\affiliation{Department of Electrical and Computer Engineering, Princeton University, Princeton, NJ 08544, USA}

\date{\today}

\begin{abstract}

Highly excited Rydberg states and their interactions play an important role in quantum computing and simulation. These properties can be predicted accurately for alkali atoms with simple Rydberg level structures. However, an extension of these methods to more complex atoms such as alkaline-earth atoms has not been demonstrated or experimentally validated. Here, we present multichannel quantum defect (MQDT) models for highly excited $^{174}$Yb and $^{171}$Yb Rydberg states with $L \leq 2$. The models are developed using a combination of existing literature data and new, high-precision laser and microwave spectroscopy in an atomic beam, and validated by detailed comparison with experimentally measured Stark shifts and magnetic moments. We then use these models to compute interaction potentials between two Yb atoms, and find excellent agreement with direct measurements in an optical tweezer array. From the computed interaction potential, we identify an anomalous F\"orster resonance that likely degraded the fidelity of previous entangling gates in $^{171}$Yb using $F=3/2$ Rydberg states. We then identify a more suitable $F=1/2$ state, and achieve a state-of-the-art controlled-Z gate fidelity of $\mathcal{F}=0.994(1)$, with the remaining error fully explained by known sources. This work establishes a solid foundation for the continued development of quantum computing, simulation and entanglement-enhanced metrology with Yb neutral atom arrays.

\end{abstract}

\maketitle

\section{Introduction}

Rydberg-mediated interactions between neutral atoms in optical tweezer arrays are enabling for quantum computing, simulation and quantum-enhanced metrology~\cite{Bernien2017probing,Scholl2021quantum,Bluvstein2022quantum,Eckner2023Realizing}. To realize high-fidelity operations, a detailed understanding of the properties of the Rydberg states and their interactions is required. Alkali atoms such as Rb or Cs can be described by simple quantum defect models, which have been validated by extensive spectroscopy~\cite{Goy1982Millimeter,Weber1987Accurate,Li2003Millimeter,Mack2011Measurement,Han2006Rb,Deiglmayr2016Precision,Peper2019Precision}. This allows the wavefunctions, and in turn, matrix elements, interaction potentials and decay rates to be computed~\cite{zimmerman1979stark,Theodosiou1984Lifetimes,Deiglmayr2016long}. Experimental studies have confirmed the predicted interaction potentials~\cite{Beguin2013Direct,Ravets2015Measurement,Anand2024dualspecies} and decay rates~\cite{Spencer1981Measurements,Spencer1982Temperature,bergstrom1986natural}. A particularly stringent test of the interaction model comes from spectroscopy of so-called macrodimer states \cite{Boisseau2002Macrodimers,Overstreet2009observation,Sassmannshausen2016Observation,Hollerith2019Quantum}.

Many recent experiments have focused on divalent alkaline-earth-like atoms, in particular Sr~\cite{Cooper2018Alkaline,Norcia2018Microscopic,Teixeira2020Preparation,Holzl2024longlived} and Yb~\cite{Wilson2022Trapping,Jenkins2022Ytterbium,Ma2022Universal,Huie2023Repetitive}. Tweezer arrays of $^{88}$Sr Rydberg atoms have been used to study many-body dynamics~\cite{Scholl2023Erasure,Shaw2024benchmarking} and entanglement-enhanced optical clocks~\cite{finkelstein2024universal,cao2024multiqubit}. On the other hand, $^{171}$Yb is ideal for use as a qubit for quantum computing, with demonstrated long coherence times for the pure nuclear spin qubit with $I=1/2$~\cite{Ma2022Universal,Jenkins2022Ytterbium,Huie2023Repetitive}, mid-circuit measurement and atom reloading~\cite{Lis2023Midcircuit,Norcia2023Midcircuit} and hardware efficient error correction strategies~\cite{Wu2022Erasure,Ma2023High}.

A challenge to the continued development of divalent atomic qubits, particularly those based on $^{171}$Yb, is the relative lack of spectroscopic information and models for the behavior of the Rydberg states. The Rydberg states of divalent atoms are more complex than alkali atoms, because of the presence of singlet and triplet Rydberg series, interactions between series converging to other ionization thresholds (\emph{i.e.}, series perturbers), and hyperfine coupling (in the case of isotopes with nuclear spin $I > 0$, such as $^{87}$Sr and $^{171}$Yb). These states can be described in the framework of multichannel quantum defect theory (MQDT)~\cite{Seaton1966Quantum,Fano1970Quantum}, which in turn allows the computation of matrix elements, interaction potentials and decay rates~\cite{Vaillant2014Multichannel,Robicheaux2018theory}. 

Single-channel approximations have been used to predict interaction potentials in both $^{174}$Yb and $^{88}$Sr, which both have no nuclear spin ($I=0$)~\cite{Vaillant2014Multichannel,Robertson2021ARC}. Moreover, relatively complete MQDT descriptions of $^{88}$Sr have been developed and used to study lifetimes and branching ratios~\cite{Vaillant2014Multichannel}, and MQDT models have been developed to fit spectroscopic data for certain Rydberg series in $^{174}$Yb~\cite{Aymar1980Highly,Aymar1984Multichannel,Ali1999Two,Lehec2018Laser,Lehec2017PhD}. The energies of certain Rydberg series in $^{87}$Sr ($I=9/2$) have also been determined~\cite{Ding2018Spectroscopy} and used to predict $C_6$ coefficients using an MDQT formalism~\cite{Robicheaux2019Calculations}. However, to the best of our knowledge, there is no comprehensive spectroscopy or MQDT model of the Rydberg states of $^{171}$Yb, and no MQDT interaction models for divalent atoms have been experimentally verified at a level approaching the precision of alkali atom models.

In this article, we present four main results. First, we present refined MQDT models for the Rydberg states of $^{174}$Yb with $L \leq 2$ (Section~\ref{sec:174Ybmain}). This set of states is sufficient to accurately predict the interactions and polarizability of the $L=0$ states that are most frequently used in experiments. This extends prior work~\cite{Camus1969spectre,Meggers1970First,Camus1980Highly,Aymar1980Highly,Aymar1984three,Neukammer1984diamagnetic,Maeda1992Optical,zerne1996lande,Lehec2017PhD,Lehec2018Laser,Niyaz2019Microwave,Okuno2022High} by incorporating new spectroscopic measurements and re-fitting all of the MQDT models simultaneously with a global fit for improved consistency and accuracy. Furthermore, we determine singlet-triplet mixing angles by comparison to experimental measurements of the static dipole polarizability, which are undetermined by the state energies alone \cite{Lu1971Spectroscopy}. We find excellent agreement with measured Rydberg state energies, polarizabilities and magnetic moments.

Next, we extend these models to describe $^{171}$Yb states with $L \leq 2$ by including the hyperfine interaction (Sec.~\ref{sec:171Ybmain}). The models are refined and validated by comparison to extensive laser and microwave spectroscopy of $^{171}$Yb Rydberg states from an atomic beam apparatus. We also test the MQDT model matrix elements by comparison to measured static dipole polarizabilities and magnetic moments, finding excellent agreement.

Third, we use the $^{171}$Yb MQDT models to compute the interaction potential for Rydberg atom pairs, which we verify against direct measurements using pairs of atoms in an optical tweezer array (Sec.~\ref{sec:RydbergRydberg_Yb171}).  We show that the $^{171}$Yb $F=3/2$ Rydberg state that was previously used to implement entangling gates~\cite{Ma2022Universal,Ma2023High} has a surprisingly small F\"orster defect ($<$\SI{10}{MHz}) over a large range of effective principal quantum number $\nu$. This gives rise to an imperfect blockade, and we conjecture that it is responsible for the discrepancy between the measured ($\mathcal{F}=0.980(1)$) and predicted ($\mathcal{F}_{th}=0.989$) gate error in Ref.~\cite{Ma2023High}.

Finally, in Sec.~\ref{sec:improved_twoqubit} we leverage these models to predict that certain $F=1/2$ states should have a larger F\"orster defect, giving rise to a cleaner blockade and improved gate fidelity. We then experimentally demonstrate improved gates with $\mathcal{F}=0.994(1)$. Importantly, the dominant remaining errors are well-understood (arising from Rydberg state decay and Doppler shifts), which is promising for future improvements in gate fidelity with increased laser power.

This work solidifies the foundation for future quantum computing, simulation and quantum-enhanced metrology applications of $^{171}$Yb Rydberg atoms. The developed MQDT models not only reproduce the energies of ytterbium Rydberg states, but also enable an accurate prediction of matrix elements, polarizabilities, and Rydberg interactions within a self-consistent framework. This comprehensive approach also establishes the basis for treating other complex atomic systems including $^{87}$Sr~\cite{Barnes2022Assembly}, and lanthanide atoms such as Ho~\cite{Hostetter2015Measurement} or Er~\cite{Trautmann2021Spectroscopy}. 
 
\section{Multichannel quantum defect model of $^{174}$Yb}
\label{sec:174Ybmain}

\begin{figure*}[t]
	\centering
    \includegraphics[width=\linewidth]{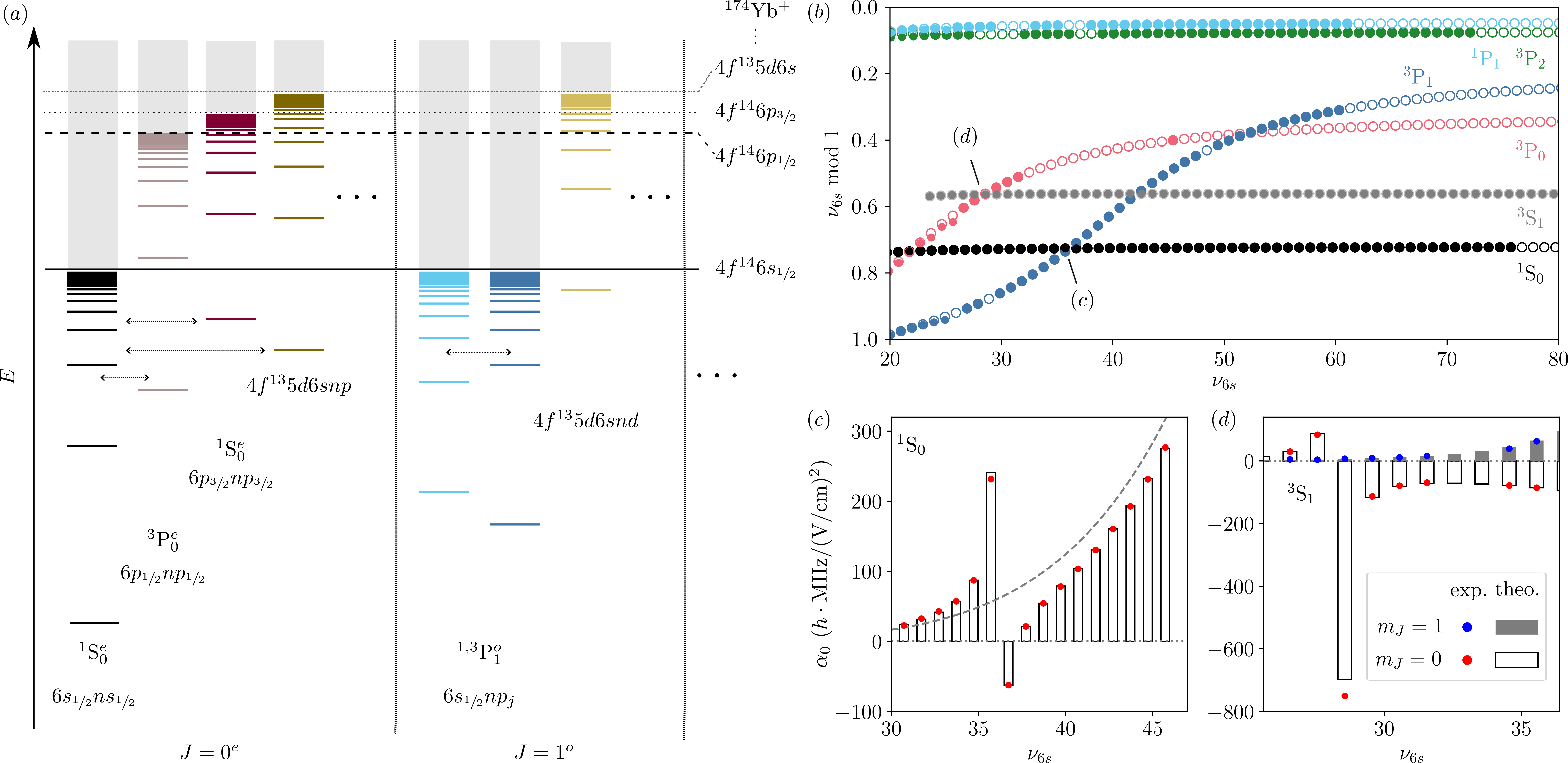}
	\caption{\label{fig:174Yb_energydiagram_perturbers_pol} $(a)$ Schematic diagram of several Rydberg series $^{174}$Yb, illustrating the principle of MQDT. Both singly-excited ($6s_{1/2}nl$) and doubly-excited (\emph{e.g.}, $6pnl$) series are included, converging to different thresholds corresponding to excited states of the $^{174}$Yb$^+$ ion (note that in this schematic diagram, the energy level spacings are not to scale, and only the subset of channels most relevant for the presented MQDT models are included). The Rydberg series are sorted by total angular momentum $J$ and parity (even/odd denoted by $e$/$o$ superscripts). $LS$ coupling term symbols are written where possible (\emph{i.e.}, $^1S_0$). Configuration interactions mix series with the same $J$ and parity, as indicated by the horizontal arrows. $(b)$ Lu-Fano-like plot of the $L=0$ and $L=1$ Rydberg series of $^{174}$Yb, showing the fractional part of the quantum defect $\left( \nu_{6s}~\textrm{mod}~1\right)$ as a function of the effective principal quantum number, $\nu_{6s}$. The open circles show the eigenstates predicted by the MQDT model, while the filled circles show experimentally determined energies. $(c)$ Static dipole polarizability $\alpha_0$ (see Eq.~(\ref{eq:staticpol})) of the $n\,^1S_0$ series close to a level crossing with the $n\,^3P_1$ series (red points: experiment, bars: MQDT prediction). The gray, dashed line indicates the expected $\nu^7$ scaling in the absence of singlet-triplet mixing.  $(d)$ Static dipole polarizability $\alpha_0$ of the $n\,^3S_1$ series near a crossing with $n\,^3P_0$. The experimental data for the polarizability of the $m_J=0$ (red) and $m_J=1$ (blue) states agrees well with the theoretical predictions from the MQDT model.}
\end{figure*}

Fig.~\ref{fig:174Yb_energydiagram_perturbers_pol}a illustrates the basic principle of MQDT, in the context of $^{174}$Yb (the most abundant isotope, with nuclear spin $I=0$). The ground state of $^{174}$Yb has the electronic configuration $[\mathrm{Xe}] 4f^{14} 6s^2$, and the states of primary interest are singly-excited states $[\mathrm{Xe}] 4f^{14} 6s nl$, where $n,l$ are the principal quantum number and orbital angular momentum of the excited electron (the Xe core and $4f^{14}$ electrons are omitted henceforth, unless otherwise noted). These states are approximately described in $LS$ coupling by the term symbol $^{2S+1}L_J$ with total electron spin $S = \{0,1\}$, $L=l$, and total electronic angular momentum $J$. In single-channel quantum defect theory, these states form isolated series converging to the  ground state of the $^{174}$Yb$^+$ ion with the electron configuration $6s^1$. The energy of the states in this series is $E_n = I_{6s}-R/\nu_{6s}^2$, where $\nu_{6s} = n-\mu$ is the effective quantum number, and $\mu(n,L,J)$ is the quantum defect that captures the interaction with the core for that series, with only weak dependence on $n$. 

Compared to lighter divalent atoms such as Sr, this picture is complicated for Yb by a relatively high density of low-lying excited states in Yb$^+$, which give rise to doubly-excited states in the energy range of interest (\emph{e.g.}, $[\mathrm{Xe}] 4f^{14} 6p nl$). These doubly-excited states are part of additional Rydberg series converging to higher thresholds (Fig.~\ref{fig:174Yb_energydiagram_perturbers_pol}a), and couple to the singly-excited Rydberg states of interest through configuration interactions, which mix series with the same $J$ and parity. This alters the energy spectrum through level repulsion (\emph{i.e.}, series perturbers), which gives rise to sharp variations in $\mu$ with $n$, and correspondingly modifies the wavefunctions needed to compute quantities of interest such as matrix elements, interactions and lifetimes. This effect can be described quantitatively using MQDT, treating the interacting series as a set of coupled channels and parameterizing the interactions with a small number of mixing angles or a $K$-matrix~\cite{Seaton1983Quantum,Cooke1985Multichannel}. We give a pedagogical introduction of the MQDT approach in Appendix~\ref{sec:MQDT}. 

Previous work in $^{174}$Yb has reported spectroscopic measurements of certain $S$, $P$, $D$, and $F$ states, along with MQDT models that adequately reproduce their energy spectrum~\cite{Aymar1980Highly,Aymar1984Multichannel,Ali1999Two,Lehec2018Laser,Wilson2022Trapping} (a comprehensive summary can be found in  Ref.~\cite{Lehec2017PhD}). We improved these models by performing additional microwave spectroscopy of some $P$ Rydberg states between $n=30-50$ (using an atomic beam apparatus, App.~\ref{sec:atomicbeam}). The resulting MQDT model is summarized in Fig.~\ref{fig:174Yb_energydiagram_perturbers_pol}b.

However, mixing between channels converging to the same threshold (\textit{i.e.}, singlet-triplet mixing) does not alter the state energies \cite{Lu1971Spectroscopy}, but can significantly alter wavefunctions and matrix elements \cite{Vaillant2015Intercombination}. To develop an MQDT model with accurate wavefunctions, we augment the traditional approach of fitting the energy levels with additional measurements of the static dipole polarizability of the atomic states, which depend directly on the matrix elements and wavefunctions. 

In Fig.~\ref{fig:174Yb_energydiagram_perturbers_pol}c, we present measurements of the static dipole polarizability (red dots) for the $^1S_0$ series, near its crossing with $^3P_1$. By adjusting the single-triplet mixing angle in the $^{1,3}P_1$ MQDT model, we can accurately capture the resonance-like feature, which deviates strongly from the usual $\nu^7$ polarizability trend~\cite{Gallagher_1994}. The final model predicts a triplet character of the $^1P_1$ Rydberg series that varies between \SI{6.30(3)}{\percent} for $n=40$ and \SI{2.96(16)}{\percent} for $n=100$, which is consistent with previous estimates based on measurements of diamagnetic shifts \cite{Neukammer1984diamagnetic}, but considerably more precise (Appendix~\ref{sec:174_13P1}). As an additional check of the predicted MQDT wavefunctions, we characterize the scalar and tensor polarizability of the $n\,^3S_1$ Rydberg series near its crossing with $^3P_0$, finding excellent agreement (Fig.~\ref{fig:174Yb_energydiagram_perturbers_pol}d). The measured polarizabilities of $6sns\,^1S_0$ and $6sns\,^3S_1$ states are summarized in the supplemental material~\cite{supplementary}.

Through a similar analysis for the $^{1,3}D_2$ using literature data for the magnetic moments of these states~\cite{zerne1996lande} we find a triplet character of $n\,^1D_2$ Rydberg states of approximately \SI{15}{\percent} (details in Appendix~\ref{sec:174_13D2}). 

Other details of the spectroscopy are presented in Appendix~\ref{sec:atomicbeam}. A complete tabulation of the model parameters and the compiled spectroscopic data are presented in the supplemental material~\cite{supplementary}. A more detailed description of the developed MQDT models is presented in Appendix~\ref{sec:APP_174}.

\begin{figure*}[tb]
	\centering
    \includegraphics[width=\linewidth]{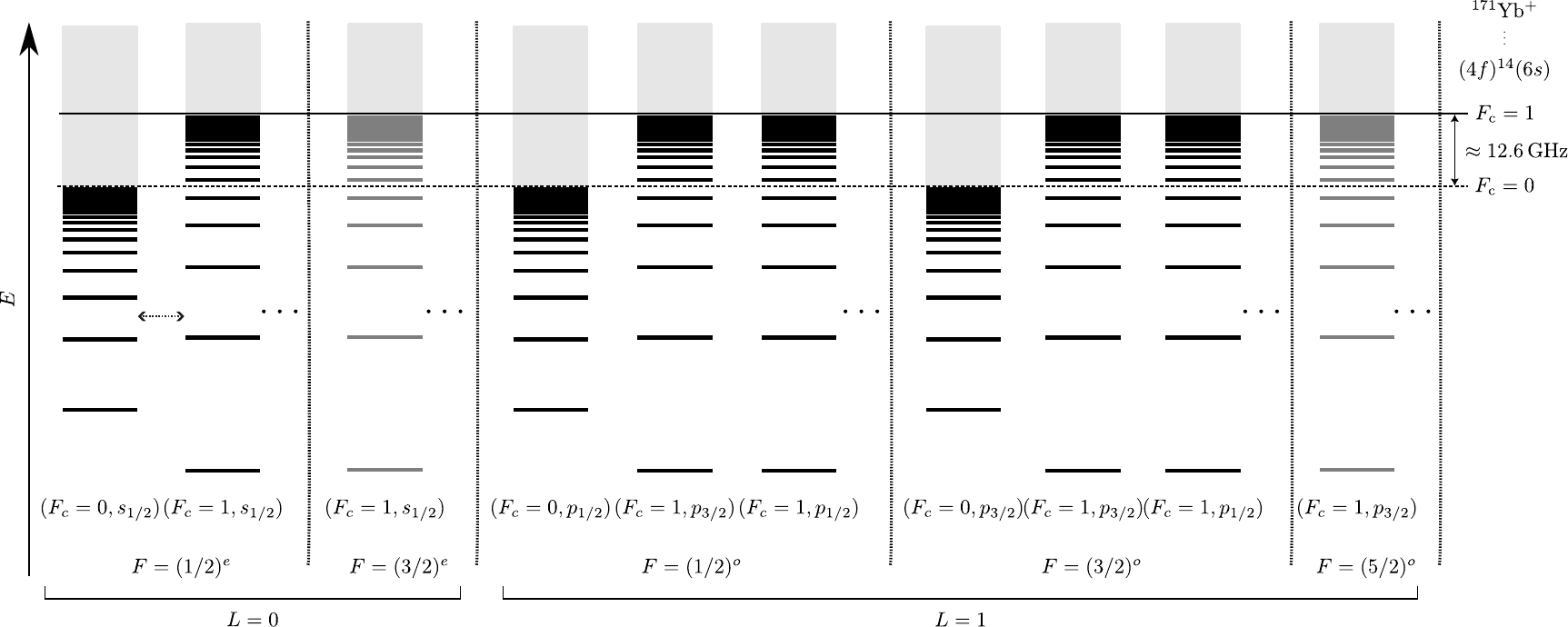}\caption{\label{fig:171Yb_energy_diagram_perturbers} Schematic energy level diagram of the $L=0$ and $L=1$ Rydberg states of $^{171}$Yb, indicating Rydberg series converging to the two hyperfine states of the $^{171}$Yb$^+$ ground state. Channels converging to electronically excited states of the ion core are not shown, but are included in the MQDT models. The series are labeled by their good quantum numbers $F$ and parity (denoted by superscript $e$, for even, and $o$ for odd).}
\end{figure*}

\section{MQDT model of $^{171}$Yb}
\label{sec:171Ybmain}

In contrast to $^{174}$Yb, $^{171}$Yb has a non-zero nuclear spin of $I=1/2$. The absence of hyperfine coupling in low-lying $J=0$ manifolds (\emph{i.e.}, $^1S_0$ or $^3P_0$) results in an ideal nuclear spin qubit~\cite{Barnes2022Assembly,Ma2022Universal,Jenkins2022Ytterbium}. Counterintuitively, the Rydberg states of $^{171}$Yb have a fairly strong hyperfine coupling, which arises from the interplay of hyperfine coupling in the Yb$^+$ core and the exchange interaction between the core and Rydberg electrons. This coupling is necessary to implement entangling gate operations on the pure nuclear spin qubit~\cite{Ma2022Universal,Ma2023High}, but also significantly complicates the description of the Rydberg series~\cite{Beigang1983Hyperfine,Sun1988Hyperfine,Ding2018Spectroscopy,Worner2003Multichannel}. We note that the role of hyperfine coupling in $^{171}$Yb qubits is almost opposite to alkali atoms, where the ground state has strong hyperfine coupling, while direct hyperfine coupling of the Rydberg state can be largely neglected~\cite{Saffman2016Quantum}.

In the context of MQDT, hyperfine effects can be represented as a coupling between multiple Rydberg series with the same parity and total angular momentum $F$, converging to different thresholds corresponding to the hyperfine-split ion core states (Fig.~\ref{fig:171Yb_energy_diagram_perturbers}). In the case of $^{171}$Yb$^+$, the $6s\,^2S_{1/2}$ ground state is split into two hyperfine states with total angular momentum $F_c=0$ and $F_c=1$, separated by $A_{HF}=12.642812$ GHz~\cite{Blatt1983precise}. For the range of $\nu$ relevant for quantum information applications ($40 \lesssim \nu \lesssim 100$), the hyperfine interaction energy is comparable to the spacing between principal quantum levels, resulting in strong channel mixing when there is more than one series with the same $F$ and parity~\cite{Liao1980Hyperfine}. For example, the $6sns$ states in $^{171}$Yb give rise to two coupled $F=1/2$ series converging to $F_c=0$ and $F_c=1$, and a single $F=3/2$ series converging to the $F_c=1$ threshold.

The effect of hyperfine-induced channel mixing has been experimentally studied previously in several species, including $^{87}$Sr \cite{Sun1988Hyperfine,Sun1989Multichannel,Sun1989Hyperfine,Ding2018Spectroscopy} and $^{135,137}$Ba \cite{neukammer1982hyperfine,Aymar1984Multichannel}. However, only a small number of measurements of $^{171}$Yb Rydberg states have been reported~\cite{Barbier1980Very,Majewski1983High,Majewski1985diploma,Ma2022Universal}.

\subsection{$L=0$ states}

\begin{figure}[tb]
	\centering
    \includegraphics[width=\linewidth]{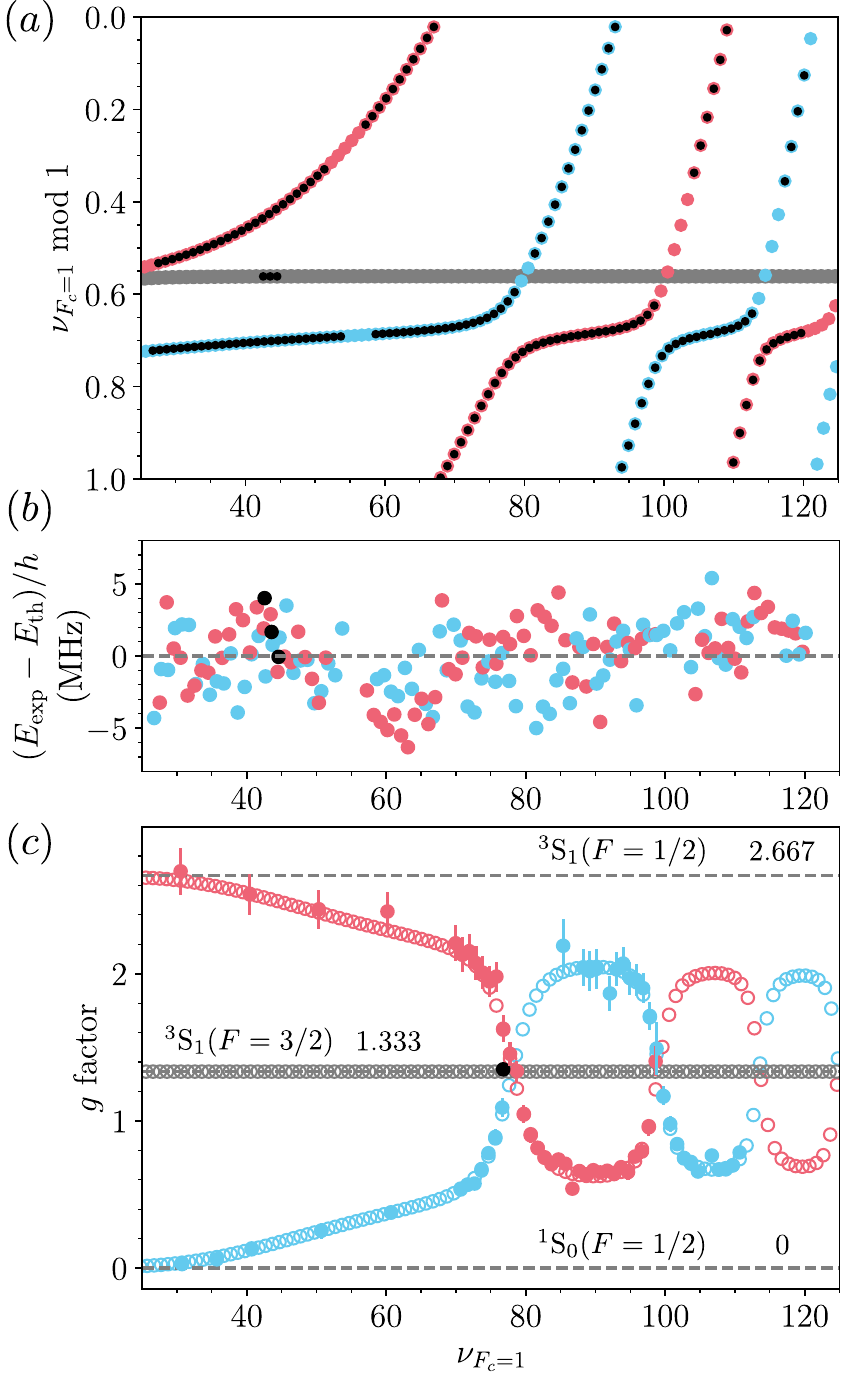}
	\caption{\label{fig:171Yb_Sstates_lufano_g} $(a)$ Lu-Fano-type plot of $^{171}$Yb $\ket{\nu,L=0,F=1/2}$ Rydberg states (red, blue) and $\ket{\nu,L=0,F=3/2}$ states (gray). The black points correspond to experimentally observed states, whereas the colored points denote bound states in the MDQT model. For $\nu<70$, we refer to the red states as ``triplet connected'' and the blue states as ``singlet connected'' based on their dominant character in $LS$ coupling. This description breaks down at higher principal quantum numbers. $(b)$ Deviation between measured and modeled bound state energies (red and blue for $F=1/2$ and black for $F=3/2$). $(c)$ Measured (filled circles, color code as in $(b)$) and predicted (empty circles) $g$-factors. The gray dashed lines indicate the Land\'e $g$ factors in pure $LS$ coupling.}
\end{figure}

There are three Rydberg series with $L=0$ in $^{171}$Yb (Fig.~\ref{fig:171Yb_energy_diagram_perturbers}). At low $\nu$, where hyperfine coupling is a small perturbation compared to the singlet-triplet splitting, these series are described in the $LS$ basis as a single $^1S_0$ series with $F=1/2$, and two $^3S_1$ series with $F=\{1/2,3/2\}$. Very close to the threshold, $jj$-coupling is more appropriate: one $F=1/2$ series results from adding the outer electron spin ($s=1/2$) to the $F_c=0$ ion core, while the other $F=1/2$ and $F=3/2$ series are connected to $F_c=1$. While these descriptions are equivalent for the $F=3/2$ series, the $F=1/2$ eigenstates cannot be simply described in either basis in between these limits, necessitating an MQDT model for these states.

We have experimentally measured the energy of most $L=0$ $F=1/2$ states with $25 < \nu < 120$ using laser spectroscopy as described in Appendix~\ref{sec:atomicbeam}. The measured $L=0$ Rydberg state energies and associated MQDT model predictions are shown in
Fig.~\ref{fig:171Yb_Sstates_lufano_g}a. The agreement is excellent: the $F=1/2$ states have a root-mean-squared deviation of \SI{2.3}{\mega\hertz} from the model, consistent with the $3\sigma$ uncertainty of \SI{10}{\mega\hertz} of the wavelength meter used to determine the laser frequencies.

In analogy to the Stark shift measurements for $^{174}$Yb in Fig.~\ref{fig:174Yb_energydiagram_perturbers_pol}c and d, we test the wavefunctions by comparing the measured and predicted magnetic moments of a subset of states, shown in Fig.~\ref{fig:171Yb_Sstates_lufano_g}c. For low $\nu$, the magnetic moments are close to the Land\'e $g$-factors for $LS$-coupled states, but they deviate significantly above $\nu = 40$.

The primary characteristic of the $F=1/2$ series is the avoided crossings in the energy spectrum. These can be understood from the channel structure of Fig.~\ref{fig:171Yb_energy_diagram_perturbers}: strong channel mixing and level repulsion occur when the separation between Rydberg levels is comparable to the ion core hyperfine splitting. For $^{171}$Yb, the $\Delta n = 1, 2$ and 3 level crossings occur at $\nu \approx 80,100$ and 115, respectively, though we note that the wavefunction character is already affected far below the first avoided crossing at $\nu = 80$. The physical mechanism for this mixing is the exchange interaction between the inner and outer electrons.

The coloring of the curves in Fig.~\ref{fig:171Yb_Sstates_lufano_g} is a guide to the eye. The transition from $LS$- to $jj$-coupled states makes it challenging to introduce an unambiguous partition of all of the $F=1/2$ states into two series. Throughout this paper, we identify specific states by $F$, $l$ and the effective principal quantum number $\nu$ specified to two decimal places, (\emph{e.g.}, $\ket{\nu=54.28, L=0, F=1/2, m_F=1/2}$). In this description, $L$ is the Rydberg electron angular momentum in the $6snl$ channels. In the specific case of the $F=1/2$ S states, we also find it convenient to refer to the states with $\nu < 70$ as being ``triplet-connected'' or ``singlet-connected'' based on their dominant character in $LS$ coupling.

Hyperfine coupling does not affect the $F=3/2$ series, which follows the same behavior as the $^3S_1$ series in $^{174}$Yb, converging to the $F_c=1$ ionization limit. We have measured the energies of several $F=3/2$ states to confirm this behavior (Fig.~\ref{fig:171Yb_Sstates_lufano_g}a).

The MQDT model parameters, tables of measured energy levels for S $F=1/2$ and S $F=3/2$, as well as Land\'e $g$ factors and values of DC polarizabilities of S $F=1/2$ Rydberg states are presented in the supplemental material~\cite{supplementary}.

\begin{figure*}[tb]
	\centering
    \includegraphics[width=\linewidth]{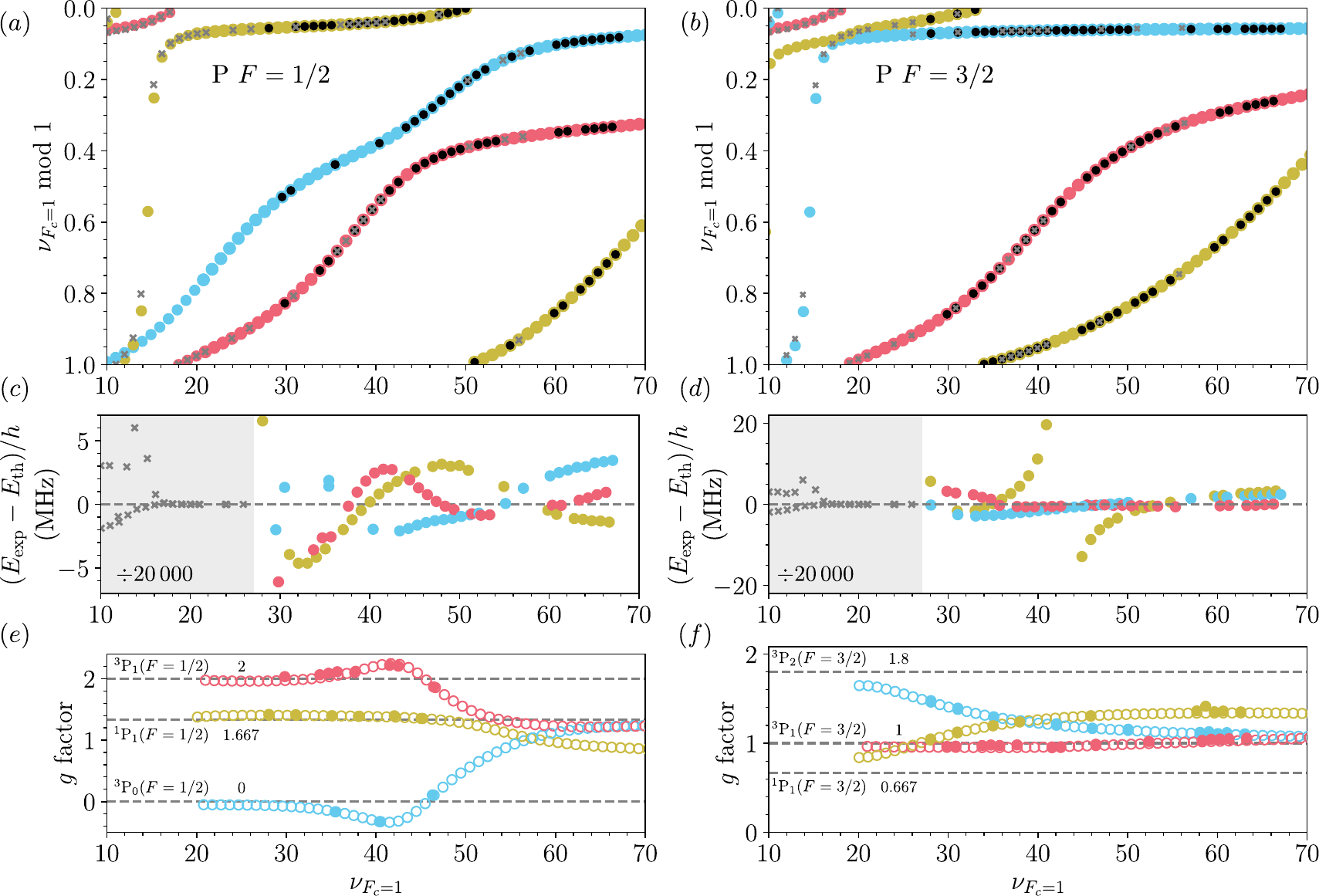}
	\caption{\label{fig:171Yb_Pstates_lufano_g} $(a)$ and $(b)$ Lu-Fano-type plot of $^{171}$Yb $\ket{\nu,L=1,F=1/2}$ and $\ket{\nu,L=1,F=3/2}$ Rydberg states, respectively. The black points correspond to experimentally observed states by microwave spectroscopy, as described in Appendix~\ref{sec:atomicbeam}. The gray crosses are three-photon laser spectroscopy data from Ref.~\cite{Majewski1985diploma}. The colored points denote the MQDT model bound states (the colors are chosen to guide the eye). $(c)$ and $(d)$ Deviation between measured and modeled bound state energies. The deviations of the laser spectroscopy measurements are divided by a factor of 20\,000, and are not shown for states that were also observed by microwave spectroscopy.  $(e)$ and $(f)$ Measured (filled circles) and predicted (empty circles) $g$-factors. The gray dashed lines indicate the Land\'e $g$ factors in pure $LS$ coupling.}
\end{figure*}

\subsection{$L=1$ states}

There are seven $L=1$ series in $^{171}$Yb. In $LS$ coupling, they can be described as $^3P_0 (F=1/2)$, $^1P_1 (F=\{1/2,3/2\})$, $^3P_1 (F=\{1/2,3/2\})$ and $^3P_2 (F=\{3/2,5/2\})$. In $^{174}$Yb, spin-orbit coupling mixes the two $J=1$ series as discussed in Section~\ref{sec:174Ybmain}; in $^{171}$Yb, hyperfine coupling additionally mixes the three $F=1/2$ series and the three $F=3/2$ series (Fig.~\ref{fig:171Yb_energy_diagram_perturbers}).

We have experimentally measured the energies of a number of $\ket{\nu,L=1,F=1/2}$ and $\ket{\nu,L=1,F=3/2}$ states, using microwave transitions from $\ket{\nu,L=0,F=1/2}$ states (for details, refer to Appendix~\ref{sec:atomicbeam}). The results are shown in Fig.~\ref{fig:171Yb_Pstates_lufano_g}, along with previously published measurements of low-$\nu$ states from three-photon laser spectroscopy~ \cite{Majewski1983High,Majewski1985diploma}. 

As in the case of the $\ket{\nu,L=0,F=1/2}$ series, we begin with the $^{174}$Yb MQDT models of $^{1,3}P_1$ and $^3P_0$ (for $F=1/2$) and $^{1,3}P_1$ and $^3P_2$ (for $F=3/2$) presented in Appendix~\ref{sec:APP_174} and refine the model by a global fit to the observed state energies of the $\ket{\nu,L=1,F=1/2}$ and $\ket{\nu,L=1,F=3/2}$ Rydberg states. A comparison of the experimental state energies and the MQDT model energies is presented in Lu-Fano-type plots in Fig.~\ref{fig:171Yb_Pstates_lufano_g}$\,(a)$ and $(b)$ for $F=1/2$ and $F=3/2$, respectively.

\begin{figure*}[tb]
	\centering
    \includegraphics[width=\linewidth]{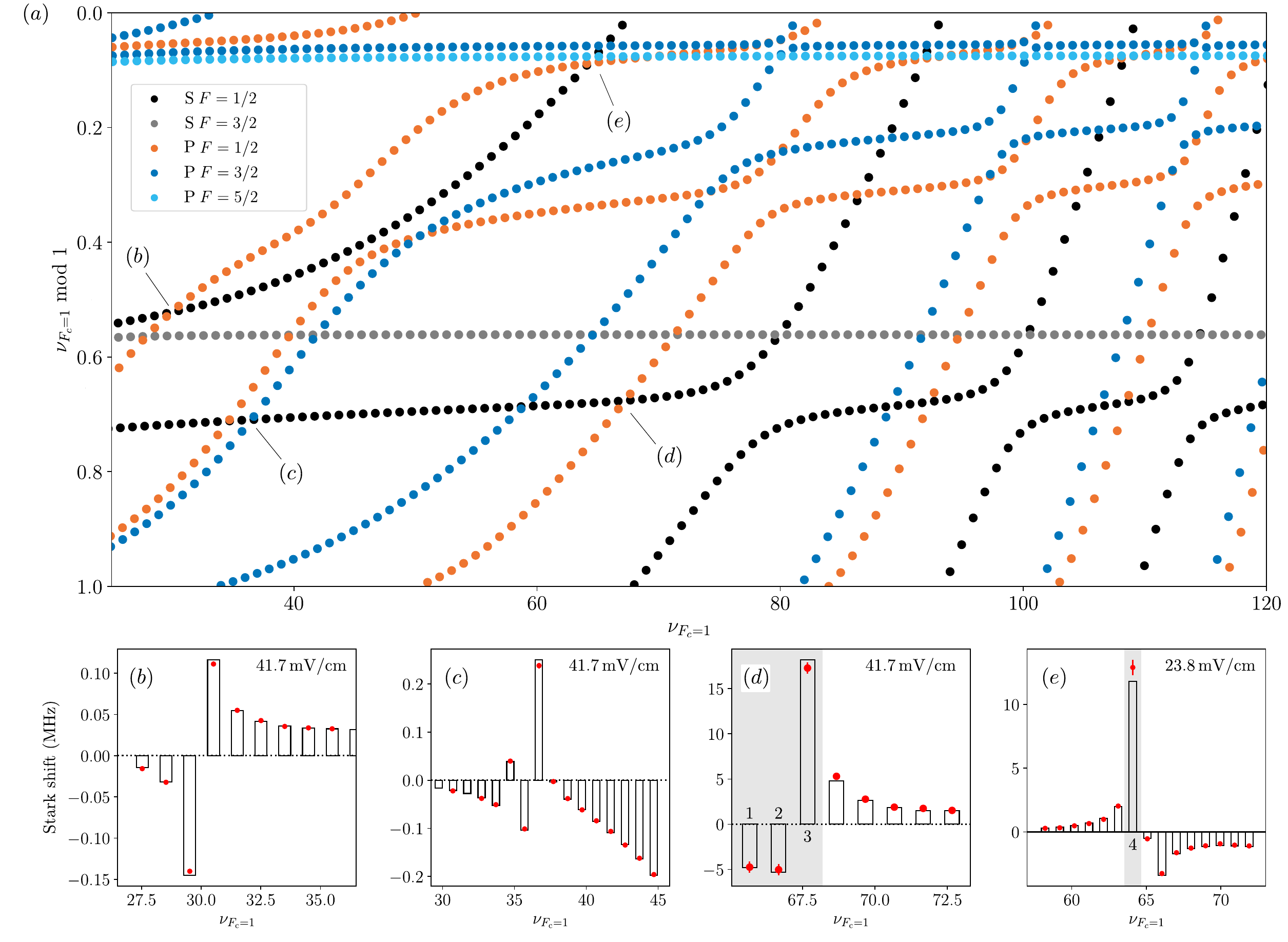}
	\caption{\label{fig:171Yb_LuFano_overview} ($a$) Combined plot summarizing the MQDT model energies for all $L=0$ and $L=1$ Rydberg series in $^{171}$Yb. ($b-e$) Measured (red) and predicted (black bars) Stark shifts of selected $L=0$ $F=1/2$ series in the vicinity of crossings with $L=1$ Rydberg series, at locations indicated by markers in panel ($a$). The Stark shift is reported at the electric field strength indicated in each panel. The gray shaded states correspond to near-degeneracies with significantly non-quadratic Stark shifts. A detailed comparison of the shifts of these states to a non-perturbative model is shown in Fig.~\ref{fig:171Yb_individual_Stark}. }
\end{figure*}

The behavior of the $L=1$ states is considerably more complex than the $L=0$ states, because the hyperfine coupling acts on states that were already strongly perturbed in $^{174}$Yb. 

We observe systematic deviations between the experimental and modeled energies. The high accuracy of the microwave measurements (\SI{100}{kHz}) makes it clear that the deviations are systematic as opposed to random. Particularly, the dispersion-like feature in the fit residuals observed near $\nu \approx43$ for $F=3/2$ suggests the existence of an additional perturbing Rydberg series. The perturbing Rydberg state could come from unaccounted hyperfine splitting of excited states of the ion core, or from interactions with odd-parity $L=3$ Rydberg channels. Based on DC polarizabilities of the measured P Rydberg states with $\nu<70$ and our ability to resolve Rydberg states with much larger DC polarizability, we conclude that the observed systematic deviations are not due to uncompensated stray electric fields. The  significant deviations for states with $\nu<16$ suggest an additional perturber in that energy range, as well.

To confirm the accuracy of the MQDT wavefunctions, we also compare the predicted magnetic moments to experimental measurements (Figs.~\ref{fig:171Yb_Pstates_lufano_g}~$(e)$ and $(f)$). As with the $L=0$ series, the moments align with the Land\'e $g$-factors for $LS$ coupling at low principal quantum number, and deviate significantly at large $\nu$ because of the combination of singlet-triplet mixing and hyperfine interaction. 

We have not directly measured any $^{171}$Yb $L=1$, $F=5/2$ states. In $LS$ coupling, this series can be described as $^3P_2\,(F=5/2)$ converging to the upper $F_c=1$ hyperfine threshold. We therefore model this series using the MQDT model for the $^3P_2$ series presented for $^{174}$Yb in the previous section. To account for isotope dependent effects, we use the MQDT model parameters as optimized from a fit to the observed $\ket{\nu,L=1,F=3/2}$ Rydberg states, which contain a contribution from the $^3P_2\,(F=3/2)$ states.

A summary of all measured $L=1$ state energies, Land\'e $g$ factors, and MQDT model parameters are presented in the supplemental material~\cite{supplementary}.

\subsection{Verification of MQDT models with Stark shifts}
\label{sec:171_starkshifts}

As with $^{174}$Yb, we probe the accuracy of Rydberg state energies and matrix elements obtained from our MQDT models by measuring the dc Stark shift of $\ket{\nu,L=0,F=1/2}$ states near degeneracies with $\ket{\nu,L=1,F=1/2}$ and $\ket{\nu,L=1,F=3/2}$ Rydberg states (Fig.~\ref{fig:171Yb_LuFano_overview}a). The measured Stark shifts are in excellent agreement with the theoretical predictions (Fig.~\ref{fig:171Yb_LuFano_overview}b--e), indicating good MQDT models for all relevant states, producing both accurate energies and channel contributions. Certain states with near-degenerate opposite parity states do not have quadratic Stark shifts even at very small electric fields, so we make the comparison between experiment and theory using the magnitude of the Stark shift at a particular field, instead of the usual static dipole polarizability. For these states, we also compare the measured field-dependent Stark shift with a non-perturbative calculation, and find excellent agreement (Fig.~\ref{fig:171Yb_individual_Stark} in App.~\ref{sec:APP_171Yb_Sstates}). The experimentally determined static dipole polarizabilities and Stark shifts are presented in the supplemental material~\cite{supplementary}. Predicted polarizability trends of the $L=0$ Rydberg states of both $^{171}$Yb and $^{174}$Yb Rydberg states are presented in Appendix~\ref{sec:polarizability}.

\section{Rydberg-Rydberg interactions in $^{171}$Yb}\label{sec:RydbergRydberg_Yb171}

A precise understanding of the interaction potential is important to realize high-fidelity gate operations. In this section, we use the developed MQDT model to predict the interaction potential for a pair of $^{171}$Yb atoms, then validate the model using direct experimental measurements in an optical tweezer array. The calculation builds on established techniques for performing similar calculations in alkali atoms, based on numerically diagonalizing the multipolar interaction Hamiltonian in a large basis of pair states~\cite{weber2017calculation,Sibalic2017ARC}. In the context of alkali atoms, this approach has been experimentally validated by spectroscopic measurements in optical tweezers~\cite{Beguin2013Direct,Ravets2015Measurement} and optical lattices~\cite{Hollerith2019Quantum}. We follow the formalism extending these techniques to states described by MQDT that was introduced previously in Refs.~\cite{Robicheaux2018theory,Robicheaux2019Calculations}.

\begin{figure}[tb]
	\centering
    \includegraphics[width=\linewidth]{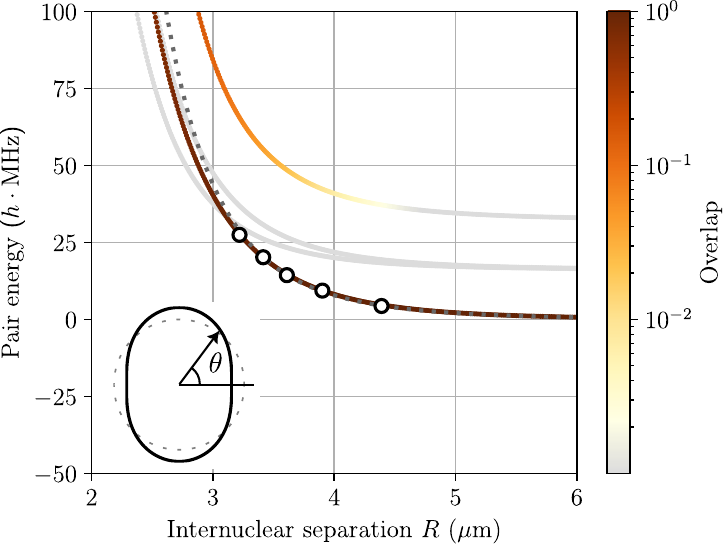}
	\caption{\label{fig:F12_54.28_interactions} Predicted pair-interaction potentials for a target Rydberg state $\ket{t} = \ket{54.28,L=0,F=1/2,m_F=-1/2}$, together with measured energy shifts (white circles). The magnetic field strength is \SI{4.88(6)}{G}, corresponding to a Zeeman splitting of \SI{16.1}{MHz}, and is oriented at $\theta = \uppi/2$ to the inter-atomic axis. The color of the curves denotes the overlap of each eigenstate with the target pair state $\ket{t}^{\otimes 2}$. The gray, dashed line shows the asymptotic $1/R^6$ scaling. \emph{Inset:} predicted angle-dependence of the $C_6$ coefficient. $\theta$ is the angle between the magnetic field and the inter-atomic axis.
 }
\end{figure}

\begin{figure*}[tb]
	\centering
    \includegraphics[width=\linewidth]{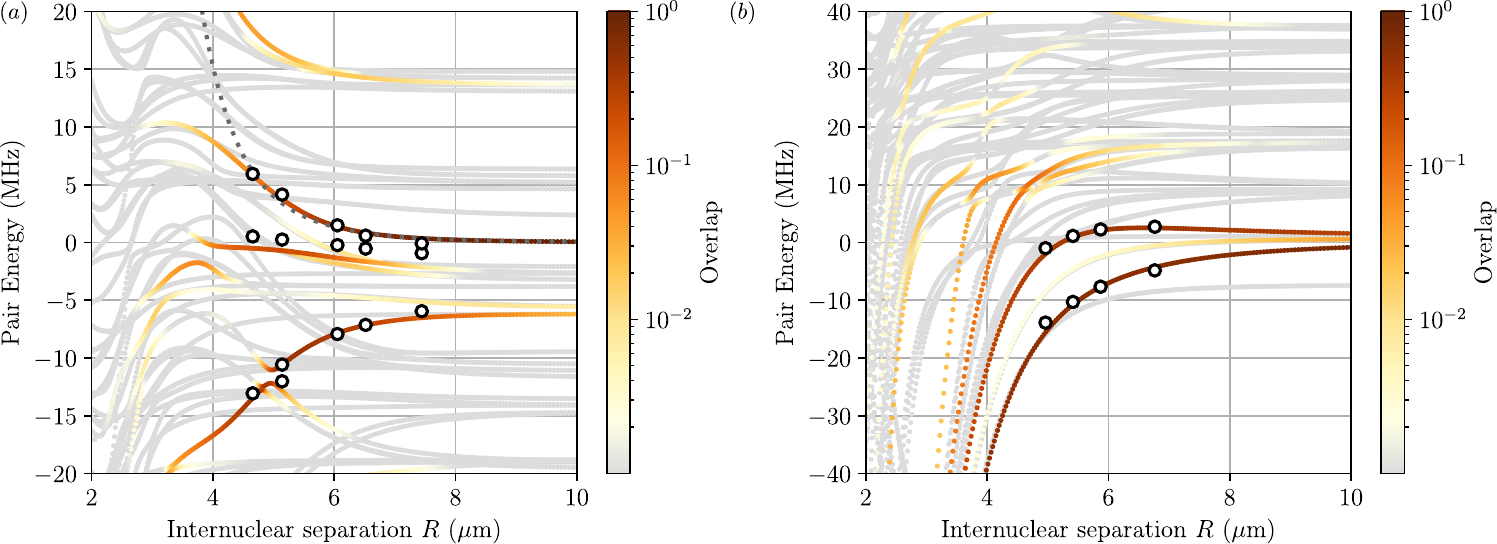}
	\caption{\label{fig:F32_54.56_interactions} Predicted pair-interaction potentials for a target state $\ket{54.56,L=0,F=3/2,m_F}$ with $(a)$ $m_F=3/2$ and $(b)$ $m_F=-3/2$, at a magnetic field of \SI{5.03(6)}{G} ($\theta=\uppi/2$). The color of the curves denotes the overlap with the target pair state. The white circles indicate experimentally observed resonances as described in the text. The gray dashed line in panel $(a)$ shows the asymptotic $1/R^6$ scaling.}
\end{figure*}

We first consider the interaction potential between a pair of atoms in a $\ket{\nu,L=0,F=1/2,m_F}$ state. For concreteness, we focus on the triplet-connected $\ket{54.28,L=0,F=1/2,m_F}$ states. The calculated pair potentials for the four combinations of $m_F$ sublevels is shown in Fig.~\ref{fig:F12_54.28_interactions}. The predicted pair potential closely follows the expected van der Waals form $V(R) = C_6/R^6$, with $C_6 \approx h \cdot \SI{34}{GHz(\mu m)^6}$. For comparison, the we note that the $C_6$ coefficient for the Rb $n S_{1/2}$ state with the most similar $\nu$ ($n=57$) is $C_6 \approx h \cdot\SI{76}{GHz(\mu m)^6}$~\cite{Sibalic2017ARC}.

We test the theoretical prediction by directly measuring the van der Waals shift using a pair of atoms in optical tweezers with separations from 3.3--$\SI{4.5}{\mu m}$. The measured energy shift is in excellent agreement with the predicted pair potential after scaling the experimental interatomic separation by a factor of 0.97 relative to the predicted separation (Fig.~\ref{fig:F12_54.28_interactions}), which we attribute to imperfect focusing of the optical system used to project the tweezer array. Additional details about the measurement technique are described in Appendix~\ref{sec:opticaltweezer}.

We note that there is a nearby $F=3/2$ $D$ state with a detuning of only \SI{+68}{MHz} at zero magnetic field. As this state is also laser-accessible from $^3P_0$, it could cause problems for blockade gates if its van der Waals interaction has the opposite sign from the target state, pushing it into resonance. We are unable to predict the $D$ state interactions, because we do not yet have an MQDT model for the $F$ states. However, we note that this problem can be avoided by using the next lowest triplet-connected $S$ state with $\nu=53.30$, where the nearest $D$ state is detuned by more than \SI{600}{MHz}.

The predicted interaction strength is anisotropic, where the $C_6$ coefficient is approximately \SI{46}{\percent} larger when the internuclear axis is aligned perpendicular to the magnetic field compared to a parallel alignment to the magnetic field (Fig.~\ref{fig:F12_54.28_interactions}, inset). In contrast, the the van der Waals interaction between alkali atoms in $S$ states is usually highly isotropic, because of the small spin-orbit coupling in the $P$ states participating in the interaction~\cite{vermersch2016anisotropy}. For example, the $C_6$ coefficient for the $50\,^2S_{1/2}$ state in Rb has an anisotropy of approximately \SI{1}{\percent}. The large anisotropy in this $^{171}$Yb state is  attributed to the large spin-orbit coupling in Yb (scaling as $Z^4$), and series perturbers that lift the degeneracy between $J$ manifolds (as shown in Fig.~\ref{fig:174Yb_energydiagram_perturbers_pol}b for the $^{174}$Yb $P$ series), which has the same effect as spin-orbit coupling. The anisotropy also results in off-diagonal $C_6$ interactions that mix different $m_F$ levels. As seen from the color scale in Fig.~\ref{fig:F12_54.28_interactions}, the $\ket{m_F=1/2,m_F=1/2}$ state acquires a significant $\ket{m_F=-1/2,m_F=-1/2}$ character once the van der Waals shift becomes comparable to the Zeeman splitting. This effect is suppressed if the interatomic spacing is parallel to the magnetic field, as a consequence of the dipole-dipole selection rules~\cite{vermersch2016anisotropy}.

We have computed the $C_6$ coefficient for all of the $F=1/2$ series, and summarized the results in Fig.~\ref{fig:171and174Yb_Interaction_overview_Sstates}. The $C_6$ coefficient for the triplet-connected $F=1/2$ series follows the expected $\nu^{11}$ scaling in the range $\nu \lesssim 65$. The singlet-connected $F=1/2$ series has a much smaller $C_6$ coefficient in this range, with the exception of isolated states with accidental hyperfine-induced F\"orster resonances. This is consistent with previous estimates of a small $C_6$ for the $^{174}$Yb $^1S_0$ series using a single-channel approximation~\cite{Vaillant2012Long}.

Next, we consider an $F=3/2$ target state. Specifically, we consider the state $\ket{54.56,L=0,F=3/2,m_F=+3/2}$, which was used to implement two-qubit gates in Ref.~\cite{Ma2023High}. The computed pair potential is shown in Fig.~\ref{fig:F32_54.56_interactions}a. Unlike the $F=1/2$ state shown in Fig.~\ref{fig:F12_54.28_interactions}, the pair potential for this $F=3/2$ state deviates strongly from the idealized $C_6/R^6$ van der Waals scaling, because of a previously unknown F\"orster resonance with a pair state made up of $L=1$ states with $F=3/2$ and $F=5/2$. At zero magnetic field, this pair state is detuned by only approximately \SI{3}{MHz} from the $\ket{54.56,L=0,F=3/2}^{\otimes 2}$ pair state. The complex series of crossings results from different magnetic sublevels being pushed into exact resonance by the interaction, in a finite magnetic field of $B\approx\SI{5}{G}$. Similar behavior is observed for the $m_F=-3/2$ states, though the resulting spectrum is slightly less complex as the Zeeman shift has the same sign as the van der Waals interaction.

We have also experimentally measured the interaction shifts for this state. We are able to identify all pair states from this complex spectrum with an overlap of more than \SI{10}{\percent} with the target state, and find good agreement for both $m_F=+3/2$ and $m_F=-3/2$ using the same rescaling factor of the interatomic separation (0.97) used in Fig.~\ref{fig:F12_54.28_interactions}. In the case of  $m_F=+3/2$, we observe deviations of 1-\SI{2}{MHz} for one of the eigenstates. This is comparable in magnitude to the fit residuals in the $F=3/2$ $P$ series MQDT model (Fig.~\ref{fig:171Yb_Pstates_lufano_g}; the $F=3/2$ state that contributes to the F\"orster resonance is in the blue-colored series in that figure).

A quantitative analysis of Rydberg interaction strengths of S $F=1/2$ and S $F=3/2$ Rydberg states is presented in App.~\ref{sec:interaction_trends}. 

\section{Improved two-qubit gates}\label{sec:improved_twoqubit}

The F\"orster resonance observed for the $\ket{54.56,L=0,F=3/2}$ state is not accidental, but rather a systematic trend: almost all $\ket{\nu,L=0,F=3/2}$ states with $\nu > 30$ have a F\"orster defect less than \SI{10}{MHz} (Fig.~\ref{fig:171Yb_pairdetuning_F12_F32}). While F\"orster resonances are sometimes sought to increase the strength of the Rydberg-Rydberg interaction at long range~\cite{Beterov2015Rydberg,Young2021Asymmetric}, such a near-degenerate resonance is problematic for two-qubit gates because it results in a large number of weakly allowed transitions near zero energy at short separations, allowing Rydberg excitation within the blockade radius~\cite{Dereviano2015Effects}. This effect is sometimes referred to as \emph{Rydberg spaghetti}. At the same time, F\"orster resonances increase the long-range tail of the interactions, preventing parallel implementation of gates in a qubit array~\cite{Levine2019Parallel}. Understanding the existence of this F\"orster resonance resolves several observations from  Ref.~\cite{Ma2023High}, including an unknown contribution to the error budget of about \SI{1}{\percent}, and the need to use a large separation ($\SI{43}{\mu m}$) between adjacent dimers to achieve the highest gate fidelity.

We now demonstrate improved gate performance using the $\ket{54.28,L=0,F=1/2}$ state. Previous demonstrations of entangling gates in $^{171}$Yb used $F=3/2$ states~\cite{Ma2022Universal,Ma2023High}. The highest reported fidelity is $\mathcal{F}=0.980(1)$, which is approximately \SI{1}{\percent} lower than the predicted fidelity of $\mathcal{F}=0.989$ based on error sources that were understood at the time~\cite{Ma2023High}. We conjecture the additional errors were the result of unwanted Rydberg excitation to nearby pair states (Fig.~\ref{fig:F32_54.56_interactions}), and that the $F=1/2$ state with a cleaner interaction potential will lead to higher fidelity.

\begin{figure}[tb]
	\centering
    \includegraphics[width=\linewidth]{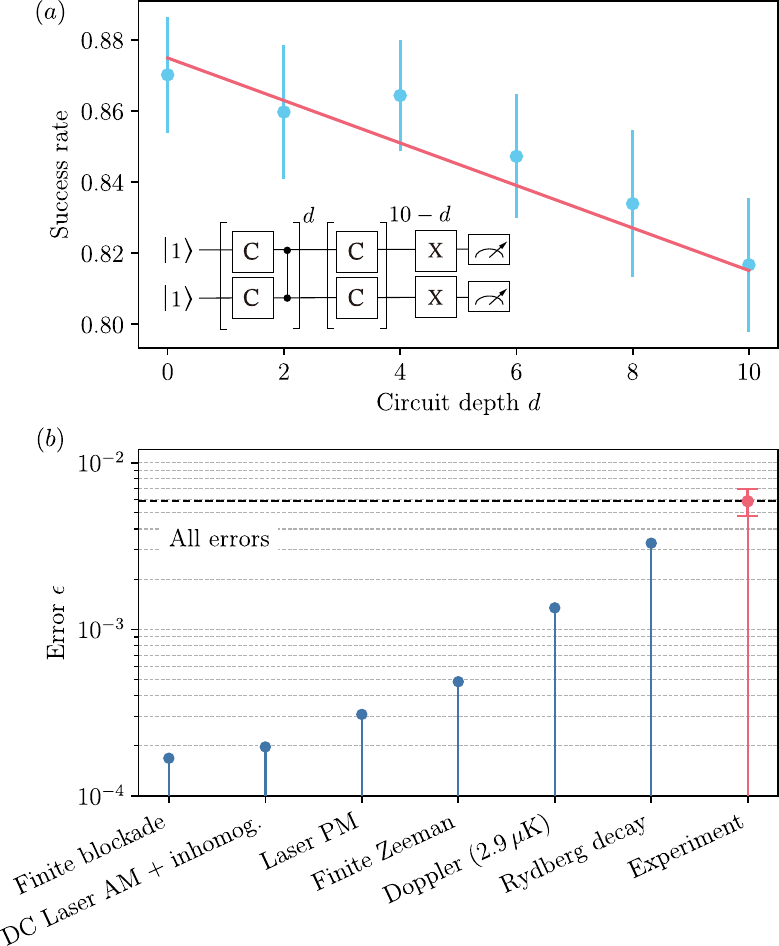}
	\caption{\label{fig:two_qubit_rb} $(a)$ Randomized circuit characterization of the time-optimal CZ gate, using a variable depth  $d$. The fitted error rate is $\epsilon=5.6(1.1)\times10^{-3}$ per two-qubit gate. $(b)$ Numerical simulation of contributions to the gate error, including the finite Rydberg state lifetime, Doppler shifts from atomic motion, off-resonant excitation of neighbouring $m_F$ sublevels in finite magnetic fields, shot-to-shot and site-to-site laser amplitude variations, finite Rydberg blockade, and fast laser phase noise (PM). The total simulated error rate with all sources applied simultaneously is $\epsilon=5.9\times10^{-3}$ (black dashed line), in good agreement with the experimentally measured error rate (red point).}
\end{figure}

We implement a two-qubit gate using the same approach as Ref.~\cite{Ma2023High}. Briefly, we prepare an array of four pairs of atoms (with a spacing of $d=\SI{2.4}{\mu m}$ between atoms within a pair, and $D=\SI{24}{\mu m}$ between pairs) and implement CZ gates in parallel using a variant of the time-optimal two-qubit gate~\cite{Jandura2022timeoptimaltwothree}. The gate is driven using a UV laser at \SI{302}{nm}, with a power of \SI{20}{mW} in a beam with a $1/e^2$ radius of $\SI{12}{\mu m}$ to achieve a Rabi frequency of $\Omega = 2\uppi \times \SI{2.5}{MHz}$ between $^3P_0$ $m_F=+1/2$ and Rydberg state with $m_F=-1/2$. Because of geometric constraints, the laser is linearly polarized perpendicular to the magnetic field, such that only half of the power contributes to the $\sigma^-$ transition that drives the gate. We estimate the fidelity using the randomized circuit characterization approach of Ref.~\cite{Evered2023High,Ma2023High}, with interleaved global single-qubit gates. With these parameters, we observe a CZ gate fidelity of $\mathcal{F}=0.994(1)$ (Fig.~\ref{fig:two_qubit_rb}).

The randomized circuit characterization involves varying the number of two-qubit gates $d$ while keeping the number of single-qubit gates and the overall sequence duration constant. This approach allows us to isolate errors specific to two-qubit operations from other sources of error in the circuit. Each randomized circuit is designed to bring the system to the final state $\ket{00}$. As a function of the circuit depth $d$, we experimentally characterize the success rate $P_{00}$ of the circuit. The experimental success rates are fitted to an exponential decay model $P_{00} = A \left(1-\epsilon\right)^ d$. Only two-qubit gate errors will contribute to $\epsilon$, while the offset at zero circuit depth $A \approx 0.87$ captures other sources of error in the circuit, including state preparation and measurement errors, and single-qubit gate errors \cite{Knill2008Randomized,Evered2023High,Ma2023High}.

This result improves on the previous best gate in $^{171}$Yb by a factor of 3.3. Importantly, the error budget for the gate in Fig.~\ref{fig:two_qubit_rb}a is now in excellent agreement with a model based on independently measured sources of error (Fig.~\ref{fig:two_qubit_rb}b) \cite{Ma2023High}. The dominant contributions are the finite lifetime of the Rydberg state (measured to be $T_r=\SI{56(4)}{\mu\second}$, contributing an error of $\epsilon=3.3\times10^{-3}$) and Doppler shifts ($\epsilon=1.4\times10^{-3}$ at an atomic temperature during the gate of $T = 2.9\,\mu$K). We also estimate the impact of other sources of error, including unwanted excitation of the other $m_F$ sublevel of the Rydberg state ($\epsilon = 4.8\times10^{-4}$), site-to-site and shot-to-shot laser intensity variation ($\epsilon =2.0\times10^{-4}$), a finite Rydberg blockade strength ($\epsilon =1.7\times10^{-4}$; the pulse is not compensated for the finite blockade~\cite{Levine2019Parallel,Jandura2022timeoptimaltwothree}), and fast laser phase noise ($\epsilon_\mathrm{PM} = 3.1\times10^{-4}$). Since the dominant errors can be suppressed by increasing the gate speed (\emph{i.e.}, with additional laser power), these results suggest that significantly higher gate fidelities are within reach for $^{171}$Yb.

\section{Discussion and conclusion}

We have presented detailed spectroscopy and modeling of the Rydberg states of both $^{174}$Yb and $^{171}$Yb with $L\leq 2$. The models are validated with experimental measurements of Stark shifts, magnetic moments and Rydberg interactions. To the best of our knowledge, this is the most comprehensive validation of an MQDT model for a complex atom, and allows key properties for Rydberg atom quantum computing and simulation to be predicted with the level of accuracy that is routine for alkali atoms. This will provide vital input for future experiments in quantum computing and simulation with Yb, or with Yb-alkali mixtures. Moreover, the calculation technique is a template for exploring other atoms in the lanthanide group, such as Ho or Er \cite{Hostetter2015Measurement,Trautmann2021Spectroscopy}.

We used this model to identify the likely cause of previously unattributed errors in entangling gates using $F=3/2$ Rydberg states. We also predicted a more suitable Rydberg state with $F=1/2$, leading to an improved gate fidelity of $\mathcal{F}=0.994(1)$, with errors reduced by a factor of 3.3 compared to the previous best gate demonstration using $^{171}$Yb~\cite{Ma2023High}.

We identify several avenues for future work. The first is including states with $L \geq 3$, which are needed to accurately predict the interactions of $D$ states, which can create blockade violations for $S$ states. The second is to extend the model to predict Rydberg state lifetimes and decay branching ratios, which will involve more careful fitting of MQDT models to low-$n$ perturbers and including matrix elements between different core electron states~\cite{Vaillant2014Multichannel}. While low-$n$ perturbers have little effect on the Rydberg character of high-$n$ states, they can have a large impact on the lifetime and decay branching ratio by providing a new decay pathway via the population of the perturbing state~\cite{bergstrom1986natural,Vaillant2014Multichannel}.

Another area of interest is the behavior of autoionizing states, of the form $6p_{1/2}nl$. These states have found use for Rydberg atom detection \cite{Lochead2013Number,Madjarov2020High}, coherent control \cite{Burgers2022Controlling,Pham2022Coherent} and in quantum error correction \cite{Wu2022Erasure}. To our knowledge, only the $^{174}$Yb $^3S_1$ autoionizing series has been characterized \cite{Burgers2022Controlling}. A systematic study, including $^{171}$Yb, would be beneficial for applications relying on use of these states.

Finally, we note that these models may have applications beyond quantum computing. Ytterbium clocks are among the most precise in the world~\cite{Ludlow2015Optical}, and Rydberg states can be used for generating entanglement to enhance precision~\cite{finkelstein2024universal,cao2024multiqubit}. It has also been proposed to use Rydberg states for \emph{in situ} absolute calibration of the temperature through the blackbody radiation spectrum~\cite{Ovsiannikov2011Rydberg}, which would warrant more precise validation of the modeled matrix elements.

\begin{acknowledgments}
We gratefully acknowledge helpful conversations with Adam Kaufman and Aruku Senoo, and Prof. Herbert Rinneberg for providing a copy of Ref.~\cite{Majewski1985diploma}.
This work was supported by the Army Research Office (W911NF-1810215), the Office of Naval Research (N00014-20-1-2426), DARPA ONISQ (W911NF-20-10021), the National Science Foundation (QLCI grant OMA-2120757, and NSF CAREER PHY-2047620), the Sloan Foundation and the Gordon and Betty Moore Foundation (Grant DOI 10.37807/gbmf12253).

\end{acknowledgments}

\appendix
\section{MQDT formalism}\label{sec:MQDT}

Energies and wavefunctions of Rydberg states are crucial for evaluating state properties ($e.g.$, Stark shifts \cite{zimmerman1979stark}). Several open-source programs have been developed for non-perturbative calculations of Stark shifts and interaction potentials for alkali metal atoms \cite{Sibalic2017ARC,weber2017calculation}, or alkaline earth atoms within a single channel quantum defect approximation~\cite{Robertson2021ARC}. However, for divalent atoms, a multichannel quantum defect theory (MQDT) treatment is necessary to obtain accurate Rydberg state energies and wavefunctions.

MQDT was introduced by Seaton~\cite{Seaton1966Quantum} and reformulated with the concept of frame transformations by Fano~\cite{Fano1970Quantum} and is established as a powerful framework for analyzing the Rydberg states of atoms \cite{Lu1971Spectroscopy,Lee1973Spectroscopy,Aymar1981Theoretical,Worner2003Multichannel,Schafer2010Millimeter,Beigang1982one,neukammer1982hyperfine,Rinneberg1983Hyperfine,Aymar1984Rydberg,Vaillant2014Multichannel,Bhattacharyya2007Odd,Hostetter2015Measurement,Trautmann2021Spectroscopy} and molecules \cite{Fano1970Quantum,Osterwalder2004high,Sprecher2014determination} with low-lying core-excited states. 

Calculations based on MQDT models predicting DC polarizabilities and Rydberg-Rydberg interaction for complex atoms have been presented previously \cite{Vaillant2015Intercombination,Robicheaux2018theory,Robicheaux2019Calculations}, but the conceptual complexity and lack of comprehensive MQDT characterization of atomic species has limited the broad adoption of this approach.

This appendix is intended to serve as a comprehensive introduction and overview to this approach, which will be useful beyond the specific case of Yb atoms. We have also made the software used to perform the computations in this work available as an open-source package, \texttt{rydcalc}~\cite{rydcalcgit}. 

\subsection{MQDT model}\label{sec>MQDT_model}

MQDT treats the short and long-range interactions of a single active electron with different configurations of the ion core. Each unique combination of angular quantum numbers of the electron--ion-core system is referred to as a channel. When the active electron is far away from the ion core, the system is well described by ``collision'' channels indexed by $i$. In this region, the interaction between the ion core and the active electron is dominated by the Coulomb interaction and the system is appropriately described in $jj$-coupling. 

The energy $E$ can be expressed with respect to the ionization limit of the $i^\mathrm{th}$ collision channel $I_i$ as 

\begin{equation}
    E = I_i - \left(hcR_M\right)/\nu_i^2\,
\end{equation}

where $\nu_i$ is the effective principal quantum number with respect to $I_i$ and $R_M=R_\infty\left(1-m_e/M\right)$ is the mass-reduced Rydberg constant, where $m_e$ is the mass of the electron and $M$ is the atomic mass.

When the active electron is close to the ion core, their interaction is dominated by non-Coulombic electrostatic interactions. Channels describing the angular momentum coupling (typically $LS$ coupling) in this regime more appropriately are called ``close-coupling'' channels indexed by $\alpha$. In this region, scattering between the electron and the ion core mixes $i$ channels with the same total angular momentum $J$ (or $F$ in the presence of a non-zero nuclear spin) and parity. The wavefunction of channel $i$ can therefore be expressed as a superposition of $\alpha$ channel wavefunctions with amplitude $\tilde{A}_\alpha$. At large electron--ion-core separations, the wavefunction at energy $E$ can be written as \cite{Lu1971Spectroscopy}

\begin{equation}
   \begin{split}
    \ket{\Psi} = \sum_i \ket{\Phi_i} \left[f(\nu_i,r)\sum_\alpha U_{i\alpha}\mathrm{cos}\left(\uppi\mu_\alpha\right)\right.&\tilde{A}_\alpha  \\ 
     - g(\nu_i,r)\sum_\alpha U_{i\alpha}\mathrm{sin}\left(\uppi\mu_\alpha\right)&\left.\tilde{A}_\alpha \right]\,,
   \end{split} \label{eq:mqdtwavefunc}
\end{equation}

where the terms in the square brackets correspond to the radial part of the active electron's wavefunction, and the spin and orbital angular momentum couplings of the outer electron and the ion core wavefunction are given by $\ket{\Phi_i}$. $f(\nu_i,r)$ and $g(\nu_i,r)$ correspond to the regular and irregular Coulomb wavefunctions \cite{Fano1970Quantum}. The short-range, non-Coulombic interaction of the Rydberg electron and the ion core are encoded in eigenchannel quantum defects $\mu_\alpha$ and the channels are coupled by the unitary transformation matrix $U_{i\alpha}$. 

For discrete bound states, the wavefunction Eq.~(\ref{eq:mqdtwavefunc}) must remain finite at large separation, resulting in the boundary condition~\cite{Fano1970Quantum,Lu1971Spectroscopy}:

\begin{equation}
    \sum_\alpha \tilde{A}_\alpha U_{i\alpha}\mathrm{sin}\left[\uppi\left(\nu_i+\mu_\alpha\right)\right]=0\label{eq:mqdt_der}
\end{equation}

This gives non-trivial solutions for the bound state energies $\nu_i$ when $\det|F_{i\alpha}|=0$ is zero, where $F_{i\alpha}$ is given by:

\begin{equation}\label{eq:detMQDT}
F_{i\alpha} = U_{i\alpha}\sin\left[\uppi\left(\mu_\alpha+\nu_i\right)\right]\,.
\end{equation}

Bound states are found at the intersection of the surface spanned by Eq.~(\ref{eq:detMQDT}) and the curve:

\begin{equation}\label{eq:nui_nuj}
    \nu_i=\left(\frac{I_i-I_j}{R_M}+\frac{1}{\nu_j^2}\right)^{-1/2}\,,
\end{equation}

where $\nu_i$ and $\nu_j$ are effective principal quantum numbers relative to the $i^\mathrm{th}$ and $j^\mathrm{th}$ ionization limit $I_i$ and $I_j$, respectively.

The MQDT model is fully specified by the unitary transformation matrix $U_{i\alpha}$ and the eigenchannel quantum defects $\mu_\alpha$. The task in determining accurate MQDT models therefore reduces to finding values of $U_{i\alpha}$ and $\mu_\alpha$, which reproduce experimental observables such as state energies, Land\'e $g$-factors or DC polarizabilities. 

\subsection{MQDT model parameters}\label{sec:mqdt_model_params}

The transformation matrix $U_{i\alpha}$ is orthogonal and can therefore be constructed from at most $N(N-1)/2$ rotation matrices, in the case of an $N$ channel model \cite{Lee1973Spectroscopy}. In practice, the mixing between channels is small and a suitable approximation can be made by composing significantly fewer rotation matrices. As noted above, the natural basis at long range is $jj$-coupled, while at short range it is $LS$-coupled. We can therefore simplify $U_{i\alpha}$ by introducing an intermediate basis $\bar\alpha$ of purely $LS$ coupled channels, and writing $U_{i\alpha}$ as the product of the $jj$--$LS$ transformation matrix $U_{i\bar\alpha}$ and a second matrix $V_{\bar\alpha\alpha}$ representing mixing between $LS$-coupled channels \cite{Lee1973Spectroscopy,Greene1991Spin}:

\begin{equation}
    U_{i\alpha} = U_{i\bar\alpha}V_{\bar\alpha\alpha}\,,
\end{equation}

where $U_{i\bar\alpha}$ corresponds to the $LS$-$jj$ frame transformation

\begin{align}
    \label{Uialpha}
    U_{i\bar\alpha} =& \langle((S_cL_c)J_c(s\ell )j)J |((S_cs)S(L_c\ell )L)J\rangle \nonumber\\
    =& [S,L,J_c,j]^{1/2}\begin{Bmatrix}
S_c & L_c & J_c\\
s & \ell & j\\
S & L & J
\end{Bmatrix}
\end{align}
from Eq.~(6.4.2) of Ref.~\cite{edmonds1996angular} and $[a,b,\dots]^{1/2}=\sqrt{(2a+1)(2b+1)\dots}$.

The second matrix $V_{\bar\alpha\alpha}$ accounts for configuration interaction by introducing couplings between the $\bar\alpha$ channels. $V_{\bar\alpha\alpha}$ is typically expressed as a series of rotations by Euler angles $\theta_{ij}$ around channels $i$ and $j$ (see Eq.~(15) of Ref.~\cite{Robaux1982Program})

\begin{equation}
\label{eq:Valpha}
    V_{\bar{\alpha}\alpha}=\prod \mathcal{R}(\theta_{ij})\,.
\end{equation}

If the $\theta_{ij}$ are small, the sequence of the rotations in Eq.~(\ref{eq:Valpha}) is not critical, however, in general $U_{i\alpha}$ will be sensitive to the chosen order.

In general, even in the absence of multi-channel interactions, the eigenchannel quantum defect is energy dependent. The energy dependence originates for example from a polarization of the ion core by the outer electron \cite{Freeman1976Core}. In practice, the eigenchannel quantum defect only slowly varies with energy and we treat energy-dependence of the eigenchannel quantum defects as

\begin{equation}
    \mu_\alpha(\epsilon) = \mu_\alpha^{(0)} + \epsilon\mu_\alpha^{(2)}+\epsilon^2\mu_\alpha^{(4)}\dots\,
\end{equation}

 and similarly we treat the energy dependence of  $\theta_{ij}$ by

\begin{equation}\label{eq:theta_energydep}
    \theta_{ij}(\epsilon) = \theta_{ij}^{(0)} + \epsilon\theta_{ij}^{(2)}\dots\,,
\end{equation}

where $\epsilon=1/\nu^2$ \cite{Lee1973Spectroscopy}. 

\subsection{MQDT models for hyperfine isotopes}

For isotopes with non-zero nuclear spin, the hyperfine interaction in the ion core has to be considered. Here, we follow a similar formalism introduced in earlier work \cite{Sun1989Multichannel,Sun1989Hyperfine,Worner2003Multichannel}. 

In the close-coupling region, the hyperfine interaction in the ion core is small compared to the exchange interaction and can be treated as perturbation to the fine structure. The close-coupling $\alpha$ channels are split into $\alpha_F = |(((S_cs)S(L_c\ell )L)JI)F\rangle$ channels, where the MQDT parameters $\mu_\alpha\approx\mu_{\alpha_F}$ and $U_{i\alpha}\approx U_{i\alpha_F}$ are only slightly affected. Therefore, we can construct the MQDT models for isotopes with non-zero nuclear spin from MQDT models obtained from even isotopes with nuclear spin $I=0$. In this work, we use the MQDT models and parameters obtained for $^{174}$Yb and introduce hyperfine coupling of the ion core with  

\begin{equation}
    U_{i_F,\alpha_F} = U_{i_F,i}U_{i,\alpha_F}
\end{equation}

where 

\begin{equation}
    U_{i,\alpha_F} = U_{i\bar\alpha}V_{\bar\alpha\alpha_F}\,,
\end{equation}

and $U_{i_F,i}$ is the frame transformation

\begin{align}
U_{i_F,i} &=\langle (((J_cI)F_cj)F|(J_cj)JI)F\rangle \nonumber\\
&=(-1)^{I+j+F_c+J}[J,F_c]^{1/2}
\begin{Bmatrix}
I & J_c & F_c\\
j & F & J
\end{Bmatrix}\,,
\end{align}

where $J$ corresponds to the total electronic angular momentum, $I$ is the nuclear spin, $F$ is the total angular momentum including nuclear spin, which uses Eq.~(6.4.2) of Ref.~\cite{edmonds1996angular}.

We find that directly applying the $^{174}$Yb MQDT parameters (\textit{i.e.} eigenchannel quantum defects $\mu_\alpha$ and rotations $\theta_{ij}$) to the $^{171}$Yb energies generally gives good agreement, but we find that a slight re-optimization of the model parameters is necessary for an improved modeling. We attribute this variation to three main reasons. First, a small isotope dependence \cite{Robicheaux2019Calculations} and hyperfine-induced mixing of channels can yield variations in the close-coupling parameters. Second, while we do not explicitly include direct coupling of the outer electron with the nucleus, this interaction can be expressed through the matrix elements of $V_{\bar\alpha \alpha}$ and $\mu_\alpha$. Therefore, to the extent that this interaction is significant, it can be incorporated into the model. We note that the magnitude of this direct coupling scales as $\nu^{-3}$ \cite{Gallagher_1994} and is at the scale of \SI{1}{MHz} at $\nu=30$, which is smaller than the current disagreement between the MQDT model and the experimentally measured state energies ($i.e.$, Fig.\ref{fig:171Yb_Pstates_lufano_g}). Therefore, attempting to predict this interaction strength and including it by hand would not significantly improve the model prediction. Third, we only consider the hyperfine splitting in the $6s\,{^2S_{1/2}}$ ground state of the $^{171}$Yb$^+$ ion. We neglect the hyperfine splitting of excited states of the $^{171}$Yb$^+$-ion core. This is likely a good approximation because the excited states are energetically far above the channels converging to the $^{171}$Yb$^+$ ground state considered in this work, but could lead to changes in the MQDT parameters.

\subsection{MQDT wavefunctions}

Once a bound state $b$ has been found with Eq.~(\ref{eq:detMQDT}) and Eq.~(\ref{eq:nui_nuj}), its wavefunctions are conveniently expressed in the $jj$ basis and Eq.~(\ref{eq:mqdtwavefunc}) is rewritten as

\begin{equation}
    \ket{\psi_b} = \ket{\Psi_b}/N_b = \sum_i \ket{\Phi_i} P_{\nu_{i,b}l_i}(r)A_{i,b}\,,\label{eq:mqdtwave_norm}
\end{equation}

where $P_{\nu_{i,b},l_i}(r)$ is the radial Coulomb function of the active electron, $A_{i,b}$ are the normalized channel contributions in terms of collision channels $i$ in $jj$-coupling \cite{Lee1973Spectroscopy}
\begin{equation}\label{eq:channelfractions_jj}
    A_{i,b}=(-1)^{l_i+1}\left(\nu_{i,b}\right)^{3/2}\sum_\alpha U_{i\alpha}\cos\left[\uppi\left(\nu_{i,b}+\mu_\alpha\right)\right]\tilde{A}_{\alpha,b}/ N_b\,,
\end{equation}

and $N_b$ ensures the normalization \cite{Seaton1966Quantum} of the wavefunction and is given by \cite{Lee1973Spectroscopy}

\begin{equation}
\begin{split}
    N_b^2 =& \sum_{i,\alpha} \nu_{i,b}^3U_{i\alpha}\cos\left[\uppi\left(\nu_{i,b}+\mu_\alpha\right)\right]\tilde A_{\alpha,b}\\
    &+\sum_\alpha \left(\frac{\mathrm{d}\mu_\alpha}{\mathrm{d}E}\right)\tilde A_{\alpha,b}^2\\
    &+\frac{1}{\uppi}\sum_{i,\alpha,\beta}\left(\frac{\mathrm{d}U_{i,\alpha}}{\mathrm{d}E}\right)U_{i,\beta}\sin\left[\uppi\left(\mu_\alpha-\mu_\beta\right)\right]\tilde A_{\alpha,b}\tilde A_{\beta,b}\,.     
\end{split}
\end{equation}

The wavefunction coefficients $\tilde{A}_{\alpha,b}$ of Eq.~(\ref{eq:mqdtwavefunc}) are obtained by evaluating

\begin{equation}\label{eq:cofactors_Aalpha}
    \tilde{A}_{\alpha,b} = C_{i\alpha} \bigg/ \left[\sum_\alpha C_{i\alpha}^2\right]^{1/2}\,,
\end{equation}
where $C_{i\alpha}$ is the cofactor of the $i^\mathrm{th}$ row and $\alpha^\mathrm{th}$ column of the matrix $F_{i\alpha}$ (see Eq.~(\ref{eq:detMQDT})), and in the evaluation of Eq.~(\ref{eq:cofactors_Aalpha}), channel $i$ can be chosen for convenience \cite{Lee1973Spectroscopy}. 

The close-coupling $\alpha$ channel fractions ($LS$ coupling) can be obtained from $A_i$ as:

\begin{equation}\label{eq:channelfractions_LS}
    A_{\bar\alpha,b}=\sum_i U_{i\bar\alpha}A_{i,b}\,.
\end{equation}

\subsection{Evaluation of matrix elements}

We evaluate the matrix elements following the procedures introduced in Ref.~\cite{Robicheaux2018theory}, which are summarized in the following.

Single atom matrix elements between state $b$ and $b'$ of operator $\hat{\zeta}$ acting on the channel function $\ket{\Phi_i}$ in Eq.~(\ref{eq:mqdtwave_norm}) that leaving the orbital angular momentum $l$ of the Rydberg electron unchanged are evaluated by

\begin{equation}\label{eq:mat}
    \zeta_{b,b'} = \sum_{i,i'}\left(A^T\right)_{b,i} \langle \Phi_i|\hat\zeta|\Phi_{i'}\rangle O_{ib,i'b'}A_{i',b'}\,
\end{equation}

where 

\begin{equation}
    O_{ib,i'b'} = \int_0^\infty P_{\nu_{i,b},l_i}(r) P_{\nu_{i',b'},l_{i'}}(r)\,dr
\end{equation}

is a radial overlap integral. The overlap integral can either be evaluated using analytic approximations \cite{Bhatti1981Analysis,Robicheaux2018theory} or numerical wavefunctions obtained from the Numerov algorithm \cite{zimmerman1979stark,Bhatti1981Analysis}.

This form of matrix elements is used in the evaluation of magnetic moments in the paramagnetic interaction Hamiltonian

\begin{equation}\label{eq:paramagnetic}
    H_\mathrm{PM} = -\vec{\mu }\cdot\vec{B} = \{\mu_B [\vec{L}_c + \vec{\ell}+g_s(\vec{S}_c+\vec{s})]-\mu_I\vec{I}\}\cdot \vec{B}
\end{equation}

using the expressions presented in Ref.~\cite{Robicheaux2018theory} for the general case of Rydberg states with hyperfine interaction in the ion core, where $\mu_B$ is the Bohr magneton, $\mu_I$ is the nuclear magnetic moment, and $g_s$ is the electron spin $g$-factor.

For the case of the multipole operator $\hat{\mathcal{Q}}^{(kq)}=r^kY_{k,q}(\Omega)$, the contribution of the core electrons is neglected and only the contribution of the Rydberg electron is considered (because of its much larger spatial extent):

\begin{equation}\label{eq:multipole}
    \mathcal{Q}_{b,b'}^{(kq)} = \sum_{i,i'} \left(A^T\right)_{b,i}\langle\Phi_i|Y_{kq}|\Phi_{i'}\rangle R_{ib,i'b'}^{(k)}A_{i',b}\,,
\end{equation}

where $R_{ib,i'b'}^{(k)}$ is the radial integral

\begin{equation}
\label{eq:radial_multipole}
    R_{ib,i'b'}^{(k)}=\int r^k P_{\nu_{i,b},l_i}(r) P_{\nu_{i',b'},l_{i'}}(r)\,dr\,.
\end{equation}

Matrix elements of the form presented in Eq.~(\ref{eq:multipole}), are used in the calculation of Stark shifts \cite{zimmerman1979stark}, Rydberg-Rydberg interactions \cite{Deiglmayr2016long} and the diamagnetic shift.

The diamagnetic interaction Hamiltonian \cite{Jenkins1939The} is given by

\begin{equation}\label{eq:diamagnetic}
    H_\mathrm{DM} = \frac{1}{8m_e}\left|\vec d\times\vec B\right|^2\,,
\end{equation}

and can be expressed in terms of spherical harmonics (compare, e.g., Ref~\cite{weber2017calculation})
 
\begin{equation}
    H_\mathrm{DM}  = \frac{e^2}{12m_e}{r}^2\sqrt{4\uppi}\left(Y_{0,0}-\sqrt{\frac{1}{5}} Y_{2,0}\right)B^2\,.
\end{equation}

Due to the multichannel nature of MQDT eigenstates, calculating matrix elements between two MQDT eigenstates as presented in Eq.~(\ref{eq:mat}) and Eq.~(\ref{eq:multipole}) requires calculating radial and angular integrals for all combinations of the electronic configurations encoded in the channel wavefunctions. We note that if a state has contributions from channels with significantly different ionization limits, the radial Coulomb functions $P_{\nu_{i,b},l_i}(r)$ of each channel will vary significantly in spatial extent because of the $\nu_{i,b}^2$ scaling of the orbital radius of the active electron. In practice, we neglect the contribution of channels with ionization limits significantly larger than the first ionization limit, since the associated wavefunctions are much more compact than those belonging to the lower limit and have comparatively small contributions to matrix elements of the form $r^k$. The only case in which radial wavefunctions corresponding to different thresholds are considered is in the case of the hyperfine-split thresholds in $^{171}$Yb.

\subsection{Numerical evaluation}

We compute state polarizabilities and interaction potentials following numerical techniques established for alkali atoms~\cite{zimmerman1979stark,Singer2005Spectroscopy,weber2017calculation,Sibalic2017ARC}, but adapted to compute matrix elements of MQDT eigenstates as described in the previous section. We review the approach here to highlight important features.

The energy of an atom in an electric and magnetic fields $F$ and $B$ is given by:

\begin{equation}
\label{eq:Ham_e}
    H = H_0 + F e r + H_\mathrm{PM} + H_\mathrm{DM}
\end{equation}

where $H_0$ encodes the energy in the absence of the field, and $H_\mathrm{PM}$ and $H_\mathrm{DM}$ are defined in Eqs.~\eqref{eq:paramagnetic} and ~\eqref{eq:diamagnetic}, respectively. To determine the energy shift for a target state $\ket{\psi_0}$, we numerically evaluate the matrix elements of Eq.~\eqref{eq:Ham_e} in a large basis of states $\{\ket{\psi_i}\}$, and then diagonalize~\cite{zimmerman1979stark}.

The matrix elements $\bra{\psi_i}r\ket{\psi_i'}$ and $\bra{\psi_i}r^2\ket{\psi_i'}$ (in $H_\mathrm{DM}$) are evaluated using Eq.~\eqref{eq:multipole}. Computing the radial integral in Eq.~\eqref{eq:radial_multipole} requires numerical wavefunctions, which we compute using the Numerov method~\cite{zimmerman1979stark}, following the implementation in the Alkali Rydberg Calculator~\cite{Sibalic2017ARC}. The Numerov method computes the radial wavefunction of a bound state from its energy, relative to the ionization limit. We note that MQDT states do not have a unique radial wavefunction: each channel may have a distinct ionization limit. Therefore, the matrix element in Eq.~(\ref{eq:multipole}) involves summing over several different wavefunctions.

The energy shift in small magnetic fields can be estimated from the diagonal elements of Eq.~\eqref{eq:Ham_e}. The Stark shift in a small electric field can be estimated from diagonalizing Eq.~\eqref{eq:Ham_e} in a small basis consisting of the several closest states with opposite parity. However, evaluating a full Stark map with multiple level crossings, or evaluating the diamagnetic shift in large fields resulting in level crossings, requires a larger basis of hundreds or thousands of states.

The Hamiltonian of two interacting Rydberg atoms in the Born-Oppenheimer approximation can be written as

\begin{equation}
    H = H^{(1)} + H^{(2)} + H_\mathrm{int}\,,
\end{equation}

where $H^{(1)}$ and $H^{(2)}$ are the Hamiltonians of the two isolated Rydberg atoms (Eq.~(\ref{eq:Ham_e})). The interaction Hamiltonian $H_\mathrm{int}$ is evaluated through an expansion over multipole terms $k_1$ and $k_2$ \cite{Rose1958electrostatic,Dalgarno1966The}

\begin{widetext}
\begin{align}
\begin{split}
    \label{eq:H_multipole}
    H_\mathrm{int} = \sum_{k_1,k_2}^\infty \frac{(-1)^{k_2}}{R^{k_1+k_2+1}}&\sqrt{\frac{\left(4\uppi\right)^3\left(2k_1+2k_2\right)!}{\left(2k_1 + 1\right)!\left(2k_2 + 1\right)!\left(2k_1+2k_2+1\right)}} \times \\
    & \sum_{p=-(k_1+k_2)}^{k_1+k_2}\sum_{p_1=-k_1}^{k_1}\sum_{p_2=-k_2}^{k_2}C_{k_1,p_1,k_2,p_2}^{k_1+k_2,p}r_1^{k_1}r_2^{k_2}Y_{k_1,p_1}(\hat{r}_1)Y_{k_2,p_2}(\hat{r}_2)Y_{k_1+k_2,p}(\hat{R}) \,,
\end{split}
\end{align}
\end{widetext}

where $\hat R$ is a vector representing the internuclear axis, $C_{k_1,p_1,k_2,p_2}^{k_1+k_2,p}=\langle k_1,p_1;k_2,p_2|k_1+k_2,p\rangle$ is a Clebsch-Gordan coefficient, and $Y_{\kappa,p}(\hat r)$ are unit-normalized spherical harmonics.

The calculations presented in this work are restricted to dipole-dipole interactions ($k_1 = k_2 = 1$), though we note that higher-order multipoles can become important at short separations~\cite{Hollerith2019Quantum}, and the $R^{-5}$ term from $k_1 = k_2 = 1$ can also dominate the asymptotically long-range interaction~\cite{Vaillant2012Long}.

As with the case of the field shift, the matrix elements are evaluated numerically using Eq.~\eqref{eq:multipole} at fixed internuclear separations $R$. The size of the required basis for computing pair interactions is much larger than for evaluating single-atom energy level shifts. Moreover, compared to alkali atoms, $^{174}$Yb has twice the number of Rydberg states ($S=0$ and $S=1$), while $^{171}$Yb has four times as many when including the nuclear spin. 

We therefore restrict the size Rydberg pair basis by only including pair states with significant contribution to the interaction potential in an energy range close to a target pair state~\cite{Deiglmayr2016long}. The Rydberg pair basis is formed in two steps. First, we build the pair basis from a set of single atom states with similar effective principal quantum number $\nu$ and orbital angular momentum $L$ as the target state. The resulting pair basis is then further truncated to pair states with a small pair-energy defect $\Delta E$ to the target pair state. Depending on the orientation of the internuclear axis with respect to external fields and the included orders of the multipole expansion, the basis size can be additionally restricted by making use of selection rules of the spherical harmonics under the conservation of certain symmetries \cite{weber2017calculation}. 
 
 $\Delta E$, $\Delta \nu$ and $\Delta L$ are then increased until convergence is observed. Typically, values of $\Delta \nu\lesssim3$ are sufficient to produce accurate interaction potentials \cite{Deiglmayr2016long,weber2017calculation}. To give a sense of scale, the computations in Fig.~\ref{fig:F32_54.56_interactions} included Rydberg pair state with $\Delta \nu<3$, $\Delta L =1$, and $\Delta E < \SI{10}{GHz}$, yielding a basis of approximately 2000 pair states. 

\begin{figure*}[tb]
	\centering
    \includegraphics[width=\linewidth]{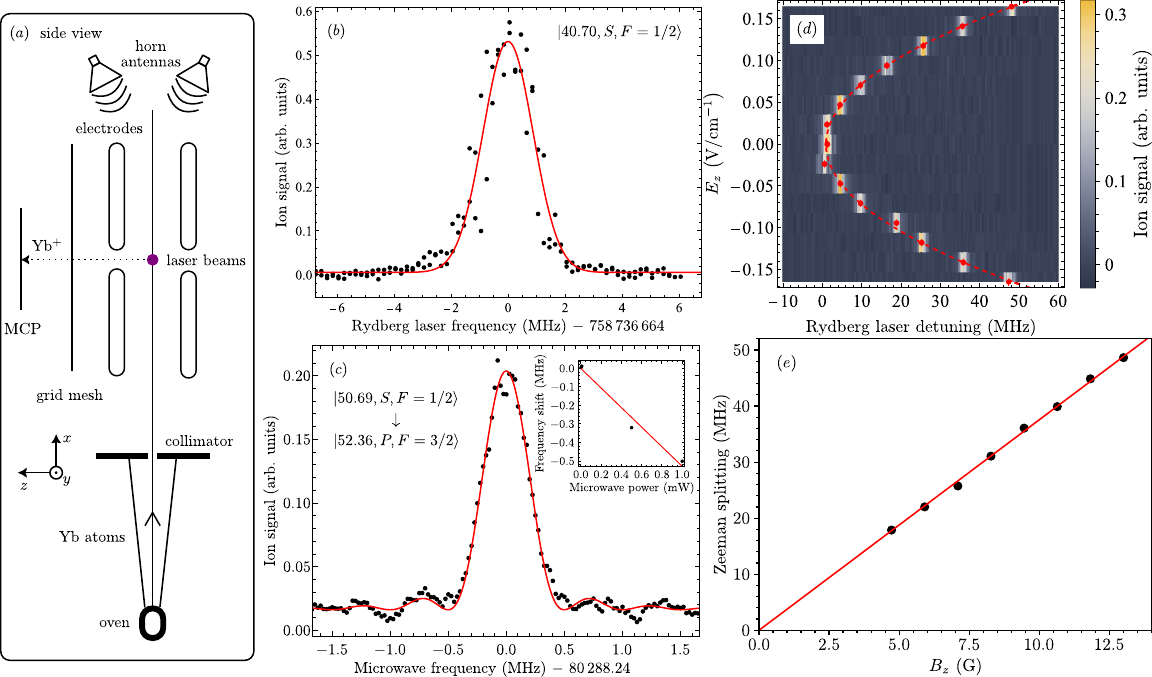}\caption{\label{fig:atomicbeam}  (a) Schematic of the atomic beam apparatus used for spectroscopy on Yb Rydberg state. See text for details. (b) Typical laser excitation spectrum, showing the $^{171}$Yb, $\ket{40.70,S,F=1/2}$ Rydberg state. Each black point corresponds to the integrated time-resolved ion signal averaged over $1\,000$ repetitions of the experiment. The obtained data is fit to a Gaussian line shape model (red line) with a full width at half maximum (FWHM) of \SI{2.1}{MHz}. (c) Typical microwave spectrum of a $\ket{50.69,S,F=1/2}\rightarrow\ket{52.36,P,3=1/2}$ transition in $^{171}$Yb. Each black point corresponds to the integrated time-resolved ion signal averaged over $1\,000$ repetitions of the experiment. The obtained data is fit to a sinc$^2$ line shape model (red line) for a \SI{2}{\mu s} rectangular microwave pulse. \textit{Inset:} Extrapolation of the transition frequency to zero microwave power. $(d)$ Typical polarizability measurement of the $\ket{62.14,S, F=1/2}$ Rydberg state, obtained by recording laser spectra in a range of electric fields $F_z$. The transition frequencies (red dots) are obtained by a least-squared fit of a Gaussian line shape model to the experimental data. The static dipole polarizability of the state is obtained by a least-squared fit of Eq.~(\ref{eq:staticpol}) (dashed red line) to observed Stark shifts. $(e)$ Zeeman splittings between the $\ket{50.69,S,F=1/2,m_F = \pm1/2}$ magnetic sub-levels obtained by laser spectroscopy in a magnetic field $B_z$. $g$ factors of the Rydberg states are obtained by a least-square fit of the experimental data to Eq.~(\ref{eq:zeeman}). }
\end{figure*}

\section{Atomic beam spectroscopy apparatus}
\label{sec:atomicbeam}

The spectroscopic data presented in this article are obtained by laser and radiofrequency (RF) spectroscopy on an atomic beam of $^{174}$Yb or $^{171}$Yb atoms (see Fig.~\ref{fig:atomicbeam}a and b). The atomic beam is generated by heating a sample of ytterbium to a temperature of \SI{310}{\celsius} in an oven with a collimator opening of approximately \SI{1}{cm}. The atomic beam is further collimated by a pinhole with a diameter of \SI{3}{mm}, approximately \SI{12}{cm} after the oven collimator and \SI{9}{cm} before the spectroscopy region.

Transitions to S and D Rydberg states are driven by a two-photon laser transition through the intermediate $6s6p\, ^1P_1$ state. The laser light needed for the transition with wavelengths of \SI{399}{nm} ($6s^2\,^1S_0\rightarrow6s6p\, ^1P_1$) and 394--\SI{399}{nm} ($6s6p\, ^1P_1\rightarrow$ Rydberg state) is generated by frequency doubling the output of tunable titanium-sapphire (Ti:sapph) lasers in a resonantly enhanced frequency doubling cavity. The isotope shifts on the $6s^2\,^1S_0\rightarrow6s6p\, ^1P_1$ transition \cite{Kleinert2016Measurement} are used for an isotope selective excitation into the Rydberg state.
To minimize Doppler shifts on the two-photon laser transitions, counter-propagating laser beams are applied at a \SI{90}{\degree} angle relative to the atomic beam (along the $y$ axis in Fig.~\ref{fig:atomicbeam}). Laser pulses with length of 1-$\SI{3}{\mu\second}$ are generated by acousto-optic modulators.
The laser frequencies are monitored by measuring the frequency of the fundamental output of the Ti:sapph laser using a \textsc{HighFinesse WS8-10} wavelength meter with a specified $3\sigma$ uncertainty of \SI{10}{\mega\hertz}. At regular intervals, the wavelength meter is calibrated to a \SI{399}{nm} laser that is frequency locked to a ultra-low-expansion cavity with known offset from the $^1S_0\rightarrow^1P_1(F=3/2)$ transition in $^{171}$Yb. 
 
Population in Rydberg states is detected by state-selective pulsed-field ionization (see, e.g., \cite{Gallagher_1994}). High-voltage (HV) ramps with a maximum voltage of \SI{2.5}{kV} are generated by switching between a low-voltage (during Rydberg spectroscopy) and a HV input using HV MOSFET switches (\textsc{Behlke} HTS 31-03-GSM) and are applied to  a set of electrodes (on the right side of the atomic beam in Fig.~\ref{fig:atomicbeam}) through a $RC$ low-pass filter ($\tau=\SI{1}{\mu s}$). The resulting maximum field of $\approx\SI{833}{\volt\per\cm}$ limits the lowest detectable Rydberg states in our setup to an effective principal quantum number $\nu\approx26$. The resulting Yb$^+$ ions are accelerated towards and detected on a time-resolved microchannel-plate (MCP) ion detector (\textsc{Hamamatsu F13446-11}) and recorded on an oscilloscope. To reduce electric field inhomogeneities caused by the high-voltage (typically \SI{-1.9}{kV}) applied to the front surface of the MCP, we place a grounded grid mesh about halfway in between the spectroscopy region and the MCP. The experiment is operated at a repetition rate of \SI{1}{kHz}, limited by the current drawn at the HV supply generating the HV field-ionization ramps.  

An exemplary laser spectrum of the $\ket{40.70,S,F=1/2}$ Rydberg state of $^{171}$Yb is presented in Fig.~\ref{fig:atomicbeam}c. Transition frequencies to the Rydberg state are determined by a least-square fit of the data to a Gaussian line shape model. 

Transition between Rydberg states are driven using RF with frequencies in the range of 10-\SI{175}{\giga\hertz}. The RF radiation is obtained from a \textsc{Windfreak Techn., LLC} \textsc{SynthHD PRO} (up to \SI{24}{GHz}) dual-channel microwave generator or by using a combination of active x4 (\textsc{Marki Microwave} \textsc{AQA-2156}, 21-\SI{56}{\giga\hertz}, \SI{20}{dBm}) and/or passive x4 (\textsc{Marki Microwave} \textsc{MMQ-40125H}, 40-\SI{175}{\giga\hertz}, \SI{0}{dBm}) frequency multipliers. The microwave generator is referenced to a stable \SI{10}{\mega\hertz} reference signal obtained from a GPS disciplined Rb atomic clock (\textsc{Stanford Research Systems} \textsc{FS725}). RF pulses with duration of 1-$\SI{8}{\mu\second}$ are obtained by amplitude modulating the output of the microwave generator before frequency multiplication using PIN absorptive modulators (\textsc{Hewlett Packard} \textsc{33008C} or \textsc{11720A}). The obtained RF radiation is coupled into the vacuum chamber through vacuum viewports using suitable horn antennas. All laser radiation is switched off for the duration of the RF pulses. The power of the RF radiation is controlled using either a digital step attenuator (\textsc{Analog Devices} ADRF5740) or a WR-10 direct-reading attenuator (\textsc{Mi-Wave} \textsc{510W/387}).  A microwave spectrum of the $33\,^1S_0\rightarrow33\,^1P_1$ transition in $^{174}$Yb is presented in Fig.~\ref{fig:atomicbeam}d. Transition frequencies between Rydberg states are determined by a least-square fit of the data to either a sinc$^2$ or Lorentzian line shape model. All stated Rydberg-Rydberg transitions have been measured at varying RF powers and extrapolated to ``zero'' RF power to remove frequency shifts caused by AC-Stark shifts. 

To control the electric field in the spectroscopy region, we apply voltages to two segmented circular electrodes (4 segments each) separated by \SI{3}{cm}. Stray electric fields are compensated at the beginning of each day, by minimizing the quadratic Stark shift of the $6s130s$~$130\,^1S_0$ Rydberg state of $^{174}$Yb. We observe day-to-day fluctuations of the required electric compensation fields of less than $\SI{3}{mV\per\cm}$. 

We determine values for the static dipole polarizabilities $\alpha_0$ of Rydberg states by measuring a shift in the transition frequencies caused by applying electric fields $F$ between the two segmented circular electrodes and subsequent fitting to

\begin{equation}
	\Delta E_\mathrm{Stark} = - \left(1/2\right)\alpha_0F^2\,.\label{eq:staticpol}
\end{equation}

When possible, we ensured that the static dipole polarizabilities are determined in a range of quadratic Stark shifts. For some cases, the Stark shift deviates from a quadratic scaling even at very small electric fields and we report values for the Stark shift at a given electric field instead.

Magnetic fields in all three spatial directions are controlled by three pairs of coils placed outside the vacuum chamber, arranged in Helmholtz configuration. The magnetic fields generated by the coils are calibrated by RF spectroscopy of the Zeeman splittings on a $n'\,^3S_1\leftarrow n\,^1S_0$ transition, and assuming a $g$-factor of the mostly unperturbed $n\,^3S_1$ Rydberg series to be $g(^3S_1)=g_s\approx2.0023$. We obtain values for Land\'e $g$ factors of Rydberg states by measuring Zeeman shifts caused by applying calibrated magnetic fields $B$ and subsequent a subsequent linear fit to

\begin{equation}
	\Delta E_\mathrm{Zeeman} = m g \mu_B B\,,\label{eq:zeeman}
\end{equation}

where we measured the Zeeman shifts at sufficiently small magnetic fields to avoid the effect of non-linear diamagnetic shifts.

\section{Measuring Rydberg interactions in optical tweezers}\label{sec:opticaltweezer}

\begin{figure}[tb]
	\centering
    \includegraphics[width=\linewidth]{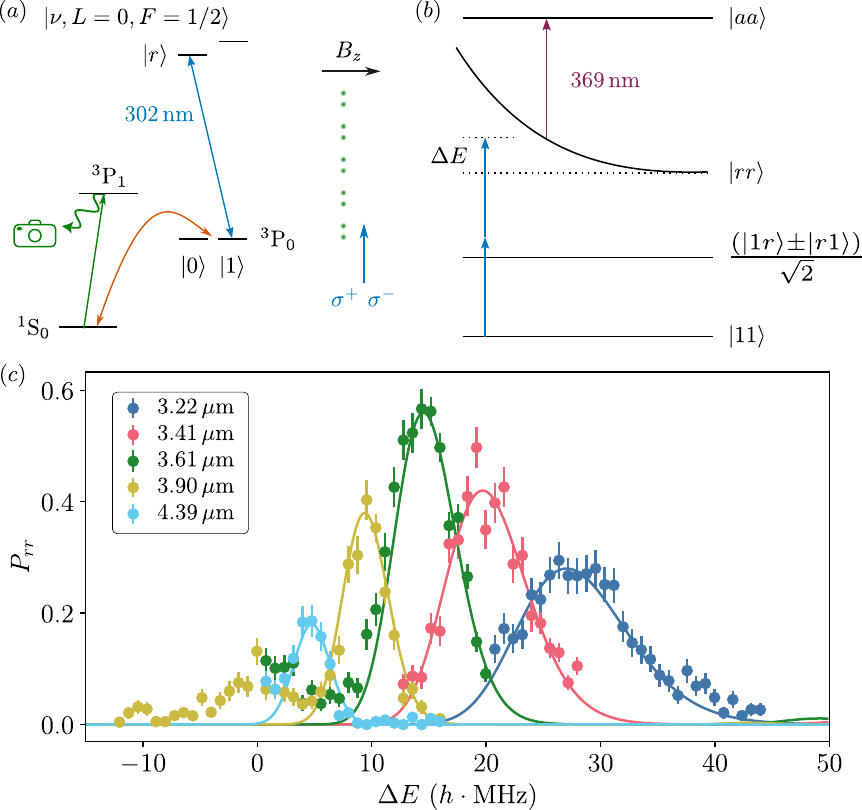}
	\caption{\label{fig:InteractionsScheme.pdf} $(a)$ Single-atom transition scheme for measuring interactions between pairs of Rydberg atoms in an optical tweezer setup \cite{Ma2023High}. The $^{171}$Yb atoms are prepared states in the $\ket{1}$ sublevel of the $^3P_0$ metastable state. Transitions into the $\ket{\nu,L=0,F}$ Rydberg state are driven by a single-photon transition (\SI{302}{nm}). For rearrangement and readout, the atoms are imaged by collecting the fluorescence on the $\ket{^1S_0}\leftrightarrow\ket{^3P_1}$  transition. $(b)$ Scheme for measuring Rydberg-Rydberg interactions between pairs of single atoms trapped in optical tweezer arrays.  $(c)$ Experimental pair-state spectra (solid dots), obtained by recording the probability of pair-wise loss from tweezers in fluorescence image as a function of two-photon detuning and tweezer separation. The experimental spectra are compared to simulated pair spectra $S(E)$ (solid lines) obtained from Eq.~(\ref{eq:pairspectrum}), as explained in the text.}
\end{figure}

We measure the distance-dependent interactions between single $^{171}$Yb Rydberg atoms trapped in optical tweezer arrays using the setup described in Ref.~\cite{Ma2023High}. Similar approaches have been previously used to measure $C_6$ coefficients of Rydberg-pair states in Rb \cite{Beguin2013Direct} and angle-dependent dipole-dipole interaction strengths \cite{Ravets2015Measurement}. We initialize pairs of atoms in the $\ket{1} \equiv \ket{m_F = +1/2}$ sublevel of the $6s6p\,^3P_0$ metastable state. Static magnetic fields are applied at \SI{90}{\degree} ($\theta=\uppi/2$) with respect to the internuclear separation. Transitions into Rydberg pair states are subsequently driven by a two-photon transition detuned by $\Delta E$ from the asymptotic case of two isolated Rydberg atoms. Pairs of atoms transferred into Rydberg states are blown out of the optical tweezers by driving a transition (\SI{369}{nm}) into an autoionizing state \cite{Burgers2022Controlling}. Remaining atoms in the $\ket{1}$ state are depumped into the $^1S_0$ ground state for fluorescence imaging. The excitation of a Rydberg-pair state is inferred by conditioning the remaining atom population after a given experimental sequence on pair-wise atom loss. An exemplary pair-state spectrum close to the $\ket{54.28,L=0,F=1/2,-1/2}^{\otimes 2}$ asymptotic pair state is given in Fig.~\ref{fig:InteractionsScheme.pdf}c. For a given optical tweezer separation, we observe an increase in pairwise atom loss when tuning the two-photon transition over a Rydberg pair state. The pair-state resonance shifts to larger two-photon detuning when reducing the tweezer separation, indicating a larger interaction between the two Rydberg atoms. We fit the observed pair-state spectra using a Gaussian line shape model. The resulting center frequency of the fits are presented in Fig.~\ref{fig:F12_54.28_interactions} and Fig.~\ref{fig:F32_54.56_interactions}.

On top of the shift of the pair-state resonance, we also observe a broadening. We attribute the observed broadening to the spatial fluctuations of the single atoms in the optical tweezers ($T\approx\SI{3}{\mu K}$). Due to the spatial fluctuations of the atoms, the linewidth of the pair-state resonance is affected by the gradient of the interaction potential, which increases towards shorter internuclear separations. To model the effect of fluctuating internuclear separations in the optical tweezers, we simulate the pair-state spectra $S(E)$ 

\begin{equation}\label{eq:pairspectrum}
    S(E)\propto\sum_\Phi \mathcal{G}(E-E_\Phi(R),\sigma_E)\mathcal{T}(R-\tilde{R},\sigma_R)\mathcal{O}_{\Phi,\Phi'}(R)\,,
\end{equation}

where the sum runs over all relevant pair state $\Phi$ coupled to the target state $\Phi'$, $E_\Phi(R)$ is the internuclear-separation-dependent pair-state energy, $\tilde{R}$ is the mean internuclear separation, $\mathcal{G}(E-E_\Phi(R),\sigma_E)$ is a function describing the experimental linewidth of the spectrum in the absence of any broadening, $\mathcal{T}(R-\tilde{R},\sigma_R)$ is a function describing the distribution of internuclear separation around a mean separation $\tilde{R}$, and $\mathcal{O}_{\Phi,\Phi'}(R)$ is the overlap coefficient.

For the case presented in Fig.~\ref{fig:InteractionsScheme.pdf}c, the line broadening can be reproduced by assuming a normal distribution in the internuclear separations with standard deviation of \SI{100}{nm}. This is close to the expected distance fluctuations of approximately \SI{50}{nm} between two atoms confined in optical tweezers with radial and axial trap frequencies of $\omega_r=\SI{60}{kHz}$ and $\omega_z=\SI{10}{kHz}$, respectively.

\section{Additional spectroscopic data and models for $^{174}$Yb}
\label{sec:APP_174}

In this appendix, we give additional details of the spectroscopic measurements and MQDT model development for $^{174}$Yb. We also make comparisons between the MQDT model and previously measured quantities, including singlet-triplet mixing in the $^{1,3}P_1$ and $^{1,3}D_2$ series, and diamagnetic shifts in the P series.

\subsection{$^{3}S_1$ and $^{1}S_0$}
\label{sec:174_S}

For the $n\,^1S_0$ Rydberg series of $^{174}$Yb, we adapt the six-channel MQDT model presented in previous work~\cite{Lehec2018Laser}. We refit the MQDT model parameters of the $^1S_0$, $^{1,3}D_2$, $^{1,3}P_1$ Rydberg series in a simultaneous, 42-parameter fit to the previously measured \cite{Camus1969spectre,Meggers1970First,Camus1980Highly,Aymar1980Highly,Aymar1984three,Maeda1992Optical,Lehec2017PhD,Lehec2018Laser} and newly measured data presented in the supplemental material~\cite{supplementary}, leveraging the higher precision microwave measurements between S, D, and P Rydberg states. The resulting MQDT models are presented in Tab.~S1, Tab.~S2, and Tab.~S3.

For the $n\,^3S_1$ with $n>28$, we adapt the single-channel quantum defect model with a Rydberg-Ritz expansion for the energy-dependent quantum defect presented in  Ref.~\cite{Wilson2022Trapping}. 

\subsection{$^{1}P_1$ and $^{3}P_1$}
\label{sec:174_13P1}

Here we introduce a six-channel MQDT model for the $^{1,3}P_1$ Rydberg series of $^{174}$Yb. Compared to the five-channel model presented in Ref.~\cite{Lehec2017PhD}, we introduce a sixth perturbing channel, which lies energetically above the first ionization limit of $^{174}$Yb. The addition of the sixth channel was necessary to accurately describe the energies of the high-$n$ states of this series. In addition, we introduce singlet-triplet mixing between the $^1P_1$ and $^3P_1$ explicitly, by introducing a rotation to the $U_{i\alpha}$ matrix, $\theta_{12}$ whose value is constrained by the Stark shift of the $^1S_0$ state, as explained in Sec.~\ref{sec:174Ybmain}.

We extend the previously measured state energies of $^{1,3}P_1$ \cite{Aymar1984three,Lehec2017PhD}, by measuring microwave transition frequencies between $n\,^{1,3}P_1$ and $n'\,^{1,3}D_2$ or $n'\,^{1}S_0$ Rydberg states in the range of $31\leq n\leq43$, as summarized in Tab.~S18 and Tab.~S19. We cross checked several transition frequencies reported in Ref.~\cite{Lehec2017PhD} and generally find good agreement within the stated error bars, with the exception of the $6s45p\,^3P_1 \leftarrow 6s44s\,^1S_0$, which deviates from our new measurement and the final theoretical prediction from the MQDT model by nearly \SI{10}{MHz}. We therefore don't include this data point in our analysis. In the fitting procedure, we optimize the 42 parameters of the $^1S_0$, $^{1,3}D_2$, $^{1,3}P_1$ MQDT models in a simultaneous fit to the data presented in the supplemental material~\cite{supplementary}. The resulting MQDT models are presented in Tab.~S1, Tab.~S2, and Tab.~S3.

\begin{figure}[tb]
	\centering
    \includegraphics[width=\linewidth]{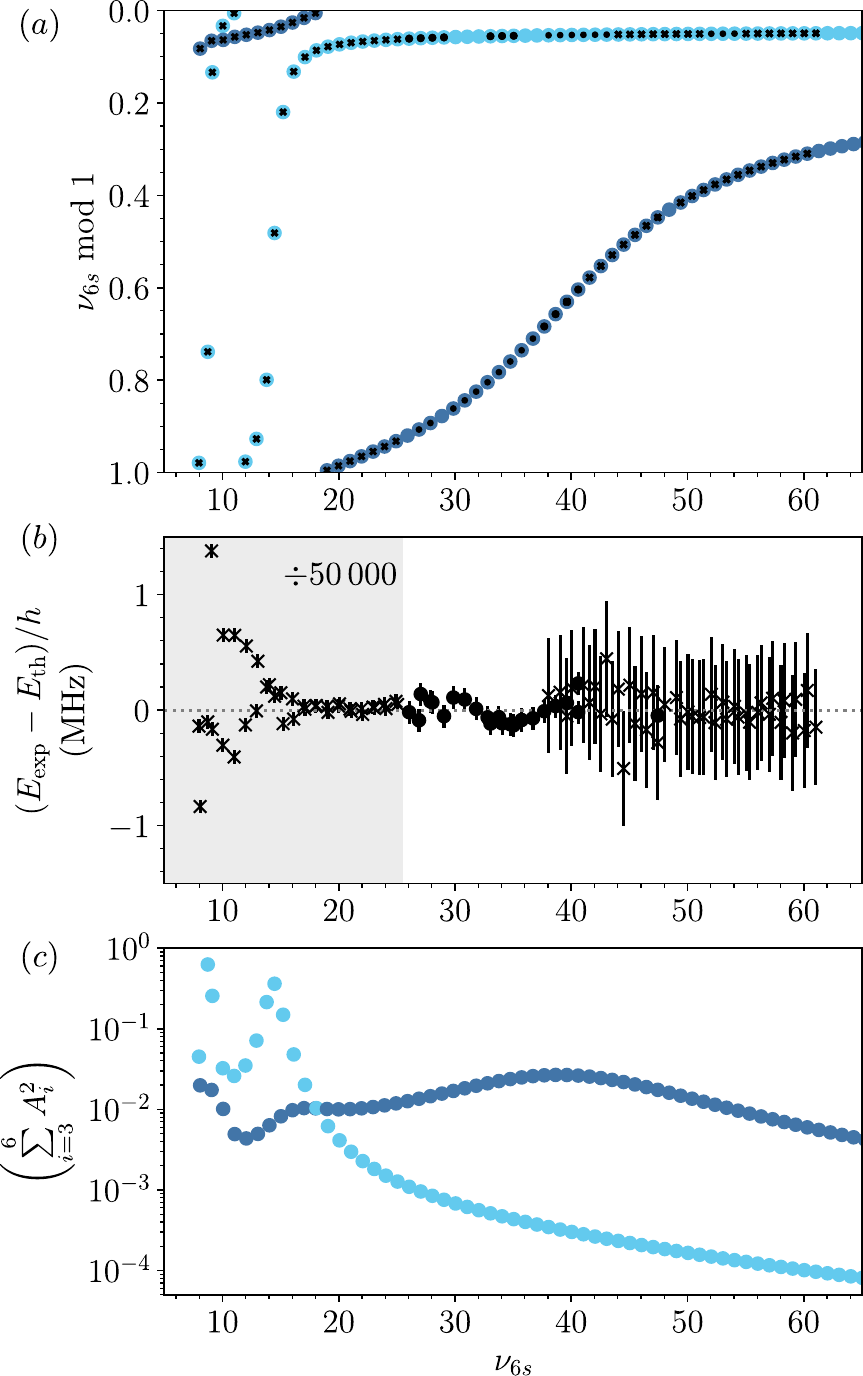}
	\caption{\label{fig:174Yb_13P1_lufano_residuals} $(a)$ Lu-Fano-type plot of the $^{174}$Yb $^{1,3}P_1$ Rydberg series. The theoretical bound states with dominant singlet and triplet character are indicated by light and dark blue dots, respectively. Experimentally observed states are indicated by black crosses (Refs.\cite{Aymar1984three,Lehec2017PhD}) and black dots (this work). $(b)$ Deviation between experimental and theoretical state energies. The energy deviations and error bars in the gray shaded area are scaled by a factor of $50\,000$ to improve visibility of the much smaller errors observed on the microwave transitions. $(c)$ Perturbing channel fraction $\sum\limits_{i=3}^6A_i^2$ of the dominantly singlet (light blue) and triplet (dark blue) Rydberg states.}
\end{figure}

The presented MQDT model correctly predicts the energies of the highly excited $n\,^{1,3}P_1$ Rydberg states (Fig.~\ref{fig:174Yb_13P1_lufano_residuals}). However, below $\nu<15$, significant deviations between the experimental and theoretical energies occur. This could be due to unaccounted perturbing channels, either directly in the low-$\nu$ regime, or in the high-$\nu$ regime, with the current MQDT model parameters overcompensating trends in the quantum defects for the more accurately determined Rydberg states at high $\nu$.

The breakdown of pure $LS$ coupling in the $^{1,3}P_1$ Rydberg channels is quantified by measuring the static dipole polarizabilities of $n\,^1S_0$ Rydberg states in the vicinity of a near degeneracy with dominantly $n\,^{3}P_1$ Rydberg states, as discussed in the main text. To obtain the best agreement between the experimental and theoretical static polarizabilities of the $n\,^{1}S_0$ Rydberg states, an energy dependent singlet-triplet mixing angle (refer to Eq.~(\ref{eq:theta_energydep})) had to be introduced. The resulting MQDT parameters are presented in the supplemental material~\cite{supplementary}.

\begin{figure}[tb]
	\centering
    \includegraphics[width=\linewidth]{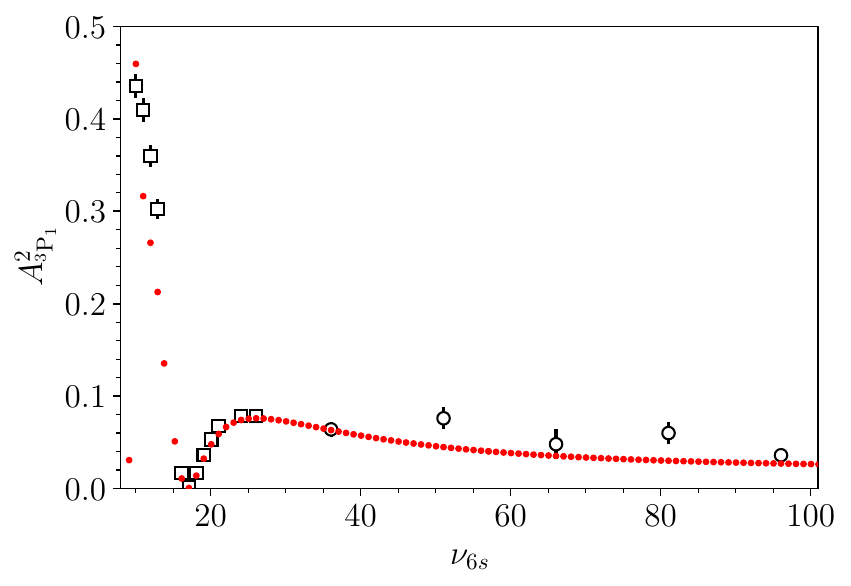}
\caption{\label{fig:174Yb_1P1_tripletchar} $6snp\,^3P_1$ character ($A_{^3P_1}^2$) in $6snp\,^1P_1$ Rydberg states of $^{174}$Yb. The open black squares correspond to experimental values obtained by measurements of the hyperfine-structure of $^{171}$Yb and $^{173}$Yb $^{1,3}P_1$ Rydberg states from Ref.~\cite{Majewski1985diploma}. The open black circles correspond to experimental values obtained by measuring diamagnetic shifts taken from Ref.~\cite{Neukammer1984diamagnetic}. The full red circles correspond to theoretical values of $A_{^3P_1}^2$ as obtained from the MQDT model presented in Tab.~S3.}
\end{figure}

\begin{figure}[tb]
	\centering
    \includegraphics[width=\linewidth]{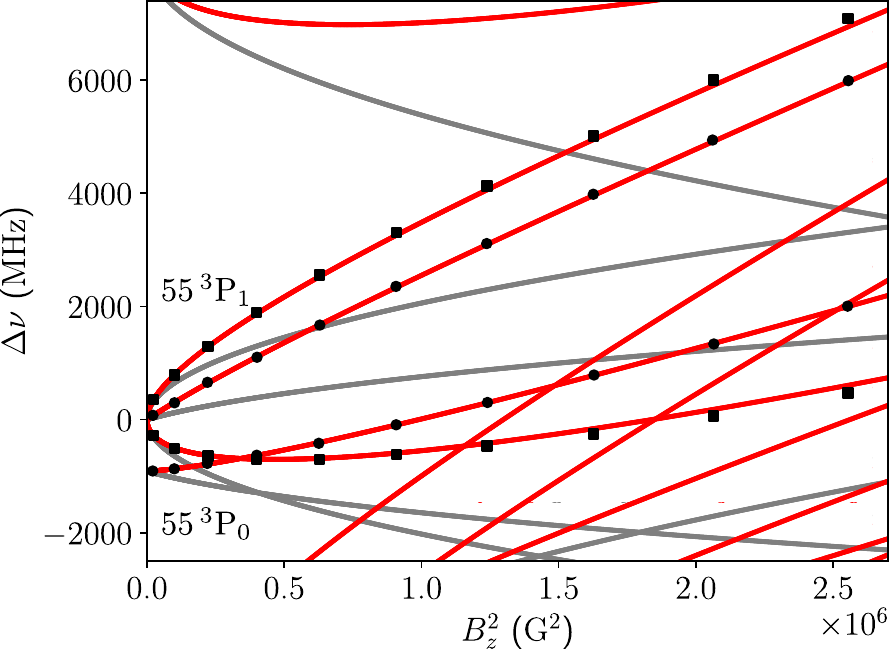}
	\caption{\label{fig:174Yb_553P1_3P0_diamagnetism.pdf} Zeeman shifts of $^{174}$Yb Rydberg states near the $55\,^3P_1$ state. The red and gray curves correspond Zeeman shift calculations with and without diamagnetism, respectively. The near degeneracy of the $55\,^{3}P_0$ and $55\,^{3}P_1$ states, leads to a strong interaction between the $55\,^{3}P_1(m_J=0)$ and $55\,^{3}P_0(m_J=0)$ sublevels. The black circles and squares correspond to experimentally observed Zeeman shifts extracted from Fig.~3 of Ref.~\cite{Neukammer1984diamagnetic} for the $m_J=0$ and $m_J=\pm1$ sublevels, respectively. The experimental Zeeman shifts from Fig.~3 of Ref.~\cite{Neukammer1984diamagnetic} are plotted directly for the $m_J=0$ states, whereas the values for $m_J\pm1$ states are obtained from the mean experimental diamagnetic shifts and theoretical predictions for the paramagnetic contribution to the Zeeman shift.}
\end{figure}

With the obtained $J=1$ (odd parity) MQDT model for $^{174}$Yb and using  Eq.~(\ref{eq:channelfractions_LS}), we can estimate the $6snp\,^3P_1$ channel contribution into the nominally $n\,^1P_1$ Rydberg states. The resulting values are presented in Fig.~\ref{fig:174Yb_1P1_tripletchar} and compared to values obtained from previous measurements of diamagnetic shifts, as presented in Ref.~\cite{Neukammer1984diamagnetic} and from previous hyperfine-structure measurements in $^{171}$Yb and $^{173}$Yb, as presented in Ref.~\cite{Majewski1985diploma}. The theoretically obtained values for the triplet contribution to the $n\,^1P_1$ Rydberg states agree well with the previously reported values between $20\leq n\leq100$, but with significantly reduced uncertainties, highlighting the sensitivity of measuring matrix elements through Stark shifts close to near degeneracies.

As an additional check of the validity of the model, we compare the predicted Zeeman and diamagnetic shifts in very large magnetic fields to a previous experimental measurement from Ref.~\cite{Neukammer1984diamagnetic}. In that work, the measured energies were fit with a phenomenological model to extract the singlet-triplet splitting for that $n$. In Fig.~\ref{fig:174Yb_553P1_3P0_diamagnetism.pdf}, we show the prediction of the MQDT model with no free parameters, finding excellent agreement.

\begin{figure}[tb]
	\centering
    \includegraphics[width=\linewidth]{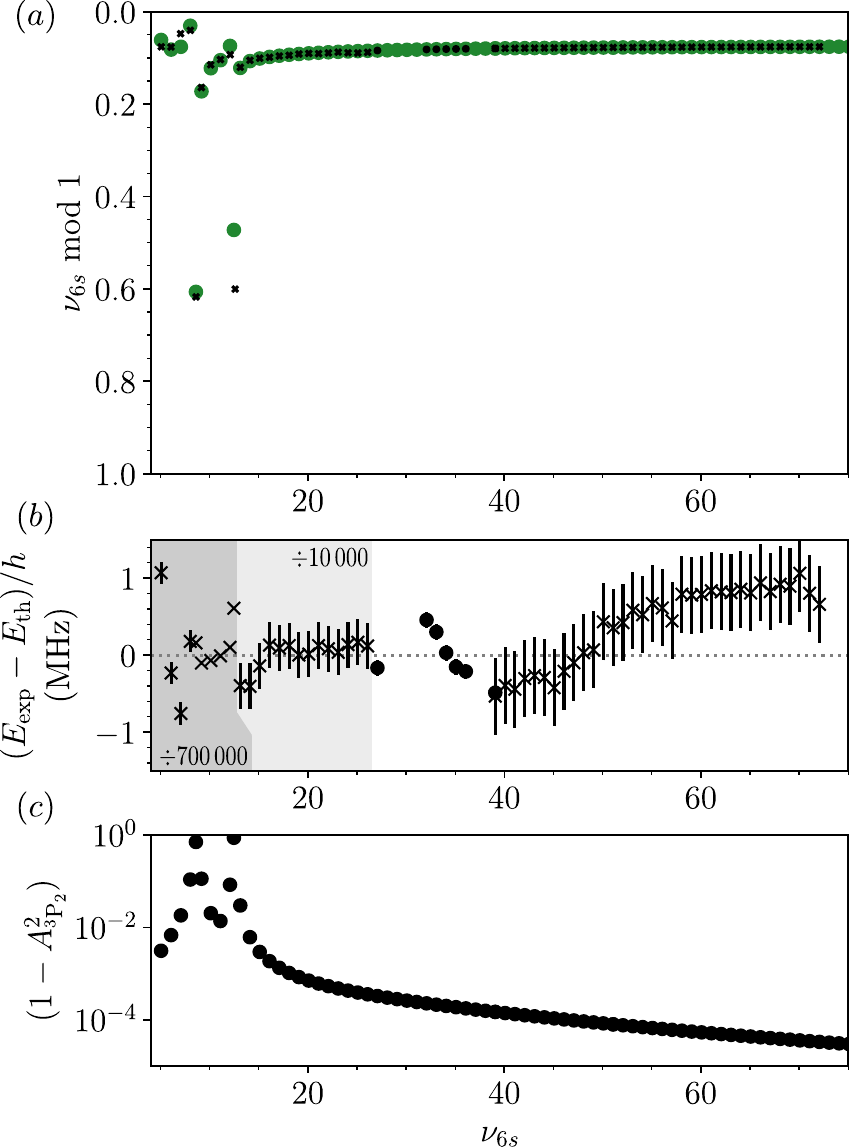}
	\caption{\label{fig:174Yb_3P2_lufano_residuals} Lu-Fano-type plot of the $^{174}$Yb $^{3}P_2$ Rydberg series. The theoretical bound states are indicated by green dots. Experimentally observed states are indicated by black dots (Refs.\cite{Aymar1984three,Lehec2017PhD}) and dots (this work). $(b)$ Deviation between experimental and theoretical state energies. The energy deviations and error bars in the light (dark) gray shaded area are scaled by a factor of $10\,000$ ($700\,000$) to improve visibility of the much smaller errors observed on the microwave transitions. $(c)$ Perturber fraction $(1-A_{^3P_2}^2)$ of the Rydberg states in the $(J=2)^o$ Rydberg series.}
\end{figure}

\subsection{$^3P_2$}

Here, we introduce a four-channel  MQDT model for $^3P_2$ Rydberg states of $^{174}$Yb. In comparison to the previously presented three-channel MQDT model from Ref.~\cite{Ali1999Two}, we introduce an additional channel with a perturbing state above the first ionization limit. The addition of the fourth channel was necessary to accurately describe the energies of the high-$n$ states of this series.

The MQDT model parameters are optimized in a weighted least-square procedure. The Rydberg state energies includes in the fitting procedure include previously measured state energies of $^{3}P_2$ by laser \cite{Aymar1984three} and microwave spectroscopy \cite{Lehec2017PhD}, as well as  newly measured microwave transition frequencies between $n\,^{3}P_2$ and $n'\,^{3}D_2$ Rydberg states in the range of $31\leq n\leq43$. The included Rydberg state energies and the resulting MQDT model parameters are presented in thesupplemental material~\cite{supplementary}.

As depicted in Fig.~\ref{fig:174Yb_3P2_lufano_residuals}$\,(a)$ and $(b)$, the newly introduced MQDT model for $^3P_2$ captures the energies of the highly excited $n\,^3P_2$ Rydberg state quantitatively, but only qualitatively captures the energies of the states in the strongly perturbed range below $\nu\approx15$.

Fig.~\ref{fig:174Yb_3P2_lufano_residuals}c, depicts the contribution of the core excited channels, which are below $10^{-3}$ for states with $\nu>20$, confirming a mostly unperturbed $^3P_2$ Rydberg series. Series perturbers can shorten the lifetime of Rydberg states even at high $n$, by introducing an additional decay channel proportional to $\sum A_i \Gamma_i$, where $\Gamma_i$ is the decay rate of the perturbing states. Since $\Gamma_i$ for low-lying atomic states can be substantially faster than for Rydberg states, even a small admixture can significantly alter the lifetime. We note that Ref.~\cite{Wilson2022Trapping} measured the lifetime of the $74\,^3P_2$ state to be $\SI{83(5)}{\mu s}$, which is close to but slightly shorter than the lifetime predicted for a P state of $^{87}$Rb with a similar quantum number (approximately $\SI{200}{\mu s}$ at $T=\SI{300}{K}$ \cite{Beterov2009Quasiclassical}). This is qualitatively consistent with the modeled small but non-zero perturbation of this series. A quantitative assessment will require predictions for matrix elements and decay rates of the perturbing states, which is a subject for future work.

\begin{figure}[tb]
	\centering
    \includegraphics[width=\linewidth]{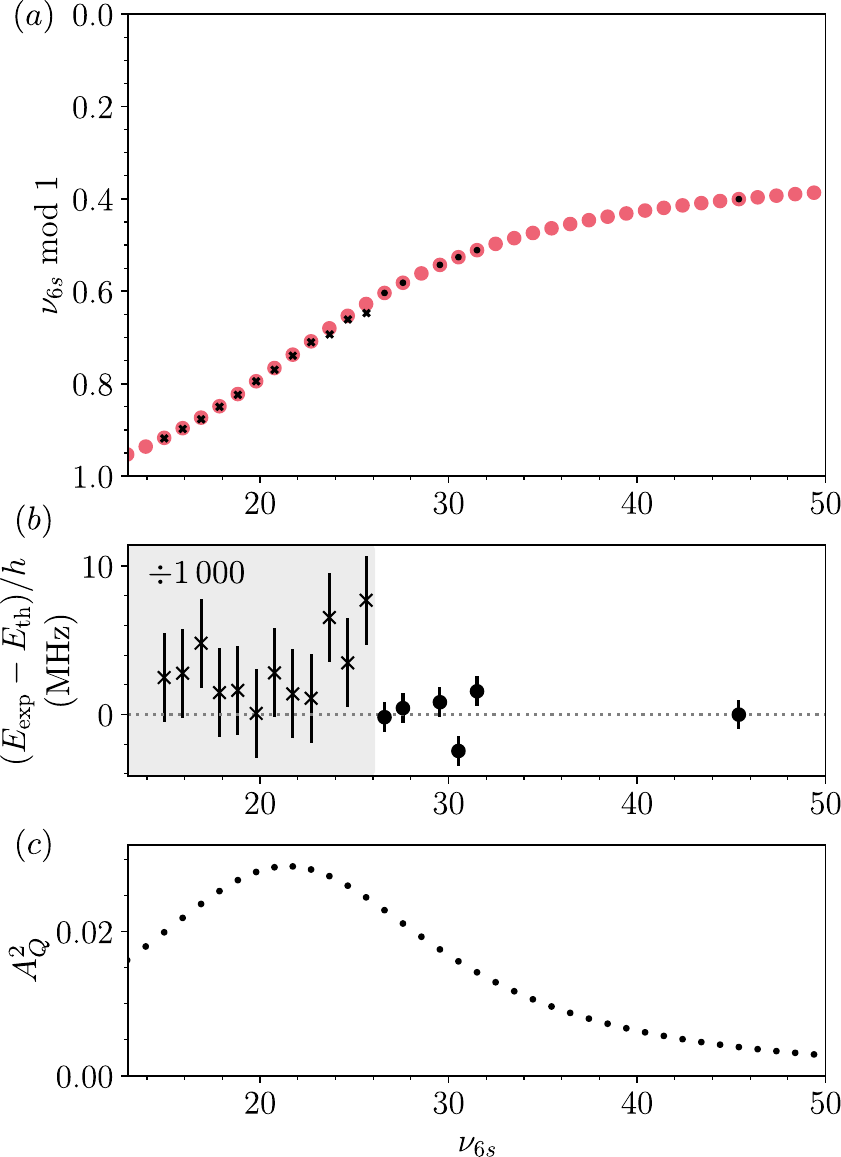}
	\caption{\label{fig:174Yb_3P0_lufano_residuals} Lu-Fano-type plot of the $^{174}$Yb $^{3}P_0$ Rydberg series. The theoretical bound states are indicated by red dots. Experimentally observed states are indicated by black crosses (Ref.\cite{Aymar1984three}) and dots (this work). $(b)$ Deviation between experimental and theoretical state energies. The energy deviations and error bars in the gray shaded area are scaled by a factor of $1\,000$ to improve visibility of the much smaller errors observed on the microwave transitions. $(c)$ $Q$ channel contribution $A_{Q}^2$ to the $n\,^3P_0$ Rydberg states.}
\end{figure}

\subsection{$^3P_0$}

Here, we present a two-channel MQDT model for the $n\,^3P_0$ Rydberg states of $^{174}$Yb. A two-channel model suffices to reproduce the observed states at high-$n$, though we believe additional channels will eventually be necessary to explain the full series. We have no direct information about the nature of the perturbing series and follow Ref.~\cite{Aymar1984three} by assigning a perturbing channel with character $4f^{13}5d6snd\,^3P_0$ (labelled $Q$). The MQDT model parameters are optimized in a weighted least-square fit to both previously measured state energies by laser spectroscopy \cite{Aymar1984three} and newly measured state energies from microwave spectroscopy, summarized in Tab.~S24 and Tab.~S23, respectively. 

The $n\,^3P_0$ Rydberg states are inaccessible by direct one-photon microwave transitions from laser accessible $n\,^{1}S_0$ and $n\,^{1,3}D_2$ Rydberg states. Instead, we perform spectroscopy on $n\,^3P_0$ Rydberg states of $^{174}$Yb, by utilizing the Autler-Townes splitting on a probe transition. To this extent, we monitor the resonant population transfer on the two-photon $n\,^3D_2\rightarrow n'\,^3S_1$ microwave transition as a function of the frequency of a simultaneously applied microwave pulse coupling the $n\,^3S_1\rightarrow n''\,^3P_0$ transition. The observed $n\,^3P_0\leftrightarrow {n'\,^3D_2}$ intervals are summarized in Tab.~S23. We recorded a total of six transitions to $n\,^3P_0$ Rydberg states in the range of  $31\leq n\leq49$, which were particularly necessary to determine the energy dependence of the strongly perturbed $^3P_0$ series.  The results of this measurement, together with previous three-photon laser spectroscopy \cite{Aymar1984three} are presented in the supplemental material~\cite{supplementary}. 

The obtained MQDT model is summarized in Tab.~S5 and presented in a Lu-Fano-like plot in Fig.~\ref{fig:174Yb_3P0_lufano_residuals}. The MQDT model represents the experimental data well within the experimental uncertainties over a range of $18\leq n\leq 50$.

The contributions of perturbing channel $Q$ into the $n\,^3P_0$ Rydberg states is presented in Fig.~\ref{fig:174Yb_3P0_lufano_residuals}c. The channel contribution is spread out over a wide range of principal quantum numbers and reaches a maximum value of approximately \SI{3}{\percent} at $n=26$. This is consistent with a significantly reduced lifetime of the $74\,^3P_0$ Rydberg state ($\SI{14(4)}{\mu s}$) observed in Ref.~\cite{Wilson2022Trapping}. Further refinement of the MQDT model for this series at low-$n$ will be useful to understand the lifetimes of this series quantitatively.

\begin{figure}[tb]
	\centering
    \includegraphics[width=\linewidth]{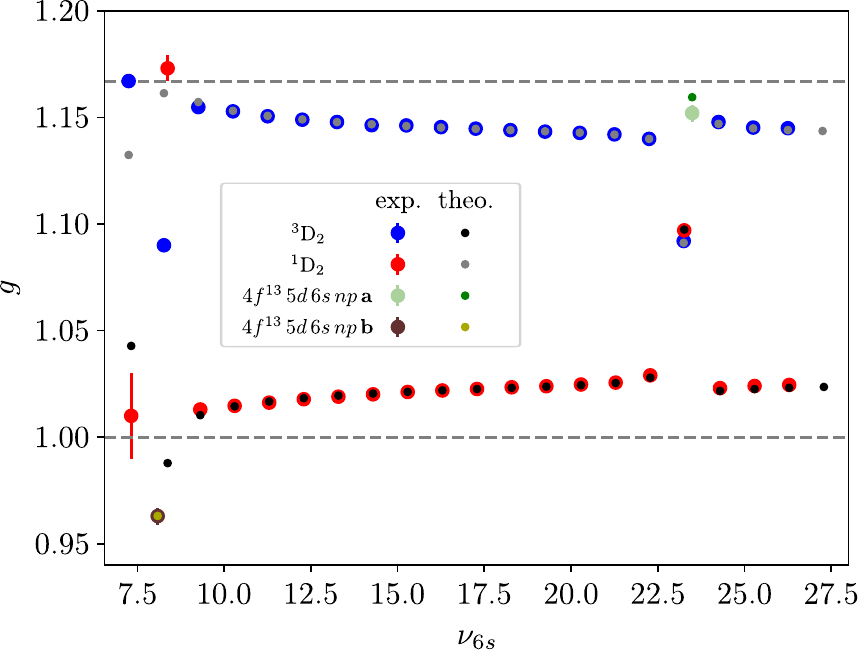}
	\caption{\label{fig:174Yb_13D2_gfactors} $g$-factors of the $^{1,3}D_2$ and its perturbers as taken from Ref.~\cite{zerne1996lande} and calculated with the MQDT presented in Tab.~S2. The gray dashed lines indicate the values for $g$ factors in pure $LS$ coupling.}
\end{figure}

\subsection{$^{1}D_2$ and $^{3}D_2$}
\label{sec:174_13D2}
Here, we present a five-channel MQDT model for the $n\,^{1,3}D_2$ Rydberg states of $^{174}$Yb. We adapt this model from Ref.~\cite{Lehec2018Laser}, but we refit the MQDT model parameters of the $^1S_0$, $^{1,3}D_2$, $^{1,3}P_1$ Rydberg series in a simultaneous, 42-parameter fit to the previously measured \cite{Camus1969spectre,Meggers1970First,Camus1980Highly,Aymar1980Highly,Aymar1984three,Maeda1992Optical,Lehec2017PhD,Lehec2018Laser} and newly measured data presented in the supplemental material~\cite{supplementary}. The resulting MQDT models are presented in Tab.~S1, Tab.~S2, and Tab.~S3. 

In addition, to account for singlet-triplet mixing, the model includes a rotation around the $^1D_2$ and $^3D_2$ channels in the $U_{i\alpha}$ matrix. To determine the magnitude of the singlet-triplet mixing, we use the $g$-factors of the $n\,^{1,3}D_2$ Rydberg series and its perturbers measured by Zerne and co-workers \cite{zerne1996lande} to obtain a value for the singlet-triplet mixing angle between the $^1D_2$ and $^3D_2$ channels. To that extent, we vary the energy-dependent mixing angle $\theta_{12}$ (Eq.~(\ref{eq:theta_energydep})), and calculate the $g$-factor in using Eq.~(22) of Ref.~\cite{Robicheaux2018theory}. In addition, we treat the $g$-factors of the perturbing states with only partially known electronic configuration as parameters of our model. To obtain the best agreement between experiment and theory, we introduce an energy dependent mixing angle, as defined in Eq.~(\ref{eq:theta_energydep}). The results of the fit are presented in Fig.~\ref{fig:174Yb_13D2_gfactors} and the MQDT model is summarized in Tab.~S2. From our MQDT model, we estimate a triplet character in the $^1D_2$ series between \SI{14.0}{\percent} and
\SI{15.5}{\percent} for $30 \leq n \leq100$. This is in agreement with an estimated triplet admixture into the $36\,^1D_2$ state of \SI{19(6)}{\percent} presented in Ref.~\cite{Maeda1992Optical}.

\begin{table}[tb]
\caption{\label{tab:174_3D1_3D3} Rydberg-Ritz expansion coefficients of the $n\,^3D_1$ and $n\,^3D_3$ Rydberg states of Yb, inferred from laser spectroscopy of D $F=3/2$ and $F=5/2$ Rydberg states of $^{171}$Yb.
}
\begin{ruledtabular}
\begin{tabular}{cS[table-format=4.8]S[table-format=4.8]}
 & \multicolumn{1}{c}{$n\,^3D_1$} & \multicolumn{1}{c}{$n\,^3D_3$} \\
\colrule
$\mu_0$  & 2.75258093 & 2.72895315 \\
$\mu_2$  & 0.3826  & -0.2065\\
$\mu_4$  & -483.1 & 220.5 \\
\end{tabular}
\end{ruledtabular}
\end{table}

\subsection{$^3D_1$ and $^3D_3$}

The $^3D_1$ and $^3D_3$ Rydberg states of $^{174}$Yb are not directly laser accessible by two-photon laser spectroscopy through the $6s6p\,^1P_1$ intermediate state. A few $^3D_1$ and $^3D_3$ Rydberg states have been observed in Ref.~\cite{Okuno2022High} and Ref.~\cite{Wilson2022PhD}. For the purposes of this work, we infer the quantum defects of the $^3D_1$ and $^3D_3$ from laser accessible D $F=3/2$ and D $F=5/2$ Rydberg states in $^{171}$Yb, as discussed in App.~\ref{sec:APP_171_Dstates}. 

The quantum defects of the two series are well described by a Rydberg-Ritz model

\begin{equation}
    \mu(n)=\mu_0+\frac{\mu_2}{\left(n-\mu_0\right)^2}+\frac{\mu_4}{\left(n-\mu_0\right)^4}
\end{equation}

for the states studied in this work with $\nu>30$. The obtained expansion coefficients for the two series are presented in Tab.~\ref{tab:174_3D1_3D3}.

The inferred quantum defects for the $n\,^3D_3$ Rydberg series, are consistent with the assignment of measurements in $^{174}$Yb presented in Ref.~\cite{Okuno2022High} to within the stated measurement uncertainties of \SI{10}{MHz}. The inferred quantum defects for the $n\,^3D_1$ Rydberg series are consistent with the assignment of measurements in $^{174}$Yb presented in Refs.~\cite{Okuno2022High,Wilson2022PhD}, but the Rydberg state energies deviate up to \SI{40}{MHz} from the values presented in Ref.~\cite{Okuno2022High}, which could be due to the limited number of D $F=3/2$ states with dominant $^3D_1$ character observed in this work.

\section{Additional spectroscopic data and models for $^{171}$Yb}\label{sec:171Yb_addtional _states}

In this appendix, we give additional details about the spectroscopic measurements and MQDT model development for $^{171}$Yb. 

\subsection{S $F=1/2$  and $F=3/2$ states}\label{sec:APP_171Yb_Sstates}

There is only a single $F=3/2$ series with $L=0$, which converges to the $F_c=1$ threshold and has $^3S_1$ character. We have measured the energy spacing between several S $F=3/2$ Rydberg states and S $F=1/2$ or D $F=5/2$ Rydberg states using microwave spectroscopy in an atomic beam. The experimental transition frequencies are presented in Tab.~S29. Due to the small number of measured transitions to S $F=3/2$ Rydberg states, we model this series using the Rydberg-Ritz model introduced for $^{174}$Yb \cite{Wilson2022Trapping} with parameters optimized in the fit to the S $F=1/2$ Rydberg states (presented in Tab.~S6). The experimental and theoretical energies of the S $F=3/2$ Rydberg series agree to within the uncertainty of the initial state energy used in the microwave spectroscopy.

\begin{figure}[tb]
	\centering
    \includegraphics[width=\linewidth]{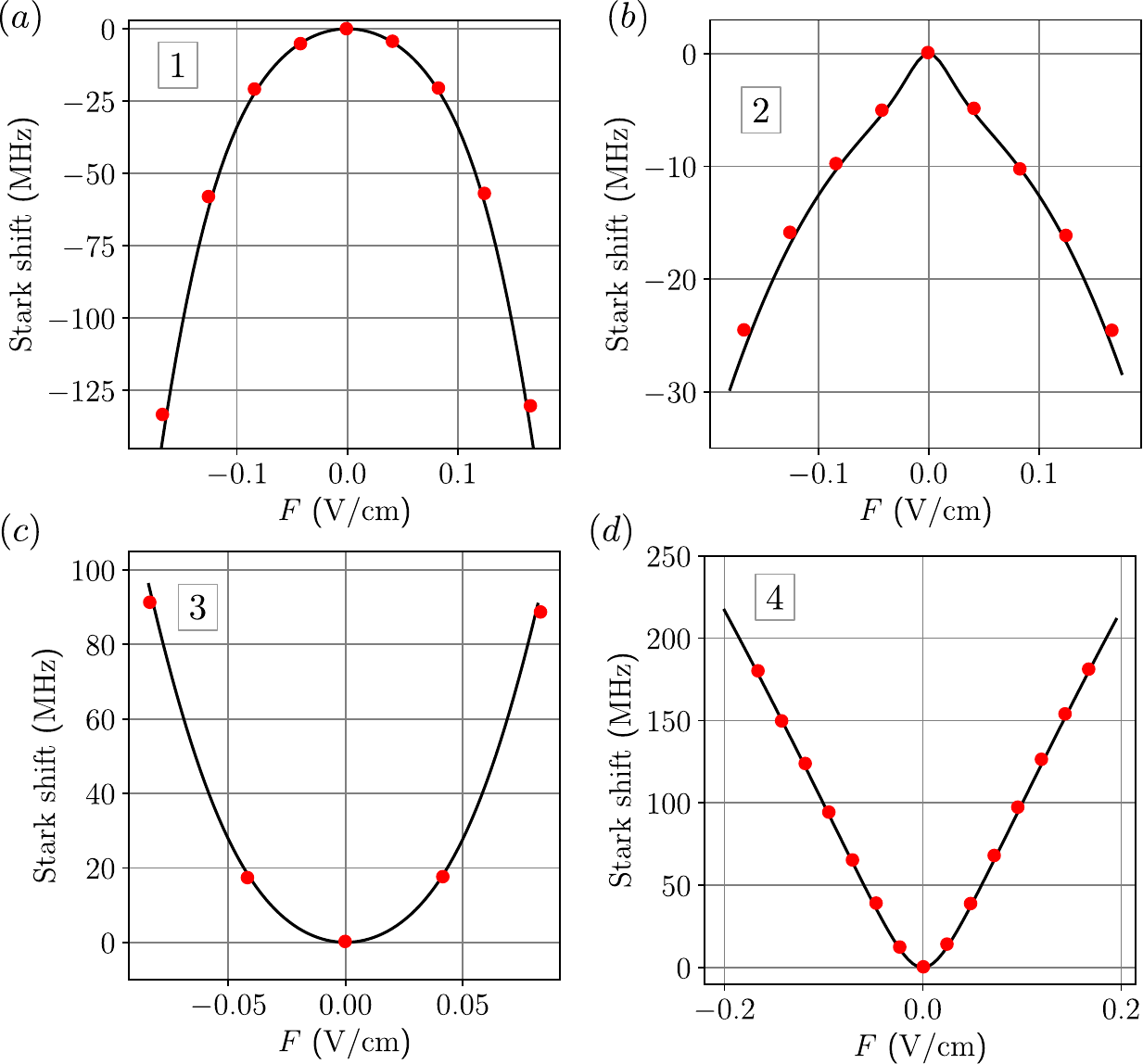}
	\caption{\label{fig:171Yb_individual_Stark} Experimental Stark shifts (red dots) of $^{171}$Yb $\ket{\nu_{F_c=1},L=0,F=1/2}$ Rydberg states with $(a)$ $\nu_{F_c=1}\approx65.68$, $(b)$ $\nu_{F_c=1}\approx66.68$, $(c)$ $\nu_{F_c=1}\approx67.68$, $(d)$ $\nu_{F_c=1}\approx64.09$ compared with theoretical Stark shifts (solid black line). The extracted experimental Stark shifts at low electric field are presented in subpanels Fig~\ref{fig:171Yb_LuFano_overview}$\,(d)$ and $(e)$ as indicated by labels 1 to 4.}
\end{figure}

In Fig.~\ref{fig:171Yb_LuFano_overview}, we presented measurements of Stark shifts of S $F=1/2$ Rydberg states. Most Rydberg states exhibit a quadratic Stark shift at small electric fields, common for non-degenerate Rydberg states, and we extract a static polarizability by a quadratic fit to the observed Stark shift. However, for cases in which the Rydberg state of interest is very nearly degenerate with other Rydberg states (shaded gray in Fig.~\ref{fig:171Yb_LuFano_overview}), the Stark shift is not quadratic even at very small electric fields. In Fig.~\ref{fig:171Yb_individual_Stark}, we present a field-dependent shift measurement together with a non-perturbative calculation, finding excellent agreement, highlighting the accuracy of the obtained Rydberg state energies and wavefunctions obtained by our MQDT treatment.

\begin{figure}[tb]
	\centering
    \includegraphics[width=\linewidth]{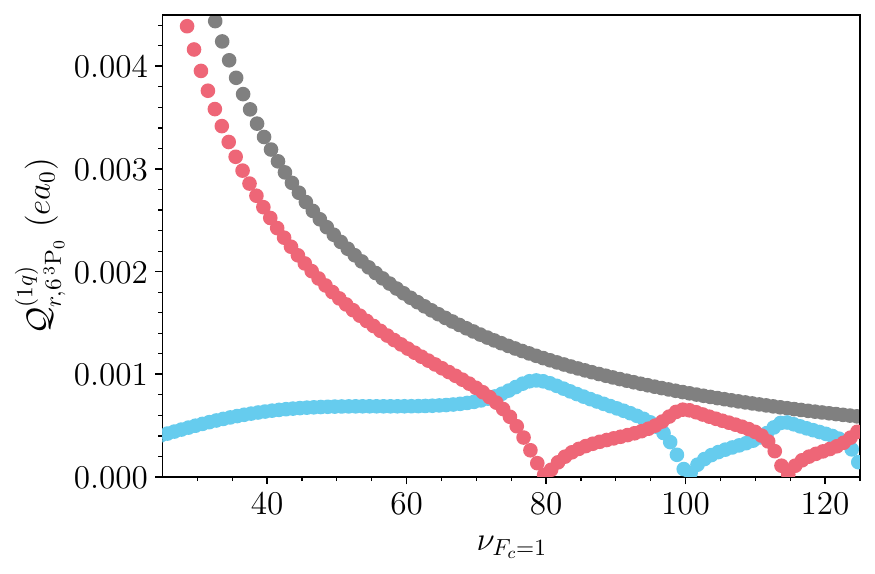}
	\caption{\label{fig:171Yb_matrixelements_F12_F32} Transition dipole matrix elements $\mathcal{Q}_{r,6\,^3P_0}$ between the $6s6p\,^3P_0(m_F=+1/2)$ metastable and Rydberg state $r$, where $r=\ket{\nu,L=0,F=3/2,m_F=+3/2}$ (gray) and $r=\ket{\nu,L=0,F=1/2,m_F=-1/2}$ (red and blue, color code as in Fig.~\ref{fig:171Yb_Sstates_lufano_g}).}
\end{figure}

Fig.~\ref{fig:171Yb_matrixelements_F12_F32} illustrates calculated transition dipole matrix elements from $6s6p\,^3P_0$ metastable state into S ($F=\{1/2,3/2\}$) Rydberg states of $^{171}$Yb, which are computed using the single active electron approximation. The matrix elements to S $(F=1/2)$ Rydberg states are generally smaller than the matrix elements to S $(F=3/2)$ Rydberg states. However, at low effective principal quantum numbers, the transition dipole matrix element to one of the two S ($F=1/2$) Rydberg series is still comparable to the S $(F=3/2)$ case, due to a dominant triplet character. At larger effective principal quantum numbers, particularly close to near degeneracies between the two S ($F=1/2$) channels, the matrix elements varies strongly with the effective principal quantum number, indicating strong hyperfine-induced singlet-triplet mixing. 

\subsection{P $F=1/2$, $F=3/2$, and $F=5/2$ states}\label{sec:APP_171_Pstates}

Here, we introduce MQDT models for P $F=1/2$, $F=3/2$, and $F=5/2$ Rydberg states of $^{171}$Yb. There are seven $L=1$ series in $^{171}$Yb. In $LS$ coupling, they can be described as $^3P_0 (F=1/2)$, $^1P_1 (F=\{1/2,3/2\})$, $^3P_1 (F=\{1/2,3/2\})$ and $^3P_2 (F=\{3/2,5/2\})$. As in the case of the $\ket{\nu,L=0,F=1/2}$, we use the $^{174}$Yb MQDT models of $^{1,3}P_1$ and $^3P_0$ (for $F=1/2$) and $^{1,3}P_1$ and $^3P_2$ (for $F=3/2$) as a basis for the $^{171}$Yb MQDT model and introduce the hyperfine coupling in the ion-core using a frame transformation (see App.~\ref{sec:mqdt_model_params}).

We optimize the MQDT model parameters for both the $F=1/2$ and $F=3/2$ series using a weighted, least-squared fitting procedure to a dataset containing Rydberg state energies obtained in this work from microwave spectroscopy (Tab.~S30 and Tab.~S33) and from previous laser spectroscopy reported Ref.~\cite{Majewski1985diploma} (Tab.~S31 and Tab.~S34). In our fitting procedure, we treat all the $^{171}$Yb MQDT model parameters originating from the same $^{174}$Yb MQDT models as a single parameter. For example, both the $^{171}$Yb P $F=1/2$ and P $F=3/2$ MQDT models contain a contribution from the $^{1,3}P_1$ MQDT model developed for $^{174}$Yb in Appendix~\ref{sec:174_13P1}. The resulting MQDT parameters for $F=1/2$ and $F=3/2$ are presented in the supplemental material~\cite{supplementary}.

The state energies reported in Tab.~S31 and Tab.~S34 are obtained from isotope shift and hyperfine splitting measurements of the $n\,^1P_1$ and $n\,^3P_{0,1,2}$ Rydberg states using three-photon laser spectroscopy, as presented in previous work~\cite{Majewski1985diploma} (reproduced in Tab.~S36 and Tab.~S37)). The isotope shift and hyperfine splittings in Ref.~\cite{Majewski1985diploma} are given with respect to $6snp\,^1P_1$ and $6snp\,^3P_1$ Rydberg states of $^{176}$Yb (reproduced in Tab.~S38). The absolute uncertainty of the $^{176}$Yb Rydberg state energies presented in Ref.~\cite{Majewski1985diploma} is stated to be \SI{4}{GHz}, but we find that the stated (relative) hyperfine splittings are more precise. To remove systematic errors on the inferred absolute energies of the $^{171}$Yb P $F=1/2$ and $F=3/2$ Rydberg states, we introduce the following treatment.

For Rydberg states with principal quantum number $n>20$, Ref.~\cite{Majewski1985diploma} reports a nearly constant isotope shift between $^{1,3}P_1$ Rydberg states of $^{174}$Yb and $^{176}$Yb. For $n>20$ therefore choose to reference the hyperfine splittings against the accurate energies obtained from the $^{174}$Yb $^{1,3}P_1$ MQDT model presented in App.~\ref{sec:174_13P1}. The remaining isotope shift between $^{171}$Yb and $^{174}$Yb Rydberg states are accounted for, by introducing a constant shift to inferred absolute energies in order to minimize the deviations to absolute energies inferred from microwave intervals to the $\ket{\nu,L=0,F=1/2}$ Rydberg states.

For low lying states $^{1,3}P_1$ of $^{174}$Yb, we observe significant deviations between experimental and predicted energies for $^{174}$Yb (App.~\ref{sec:174_13P1}). For $n\leq20$, we therefore choose to infer absolute energies of the observed $P$ Rydberg states in $^{171}$Yb by referencing the hyperfine splitting measurements to the observed absolute energies of $^{176}$Yb, as presented in Ref.~\cite{Majewski1985diploma} (reproduced in Tab.~S38). When applying this treatment to Rydberg state with $n>20$, we observe a systematic shift of the inferred energies compared to the more accurate absolute state energies obtained from microwave intervals to the $\ket{\nu,L=0,F=1/2}$ Rydberg states. The mean deviation is approximately \SI{1.4}{GHz} and is well within the \SI{4}{GHz} uncertainty stated for the $^{176}$Yb term values reported in Ref.~\cite{Majewski1985diploma}. In an attempt to remove this systematic shift, we add a constant energy offset to all $^{171}$Yb $P$ Rydberg states that were referenced to the $^{176}$Yb energies.

The resulting inferred absolute energies of $\ket{\nu,L=1,F=1/2}$ and $\ket{\nu,L=1,F=3/2}$ Rydberg states are presented in Tab.~S31 and Tab.~S34. 

\begin{figure*}[tb]
	\centering
    \includegraphics[width=\linewidth]{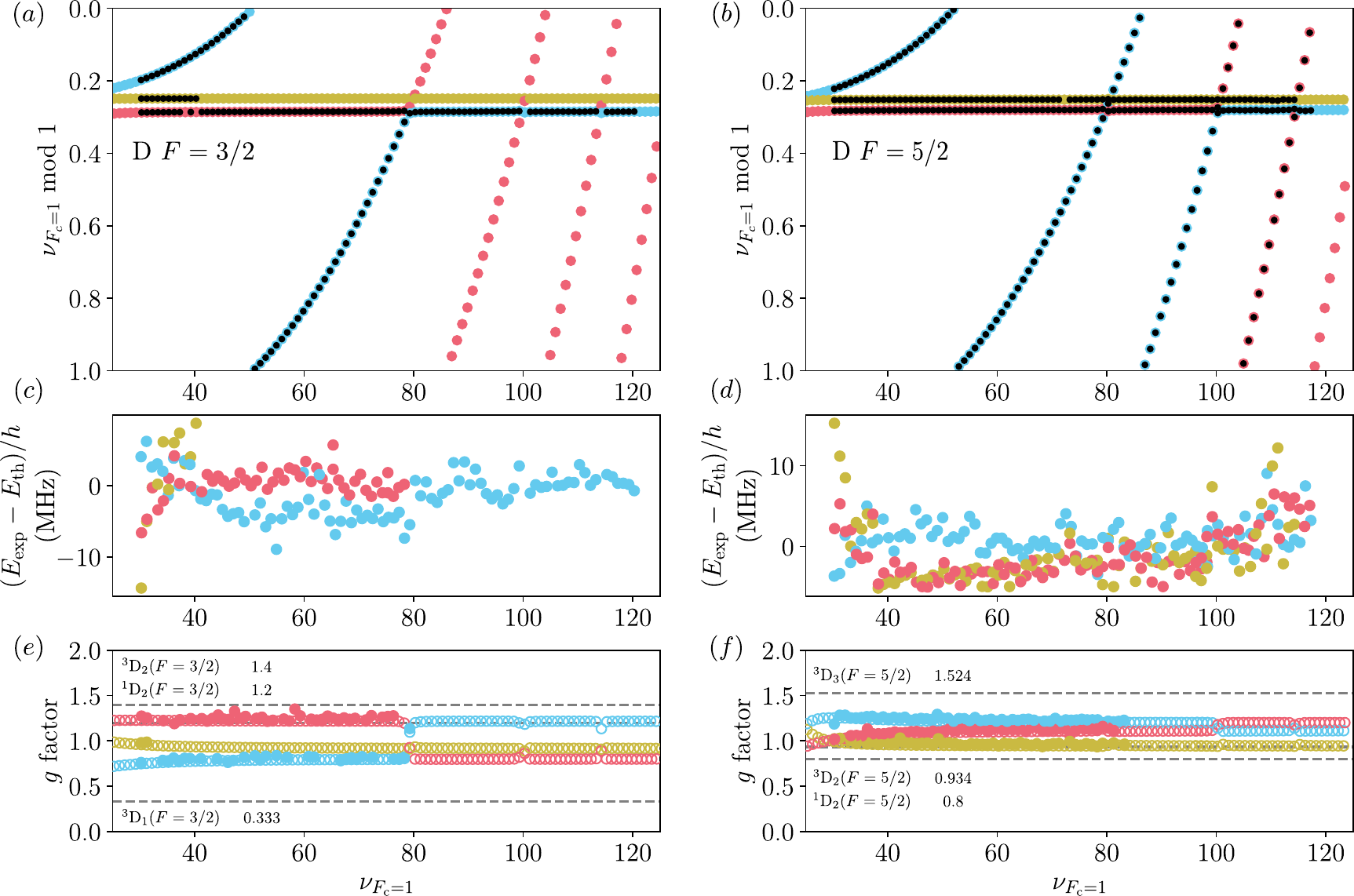}
	\caption{\label{fig:171Yb_Dstates_energy_gfactor} Lu-Fano-type plot of the $^{171}$Yb $(a)$ $\ket{\nu,L=2,F=3/2}$ and $(b)$ $\ket{\nu,L=2,F=5/2}$ Rydberg states. The experimentally observed Rydberg states are indicated by black dots and compared to theoretically obtained energies from the MQDT model (colored dots) presented in Tab.~S9 and Tab.~S10. $(c)$ and $(d)$ Deviations between the experimentally observed and theoretically calculated state energies. $(d)$ and $(e)$ Experimentally observed (full circles) and theoretically calculated $g$-factors of the $\ket{\nu,L=2,F=3/2}$ and $\ket{\nu,L=2,F=5/2}$ Rydberg states, respectively. Color coding as in $(a)$ and $(b)$. The gray dashed lines indicate the values of Land\'e $g$ factors in pure $LS$ coupling.}
\end{figure*}

\subsection{D $F=1/2$, $F=3/2$, $F=5/2$, and $F=7/2$}\label{sec:APP_171_Dstates}

There are eight $L=2$ series in $^{171}$Yb. In $LS$ coupling, they can be described as $^3D_1 (F=\{1/2,3/2\})$, $^1D_2 (F=\{3/2,5/2\})$, $^3D_2 (F=\{3/2,5/2\})$ and $^3D_3 (F=\{5/2,7/2\})$. In $^{174}$Yb, spin-orbit coupling mixes the two $J=2$ series as discussed in Section~\ref{sec:174Ybmain}; in $^{171}$Yb, hyperfine coupling additionally mixes the three $F=3/2$ series and the three $F=5/2$ series.

As in the case of the $\ket{\nu,L=0,F=1/2}$ Rydberg states, we measure the energies of $\ket{\nu,L=2,F=3/2}$ and $\ket{\nu,L=2,F=5/2}$ Rydberg states by laser spectroscopy by a two-photon transition via the  $6s6p\,^1P_1(F=3/2)$ intermediate state. Transition frequencies to all measured $\ket{\nu,L=2,F=3/2}$ and $\ket{\nu,L=2,F=5/2}$ states are summarized in the supplemental material~\cite{supplementary} and plotted on a Lu-Fano-type plot in Fig.~\ref{fig:171Yb_Dstates_energy_gfactor}a.

We use a least-square procedure to obtain MQDT parameters for the $F=3/2$ and $F=5/2$ series. As in the case of the P Rydberg states, we treat the MQDT parameters that originate from the $^{174}$Yb models and occur in both the $F=3/2$ and $F=5/2$ MQDT models as a single parameter. The developed MQDT models describe both the Rydberg state energies and the measured $g$-factors (summarized in the supplemental material~\cite{supplementary}) well over nearly the entire energy range. Significant deviations for both $F=3/2$ and $F=5/2$ occur towards low effective principal quantum numbers. These deviations could arise from unaccounted perturbers in the MQDT or from channel interactions with additional $(F=3/2)^e$ Rydberg series. In the case of $F=5/2$, we additionally observe slight deviations between experiment and theory, that could be caused by uncompensated stray electric fields. 

There is only a single D Rydberg series with $F=1/2$ ($F=7/2$), converging to the $F_c=1$ threshold with $^3D_1$ ($^3D_3$) character. Because we use the $6s6p\,^1P_1(F=3/2)$ (dominantly singlet character) state as an intermediate state, the $\ket{\nu,L=2,F=1/2}$ (dominantly triplet character) and the $\ket{\nu,L=2,F=7/2}$ Rydberg states are inaccessible by direct laser spectroscopy. However, the fitted $F=3/2$ and $F=5/2$ series MQDT models include a Rydberg-Ritz model for the $^3D_1$ and $^3D_3$ series (Tab.~S9 and Tab.~S10). Therefore, we use this model to predict the positions of the $F=1/2$ and $F=7/2$ series.

\begin{figure}[tb]
	\centering
    \includegraphics[width=\linewidth]{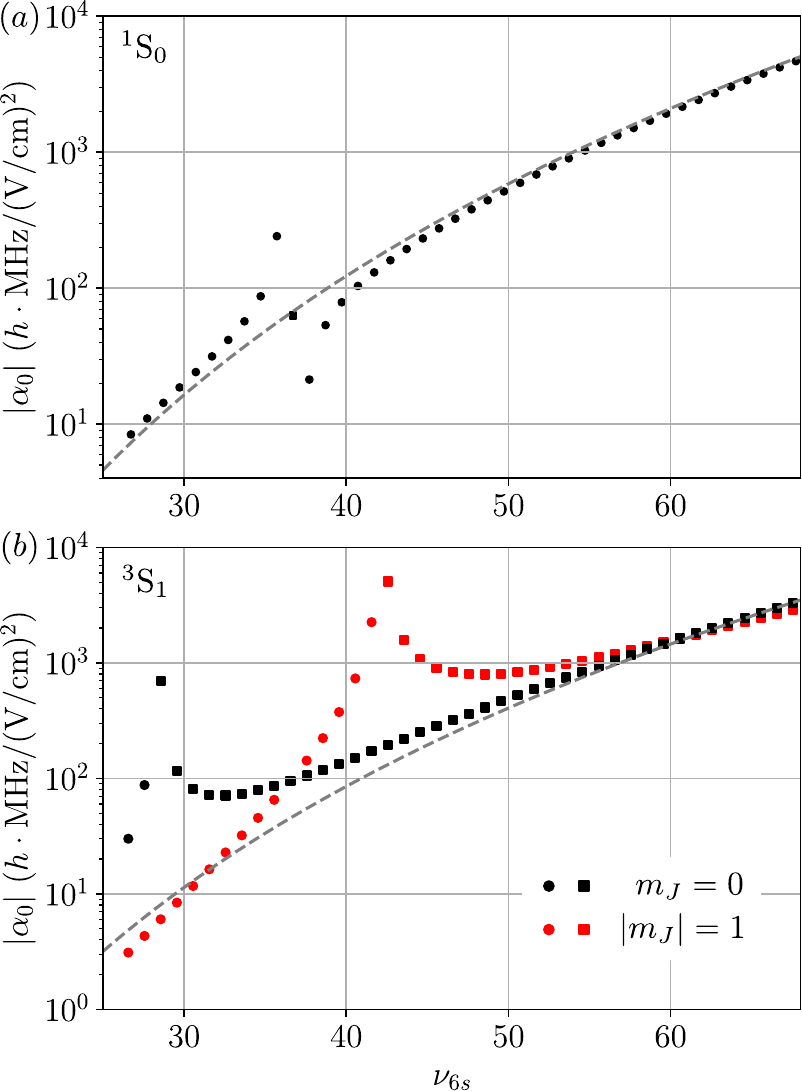}
\caption{\label{fig:174Yb_poloverview} Predicted absolute static dipole polarizabilities of Rydberg series of $^{174}$Yb for $(a)$ $6sns\,^{1}S_0$ and $(b)$ $6sns\,^{3}S_1$. Circles indicate positive values of $\alpha_0$, whereas squares indicate negative values of $\alpha_0$. The gray, dashed lines in $(a)$ and $(b)$ serve as a guide to the eye, indicating a $\nu_{6s}^7$ scaling.}
\end{figure}

\begin{figure}[tb]
	\centering
    \includegraphics[width=\linewidth]{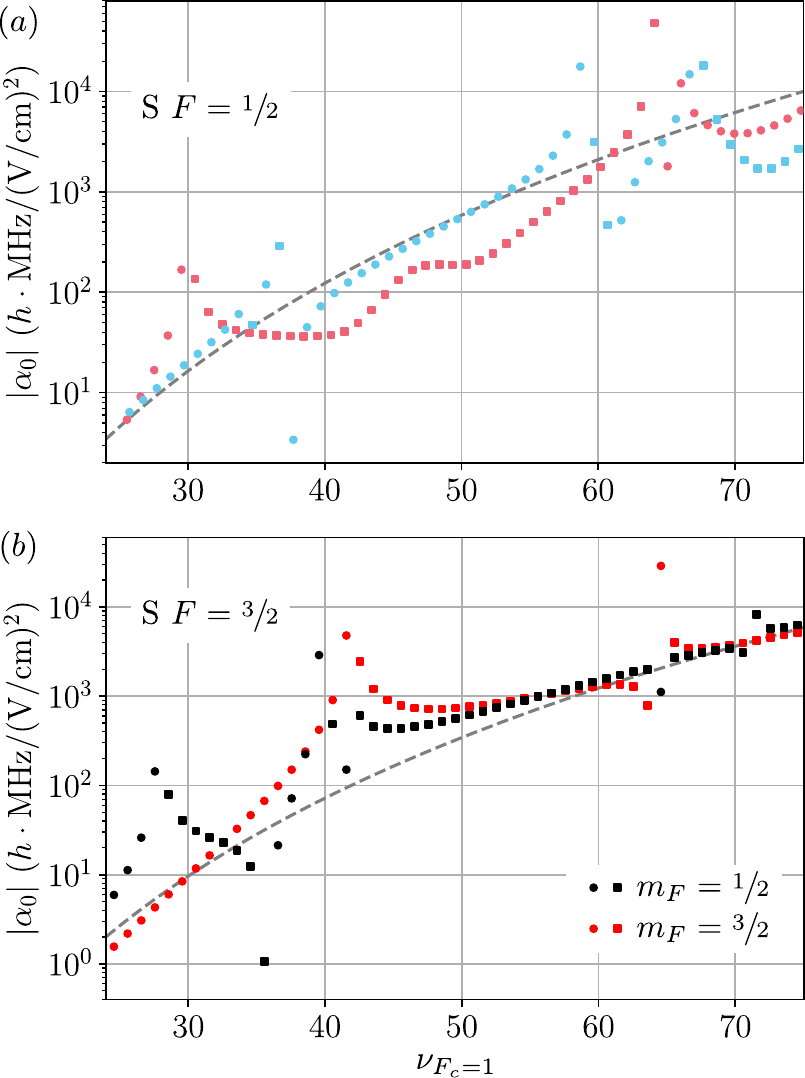}
\caption{\label{fig:171Yb_polarization_SF12_F32_ndependence} Predicted absolute static dipole polarizabilities of Rydberg series of $^{171}$Yb with $(a)$ S~$F=1/2$ (color code as in Fig.~\ref{fig:171Yb_Sstates_lufano_g}) and $(b)$ S~$F=3/2$. Circles indicate positive values of $\alpha_0$, whereas squares indicate negative values of $\alpha_0$. The gray, dashed lines in $(a)$ and $(b)$ serve as a guide to the eye, indicating a $\nu_{F_c=1}^7$ scaling.}
\end{figure}

\section{Polarizability trends in $^{174}$Yb and $^{171}$Yb S Rydberg states}\label{sec:polarizability}

In Fig.~\ref{fig:174Yb_poloverview} and Fig.~\ref{fig:171Yb_polarization_SF12_F32_ndependence} we present predicted polarizabilities for $L=0$ states of both $^{174}$Yb and $^{171}$Yb. We note several interesting features. First, the $^{174}$Yb $^3S_1$ polarizability is positive at low-$n$ and negative at high-$n$, as the strongly perturbed $^3P_0$ and $^3P_1$ series cross the $^3S_1$ series in energy. Second, the $^{171}$Yb polarizabilities have a much more complex behavior, with multiple sign changes resulting from a number of resonances.

\section{Rydberg-Rydberg-interaction trends}
\label{sec:interaction_trends}
\begin{figure}[tb]
	\centering
    \includegraphics[width=\linewidth]{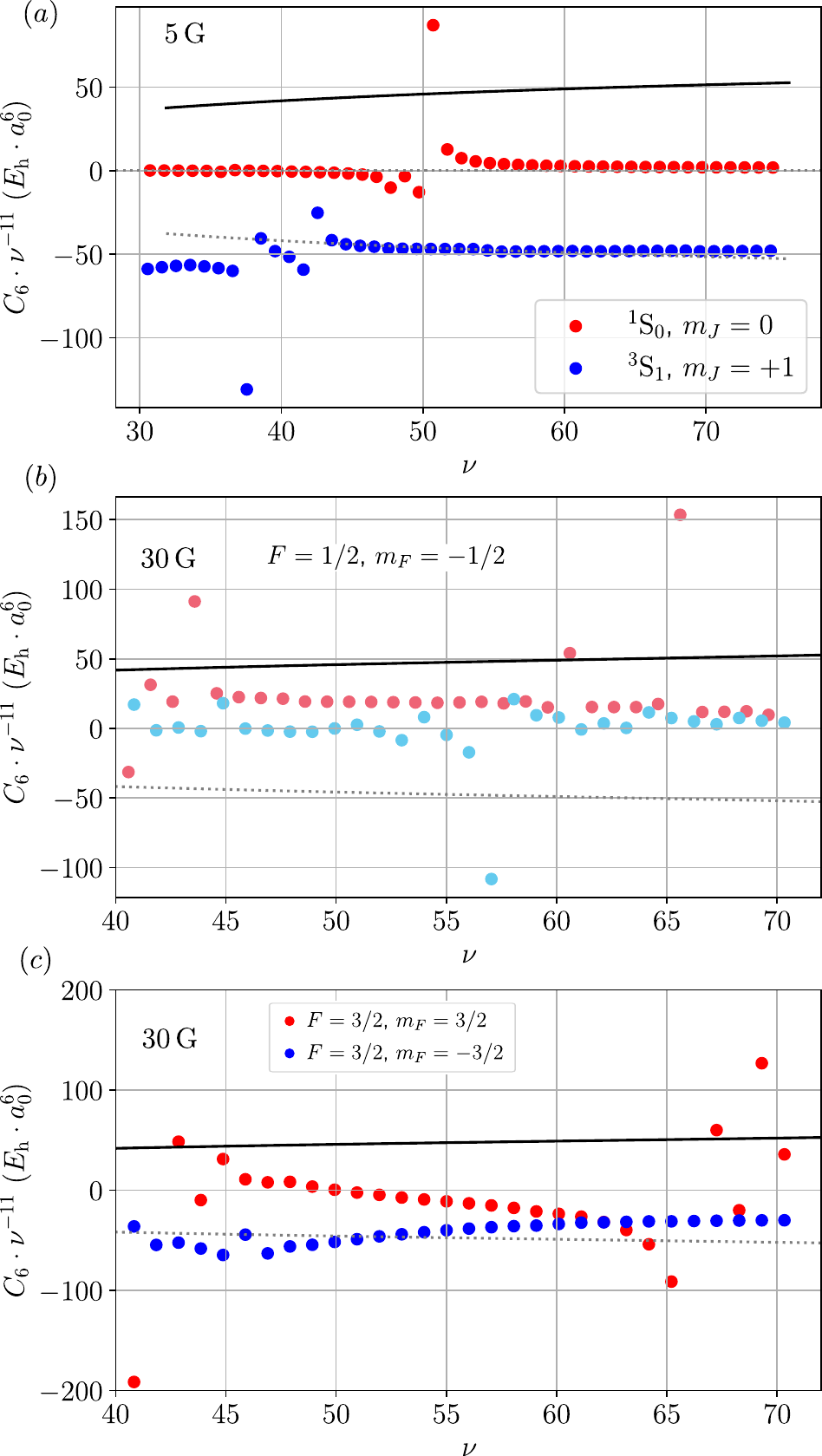}
	\caption{\label{fig:171and174Yb_Interaction_overview_Sstates}  Predicted scaled $C_6$ coefficients of $(a)$ $^{174}$Yb $^1S_0$ (red dots) and $^3S_1$ (blue dots) Rydberg states at a magnetic field of \SI{5}{G} ($\theta=\uppi/2$), $(b)$ and $(c)$ $^{171}$Yb $\ket{\nu,L=0,F=1/2}$ and $\ket{\nu,L=0,F=3/2}$ Rydberg states at a magnetic field of \SI{30}{G} ($\theta=\uppi/2$). $\theta$ is the angle between the magnetic field and the inter-atomic axis. For comparison, the scaled $C_6$ coefficients of $n\,^{2}S_{1/2}$ Rydberg states of Rb are given by solid ($C_6$) and dotted ($-C_6$) black lines. The color code in $(b)$ corresponds to the color code in Fig.~\ref{fig:171Yb_Sstates_lufano_g}.}
\end{figure}

\begin{figure}[tb]
	\centering
    \includegraphics[width=\linewidth]{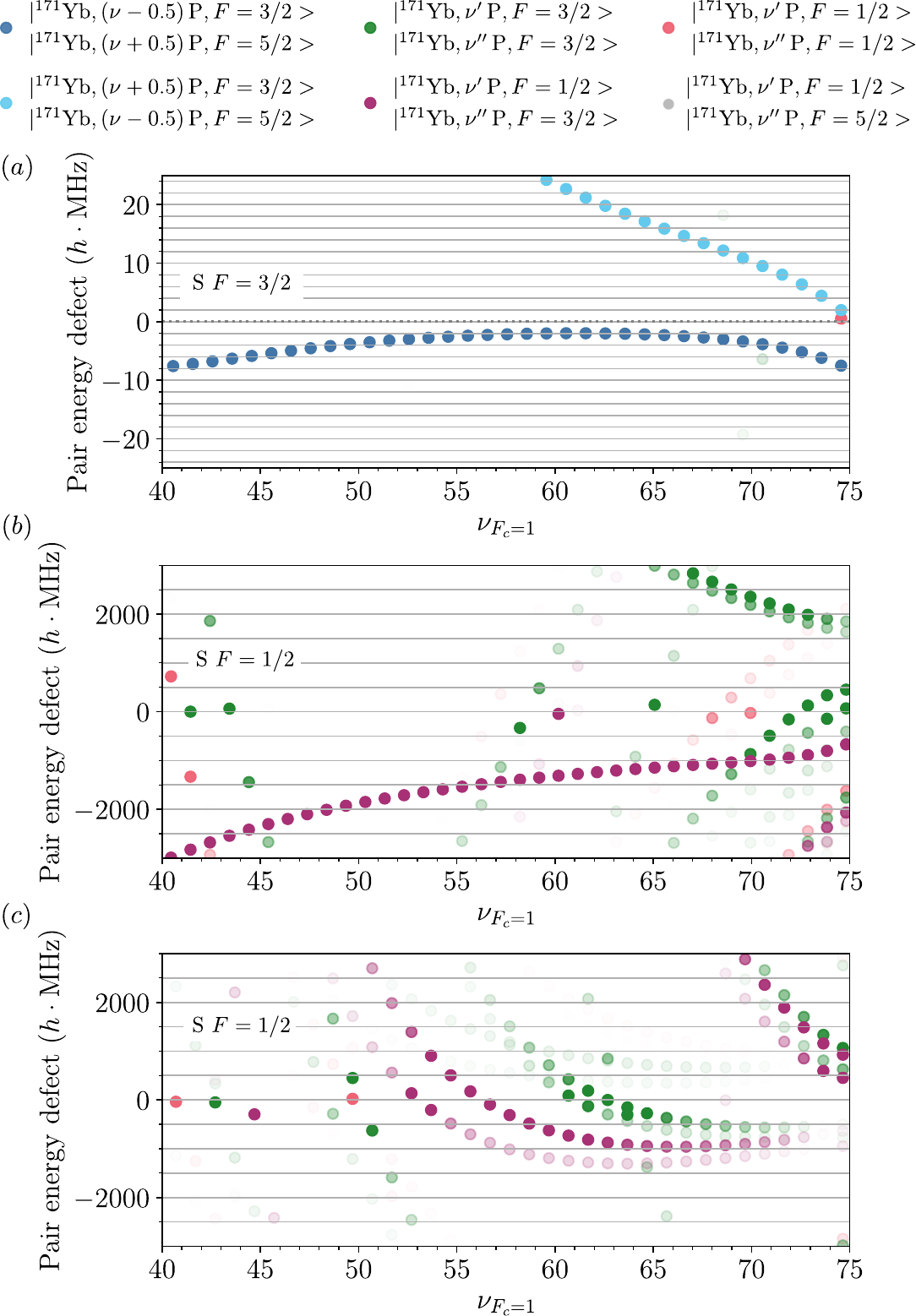}
	\caption{\label{fig:171Yb_pairdetuning_F12_F32} $^{171}$Yb pair energy defects between $\ket{\nu_{F_c=1},L=0,F}\ket{\nu_{F_c=1},L=0,F}$ and the nearest dipole coupled $\ket{\nu_{F_c=1},L=1,F}\ket{\nu_{F_c=1},L=1,F'}$ states for $(a)$ $F=3/2$, and $(b)$ and $(c)$ $F=1/2$. The color of the markers indicates the combination of hyperfine states of the $\ket{\nu,L=1,F}\ket{\nu',L=1,F'}$ pair state according to the legend. The opacity encodes $\left(\frac{(d_1d_2)^2}{\Delta E}\right)\left(\frac{1}{\nu_{F_c=1}}\right)^8$, where $d_i$ are the dipole matrix elements between the involved Rydberg states and $\Delta E$ corresponds to the pair defect.}
\end{figure}

The van der Waals coefficients $C_6$ for the $L=0$ Rydberg states of $^{174}$Yb and $^{171}$Yb Rydberg states are presented in Fig.~\ref{fig:171and174Yb_Interaction_overview_Sstates}. In $^{174}$Yb, the $C_6$ coefficients of $6sns\,^1S_0$ Rydberg series are unusually small, consistent with the single-channel calculations of Ref.~\cite{Vaillant2012Long} (we additionally note a F\"orster resonance around $\nu\approx50$, which arises from the inclusion of singlet-triplet mixing in the $^{1,3}P_1$ Rydberg series). Similar F\"orster resonances have also been predicted for $^3D_2$ and $^{1,3}F_3$ states of $^{88}$Sr \cite{Vaillant2015Intercombination}. In contrast, the $C_6$ coefficients of $6sns\,^3S_1$ Rydberg states of $^{174}$Yb are larger, consistent with previous estimates~\cite{Wilson2022Trapping} and observations~\cite{Burgers2022Controlling}, but calculated here for the first time. The $C_6$ coefficients of this series are the same magnitude as for the $ns_{1/2}$ Rydberg states of rubidium, but with opposite sign (\emph{i.e.}, attractive).

As in the case for $^1S_0$ Rydberg states of 
$^{174}$Yb, the singlet connected $\ket{\nu,L=0,F=1/2}$ Rydberg states of $^{171}$Yb (blue dots in Fig.~\ref{fig:171and174Yb_Interaction_overview_Sstates}) have small $C_6$ coefficients in the range $40\leq\nu\leq70$. The the triplet connected $\ket{\nu,L=0,F=1/2,m_F}$ states have $C_6$ coefficients that are approximately half the magnitude of Rb, with the same sign (\emph{i.e.}, repulsive).

As pointed out in Section~\ref{sec:RydbergRydberg_Yb171}, $\ket{\nu,L=0,F=3/2}\ket{\nu,L=0,F=3/2}$ Rydberg pair states of $^{171}$Yb have small F\"orster defects over a large range of effective principal quantum number $\nu$ (Fig.~\ref{fig:171Yb_pairdetuning_F12_F32}). It is surprising that the F\"orster defect is so small over such a large range of $\nu$, which results from a combination of series perturbers and hyperfine coupling. To calculate a meaningful asymptotic $C_6$ coefficient, we calculate the pair potentials under a \SI{30}{G} magnetic field to lift the degeneracy somewhat (Fig.~\ref{fig:171and174Yb_Interaction_overview_Sstates}c). The extracted $C_6$ coefficients are not meaningful at short separations, which are dominated by resonant dipole-dipole interaction, as shown in  Fig.~\ref{fig:F32_54.56_interactions}.

\pagebreak
\widetext
\clearpage % added by PV
~\vspace{2cm} % added by JG
\begin{center}
\textbf{\large Supplemental Material}
\end{center}
%%%%%%%%%% Merge with supplemental materials %%%%%%%%%%
%%%%%%%%%% Prefix a "S" to all equations, figures, tables and reset the counter %%%%%%%%%%
\setcounter{equation}{0}
\setcounter{figure}{0}
\setcounter{table}{0}
\setcounter{section}{0}
\makeatletter
\renewcommand{\thefigure}{S\arabic{figure}}
\renewcommand{\thetable}{S\arabic{table}}
\renewcommand{\thesection}{S\Roman{section}}
\renewcommand{\thesubsection}{\Alph{subsection}}
\renewcommand{\bibnumfmt}[1]{[S#1]}
\renewcommand{\citenumfont}[1]{S#1} % might need to play with \renew v \new
%%%%%%%%%% Prefix a "S" to all equations, figures, tables and reset the counter %%%%%%%%%%

\onecolumngrid
\renewcommand\appendixname{}
\section{MQDT model parameters}
\label{sec:MQDTparams}
\subsection{$^{174}$Yb}

\begin{table*}[ht]
\caption{\label{tab:174MQDT_1S0} Six-channel MQDT model parameters for the $6sns\,^{1}S_0$ series of $^{174}$Yb obtained from fit to spectroscopic data presented in Tabs.~\cref{tab:174Spec_1S0,tab:174Spec_13D2,tab:174Spec_Microwave_DD,tab:174Spec_Microwave_SS,tab:174Spec_Microwave_SD,tab:174YbSpec_3P1,tab:174Spec_Microwave_13p1_S,tab:174Spec_Microwave_13p1_D}. The electronic configuration of the perturbing Rydberg channels $4f^{13}5d6snp$ is abbreviated as $5d$. The rotations $\mathcal{R}(\theta_{ij})$ are applied in the order $\mathcal{R}(\theta_{12})\mathcal{R}(\theta_{13})\mathcal{R}(\theta_{14})\mathcal{R}(\theta_{34})\mathcal{R}(\theta_{35})\mathcal{R}(\theta_{16})$.
}
\begin{ruledtabular}
\begin{tabular}{ccccccc}
$i$,$\bar\alpha$,$\alpha$ & 1 & 2 & 3& 4& 5 & 6\\
\colrule
$\ket{i}$ & $(6s_{1/2})(ns_{1/2})$& $5d$ \textbf{a} & $(6p_{3/2})(np_{3/2})$ & $5d$ \textbf{b} & $(6p_{1/2})(np_{1/2})$ & $5d$ \textbf{c} \\ 
$I_i$ ($\mathrm{cm}^{-1}$) & $50\,443.070393$ & $83\,967.7$ & $ 80\,835.39$ & $83\,967.7$ & $77\,504.98$& $83\,967.7$\\
$\ket{\bar\alpha}$ & $6sns\,^1S_0$ & $5d$ \textbf{a} & $6pnp\,^1S_0$ & $5d$ \textbf{b} & $6pnp\,^3P_0$ & $5d$ \textbf{c} \\ 
$\mu_\alpha^{(0)}$ & 0.355097325 & 0.204537279 & 0.116394359& 0.295432196&0.25765161 & 0.155807042 \\
$\mu_\alpha^{(2)}$ & 0.278368431 & 0.0 & 0.0 & 0.0 & 0.0 & 0.0\\
\colrule
$V_{\bar\alpha\alpha}$  & $\theta_{12}=0.12654859$ & $\theta_{13}=0.30010744$ & $\theta_{14}=0.05703381$ & $\theta_{34}=0.11439805$  & $\theta_{35}=0.09864375$ & $\theta_{16}=0.14248210$  \\
\colrule
$U_{i\bar\alpha}$   & 1&0&0&0&0&0\\
                    & 0&1&0&0&0&0\\
                    & 0&0&$-\sqrt{2/3}$&0&$\sqrt{1/3}$&0\\
                    & 0&0&0&1&0&0\\
                    & 0&0&$\sqrt{1/3}$&0&$\sqrt{2/3}$&0\\
                    & 0&0&0&0&0&1\\

\end{tabular}
\end{ruledtabular}
\end{table*}

\begin{table*}[ht]
\caption{\label{tab:174MQDT_13D2} Five-channel MQDT model parameters for the $6snd\,^{1,3}D_2$ series of $^{174}$Yb obtained from a global fit to spectroscopic data presented in Tabs.~\cref{tab:174Spec_1S0,tab:174Spec_13D2,tab:174Spec_Microwave_DD,tab:174Spec_Microwave_SS,tab:174Spec_Microwave_SD,tab:174YbSpec_3P1,tab:174Spec_Microwave_13p1_S,tab:174Spec_Microwave_13p1_D}. The electronic configuration of the perturbing Rydberg channels $4f^{13}5d6snp$ is abbreviated as $5d$. The rotations $\mathcal{R}(\theta_{ij})$ are applied in the order $\mathcal{R}(\theta_{12}(\epsilon))\mathcal{R}(\theta_{13})\mathcal{R}(\theta_{14})\mathcal{R}(\theta_{24})\mathcal{R}(\theta_{15})\mathcal{R}(\theta_{25})$. The channel-$g$-factors $g^*$ are obtained in $LS$ coupling for channels 1 and 2, and by a fit to experimentally observed $g$ factors \cite{Szerne1996lande} for the remaining channels.}
\begin{ruledtabular}
\begin{tabular}{cccccc}
$i$,$\bar\alpha$,$\alpha$ & 1 & 2 & 3& 4& 5 \\
\colrule
$\ket{i}$ & $(6s_{1/2})(nd_{5/2})$&  $(6s_{1/2})(nd_{3/2})$ & $5d$ \textbf{a} & $5d$ \textbf{b} & $(6p)(np)$ \\ 
$I_i$ ($\mathrm{cm}^{-1}$) & $50\,443.070393$ & $50\,443.070393$ & $83\,967.7$ &  $83\,967.7$ & $79\,725.35$\\
$\ket{\bar\alpha}$ & $6snd\,^1D_2$ &  $6snd\,^3D_2$ & $5d$ \textbf{a} &  $5d$ \textbf{b} & $6pnp\,^1D_2$  \\ 
$\mu_\alpha^{(0)}$ & 0.729500971 & 0.75229161 & 0.196120406& 0.233821165&0.152890218  \\
$\mu_\alpha^{(2)}$ & $-0.0284447537$ & 0.0967044398 & 0.0 & 0.0 & 0.0\\
\colrule
$V_{\bar\alpha\alpha}$  & $\theta_{12}^{(0)}=0.21157531$ & $\theta_{13}=0.00522534111$ & $\theta_{14}=0.0398754262$ & $\theta_{24}=-0.00720265975$  & $\theta_{15}=0.104784389$   \\
& $\theta_{12}^{(2)}=-15.3844$&  $\theta_{25}=0.0721775002$ & & & \\
\colrule
$g^*$ & 1  & 1.1670 &1.1846 & 0.9390& 1.2376 \\
\colrule
$U_{i\bar\alpha}$   & $\sqrt{3/5}$&$\sqrt{2/5}$&0&0&0\\
                    & $-\sqrt{2/5}$&$\sqrt{3/5}$&0&0&0\\
                    & 0&0&1&0&0\\
                    & 0&0&0&1&0\\
                    & 0&0&0&0&1\\

\end{tabular}
\end{ruledtabular}
\end{table*}

\begin{table*}[ht]
\caption{\label{tab:174MQDT_13P1} Six-channel MQDT model parameters for the $6snp\,^{1,3}P_1$ series of $^{174}$Yb obtained from fit to spectroscopic data presented in Tabs.~\cref{tab:174Spec_1S0,tab:174Spec_13D2,tab:174Spec_Microwave_DD,tab:174Spec_Microwave_SS,tab:174Spec_Microwave_SD,tab:174YbSpec_3P1,tab:174Spec_Microwave_13p1_S,tab:174Spec_Microwave_13p1_D}. The electronic configuration of the perturbing Rydberg channels $4f^{13}5d6snd$ is abbreviated as $5d$. The rotations $\mathcal{R}(\theta_{ij})$ are applied in the order $\mathcal{R}(\theta_{12}(\epsilon))\mathcal{R}(\theta_{13})\mathcal{R}(\theta_{14})\mathcal{R}(\theta_{15})\mathcal{R}(\theta_{16})\mathcal{R}(\theta_{23})\mathcal{R}(\theta_{24})\mathcal{R}(\theta_{25})\mathcal{R}(\theta_{26})$.
}
\begin{ruledtabular}
\begin{tabular}{ccccccc}
$i$,$\bar\alpha$,$\alpha$ & 1 & 2 & 3& 4& 5 & 6\\
\colrule
$\ket{i}$ & $(6s_{1/2})(np_{3/2})$& $(6s_{1/2})(np_{1/2})$ & $5d$ \textbf{a} & $5d$ \textbf{b}& $5d$ \textbf{c}& $5d$ \textbf{d}\\ 
$I_i$ ($\mathrm{cm}^{-1}$) & $50\,443.070393$ & $50\,443.070393$ & $83\,967.7$ & $83\,967.7$ & $83\,967.7$& $83\,967.7$\\
$\ket{\bar\alpha}$ & $6snp\,^1P_1$ & $6snp\,^3P_1$ & $5d$ \textbf{a} & $5d$ \textbf{b}& $5d$ \textbf{c}& $5d$ \textbf{d}\\
$\mu_\alpha^{(0)}$ & 0.92271098 & 0.98208719 & 0.22851720& 0.20607759&0.19352751 & 0.18153094 \\
$\mu_\alpha^{(2)}$ & 2.6036257 & $-5.4562725$ & 0.0 & 0.0 & 0.0 & 0.0\\
\colrule
$V_{\bar\alpha\alpha}$  & $\theta_{12}^{(0)}=-0.08410871$ & $\theta_{12}^{(2)}=120.37555$  & $\theta_{12}^{(4)}=-9\,314.23$  &$\theta_{13}=-0.07318156$ & $\theta_{14}=-0.06651977$ & $\theta_{15}=-0.02210989$    \\
    & $\theta_{16}=-0.10451698$ & $\theta_{23}=0.02477048$ &  $\theta_{24}=0.05765807$ & $\theta_{25}=0.08606276$ & $\theta_{26}=0.04994363$  \\  
\colrule
$U_{i\bar\alpha}$   & $\sqrt{2/3}$&$\sqrt{1/3}$&0&0&0&0\\
                    & $-\sqrt{1/3}$&$\sqrt{2/3}$&0&0&0&0\\
                    & 0&0&1&0&0&0\\
                    & 0&0&0&1&0&0\\
                    & 0&0&0&0&1&0\\
                    & 0&0&0&0&0&1\\

\end{tabular}
\end{ruledtabular}
\end{table*}

\begin{table*}[ht]
\caption{\label{tab:174MQDT_3P2} Four-channel MQDT model parameters for the $6snp\,^{3}P_2$ series of $^{174}$Yb obtained from fit to spectroscopic data presented in Tabs.~\cref{tab:174YbSpec_3P2,tab:174Spec_Microwave_3p2}. The electronic configuration of the perturbing Rydberg channels $4f^{13}5d6snd$ is abbreviated as $5d$. The rotations $\mathcal{R}(\theta_{ij})$ are applied in the order $\mathcal{R}(\theta_{12})\mathcal{R}(\theta_{13})\mathcal{R}(\theta_{14})$}
\begin{ruledtabular}
\begin{tabular}{ccccc}
$i$,$\bar\alpha$,$\alpha$ & 1 & 2 & 3 & 4\\
\colrule
$\ket{i}$ & $(6s_{1/2})(np_{3/2})$& $5d$ \textbf{a} & $5d$ \textbf{b} & $5d$ \textbf{c} \\ 
$I_i$ ($\mathrm{cm}^{-1}$) & $50\,443.070393$ & $83\,967.7$ & $83\,967.7$ & $83\,967.7$\\
$\ket{\bar\alpha}$ & $6snp\,^3P_2$ & $5d$ \textbf{a} & $5d$ \textbf{b} & $5d$ \textbf{c}\\
$\mu_\alpha^{(0)}$ & 0.925121305 & 0.230133261 & 0.209638118 & 0.186228192   \\
$\mu_\alpha^{(2)}$ & $-2.73247165$ & 0.0 & 0.0 & 0.0  \\
$\mu_\alpha^{(4)}$ & 74.664989 & 0.0 & 0.0 & 0.0  \\
\colrule
$V_{\bar\alpha\alpha}$ & $\theta_{12}=0.0706666127$ & $\theta_{13}=0.0232711158$ & $\theta_{14}=-0.0292153659$ &  \\
\colrule
$U_{i\bar\alpha}$    & 1&0 & 0 & 0\\
                     & 0&1 & 0 & 0\\
                     & 0 & 0& 1& 0\\
                     & 0 & 0& 0& 1\\
\end{tabular}
\end{ruledtabular}
\end{table*}

\begin{table*}[ht]
\caption{\label{tab:174MQDT_3P0} Two-channel MQDT model parameters for the $6snp\,^{3}P_0$ series of $^{174}$Yb obtained from fit to spectroscopic data presented in Tabs.~\cref{tab:174Spec_Microwave_3p0,tab:174YbSpec_3P0}.  We follow Ref.~\cite{SAymar1984three} and presume a perturbing channel of character $4f^{13}5d6snd\,^3P_0$ (labelled $Q$) with an ionization limit of  as $\SI{83967.7}{\mathrm{cm}^{-1}}$.}
\begin{ruledtabular}
\begin{tabular}{ccc}
$i$,$\bar\alpha$,$\alpha$ & 1 & 2\\
\colrule
$\ket{i}$ & $(6s_{1/2})(np_{1/2})$& $4f^{13}5d6snd$ \\ 
$I_i$ ($\mathrm{cm}^{-1}$) & $50\,443.070393$ & $83\,967.7$\\
$\ket{\bar\alpha}$ & $6snp\,^3P_0$ & $Q$\\
$\mu_\alpha^{(0)}$ & 0.95356884 & 0.19845903 \\
$\mu_\alpha^{(2)}$ & $-0.28602498$ & 0.0 \\
\colrule
$V_{\bar\alpha\alpha}$ & $\theta_{12}=0.16328854$ &  \\
\colrule
$U_{i\bar\alpha}$    & 1&0\\
                     & 0&1\\

\end{tabular}
\end{ruledtabular}
\end{table*}

\clearpage
\onecolumngrid
\subsection{$^{171}$Yb}
\begin{turnpage}
\begin{table}[ht]
\caption{\label{tab:171MQDT_S_F12} Seven-channel MQDT model parameters for the $\ket{\nu,L=0,F=1/2}$ series of $^{171}$Yb obtained from fit to spectroscopic data presented in Tab.~\ref{tab:171Spec_S_F12}. The electronic configuration of the perturbing Rydberg channels $4f^{13}5d6snp$ is abbreviated as $5d$. The rotations $\mathcal{R}(\theta_{ij})$ are applied in the order $\mathcal{R}(\theta_{12})\mathcal{R}(\theta_{13})\mathcal{R}(\theta_{14})\mathcal{R}(\theta_{34})\mathcal{R}(\theta_{35})\mathcal{R}(\theta_{16})$.
}
\begin{ruledtabular}
\begin{tabular}{cccccccc}
$i$,$\bar\alpha$,$\alpha$ & 1 $(F_c=0)$& 2 & 3& 4& 5 & 6 & 7 $(F_c=1)$ \\
\colrule
$\ket{i_F}$ & $(6s_{1/2})(ns_{1/2})$ & $5d$ \textbf{a} & $(6p_{3/2})(np_{3/2})$ & $5d$ \textbf{b} & $(6p_{1/2})(np_{1/2})$ & $5d$ \textbf{c} & $(6s_{1/2})(ns_{1/2})$ \\ 
$I_i$ ($\mathrm{cm}^{-1}$) & $50\,442.795744$ & $83\,967.7$ & $ 80\,835.39$ & $83\,967.7$ & $77\,504.98$& $83\,967.7$ & $50\,443.217463$\\
$\ket{\bar\alpha}$ & $6sns\,^1S_0$ & $5d$ \textbf{a} & $6pnp\,^1S_0$ & $5d$ \textbf{b} & $6pnp\,^3P_0$ & $5d$ \textbf{c} & $6sns\,^3S_1$ \\ 
$\mu_\alpha^{(0)}$ & 0.357519763 & 0.203907536 & 0.116803536& 0.286731074&0.248113946 & 0.148678953 &  $\mu_{^3S_1}^{rr}$\\
$\mu_\alpha^{(2)}$ & 0.298712849 & 0.0 & 0.0 & 0.0 & 0.0 & 0.0 &\\
\colrule
$\mu_{^3S_1}^{rr}$ & $\mu_{^3S_1}^{(0)}=0.438426851$ & $\mu_{^3S_1}^{(2)}=3.91762642$ &  $\mu_{^3S_1}^{(4)}=-10612.6828$& $\mu_{^3S_1}^{(6)}=8017432.38$& $\mu_{^3S_1}^{(8)}=-2582622910.0$& &\\
\colrule
$V_{\bar\alpha\alpha}$  & $\theta_{12}=0.131810463$ & $\theta_{13}=0.297612147$ & $\theta_{14}=0.055508821$ & $\theta_{34}=0.101030515$  & $\theta_{35}=0.102911159$ & $\theta_{16}=0.137723736$ &  \\
\colrule
$U_{i\bar\alpha}$   & $1/2$&0&0&0&0&0&$\sqrt{3}/2$\\
                    & 0&1&0&0&0&0&0\\
                    & 0&0&$-\sqrt{2/3}$&0&$\sqrt{1/3}$&0&0\\
                    & 0&0&0&1&0&0&0\\
                    & 0&0&$\sqrt{1/3}$&0&$\sqrt{2/3}$&0&0\\
                    & 0&0&0&0&0&1&0\\
                    & $\sqrt{3}/2$&0&0&0&0&0&$-1/2$\\

\end{tabular}
\end{ruledtabular}
\end{table}

\begin{table}[ht]
\caption{\label{tab:171MQDT_P_F12} Eight-channel MQDT model parameters for the $\ket{\nu,L=1,F=1/2}$ series of $^{171}$Yb obtained from fit to spectroscopic data presented in Tab.~\cref{tab:171Spec_P_F12,tab:171Spec_P_F12_Laser,tab:171Spec_P_F32,tab:171Spec_P_F32_Laser}. The electronic configuration of the perturbing Rydberg channels $4f^{13}5d6snd$ is abbreviated as $5d$. The rotations $\mathcal{R}(\theta_{ij})$ are applied in the order $\mathcal{R}(\theta_{12}(\epsilon))\mathcal{R}(\theta_{27})\mathcal{R}(\theta_{13})\mathcal{R}(\theta_{14})\mathcal{R}(\theta_{15})\mathcal{R}(\theta_{16})\mathcal{R}(\theta_{23})\mathcal{R}(\theta_{24})\mathcal{R}(\theta_{25})\mathcal{R}(\theta_{26})\mathcal{R}(\theta_{78})$.
}
\begin{ruledtabular}
\begin{tabular}{ccccccccc}
$i$,$\bar\alpha$,$\alpha$ & 1 $(F_c=1)$ & 2 $(F_c=1)$& 3& 4& 5 & 6 & 7 $(F_c=0)$& 8\\
\colrule
$\ket{i}$ & $(6s_{1/2})(np_{3/2})$& $(6s_{1/2})(np_{1/2})$& $5d$ \textbf{a} & $5d$ \textbf{b}& $5d$ \textbf{c}& $5d$ \textbf{d}& $(6s_{1/2})(np_{1/2})$&$5d$ \textbf{e} \\ 
$I_i$ ($\mathrm{cm}^{-1}$) & $50\,443.217463$ & $50\,443.217463$ & $83\,967.7$ & $83\,967.7$ & $83\,967.7$& $83\,967.7$ &$50\,442.795744$ &$83\,967.7$\\
$\ket{\bar\alpha}$ & $6snp\,^1P_1$ & $6snp\,^3P_1$ & $5d$ \textbf{a} & $5d$ \textbf{b}& $5d$ \textbf{c}& $5d$ \textbf{d} & $6snp\,^3P_0$ & $5d$ \textbf{e}\\
$\mu_\alpha^{(0)}$ & 0.921706585 & 0.979638580 & 0.228828720& 0.205484818&0.193528629 & 0.181385000 & 0.953071282 & 0.198445928 \\
$\mu_\alpha^{(2)}$ & 2.56569459 & $ -5.239904224$ & 0.0 & 0.0 & 0.0 & 0.0  & 0.131025247& 0.0\\
\colrule
 & $\theta_{12}^{(0)}$& $\theta_{12}^{(2)}$ & $\theta_{12}^{(4)}$ & $\theta_{27}$ &$\theta_{13}$ &$\theta_{14}$ & $\theta_{15}$ & $\theta_{16}$  \\ 
$V_{\bar\alpha\alpha}$  & $-0.087127227$ & $135.400009$& $-12985.0162$& $-0.001430175$ &$-0.073904060$ & $-0.063632668$ & $-0.021924569$  & $-0.106678810$    \\
\colrule

 &$\theta_{23}$&$\theta_{24}$ &$\theta_{25}$ &$\theta_{26}$ & $\theta_{78}$& & &   \\

   & $0.032556999$  & $0.054105142$ & $0.086127672$ & $0.053804487$ & $0.163043619$ & & \\ 
\colrule
$U_{i\bar\alpha}$   & $-\sqrt{2/3}$&$-\sqrt{1/3}$&0&0&0&0& 0&0\\
                    &$1/\left(2\sqrt{3}\right)$ &$-\sqrt{1/6}$&0&0&0&0& $\sqrt{3}/2$&0\\
                    & 0&0&1&0&0&0&0&0\\
                    & 0&0&0&1&0&0&0&0\\
                    & 0&0&0&0&1&0&0&0\\
                    & 0&0&0&0&0&1&0&0\\
                    & $-1/2$&$1/\sqrt{2}$&0&0&0&0& $1/2$&0\\
                    & 0&0&0&0&0&0& 0&1\\
\end{tabular}
\end{ruledtabular}
\end{table}

\begin{table}
\caption{\label{tab:171MQDT_P_F32} Ten-channel MQDT model parameters for the $\ket{\nu,L=1,F=3/2}$ series of $^{171}$Yb obtained from global fit to spectroscopic data presented in Tab.~\cref{tab:171Spec_P_F12,tab:171Spec_P_F12_Laser,tab:171Spec_P_F32,tab:171Spec_P_F32_Laser}. The electronic configuration of the perturbing Rydberg channels $4f^{13}5d6snd$ is abbreviated as $5d$. The rotations $\mathcal{R}(\theta_{ij})$ are applied in the order $\mathcal{R}(\theta_{12}(\epsilon))\mathcal{R}(\theta_{13})\mathcal{R}(\theta_{14})\mathcal{R}(\theta_{15})\mathcal{R}(\theta_{16})\mathcal{R}(\theta_{23})\mathcal{R}(\theta_{24})\mathcal{R}(\theta_{25})\mathcal{R}(\theta_{26})\mathcal{R}(\theta_{78})\mathcal{R}(\theta_{79})\mathcal{R}(\theta_{7\,10})$.
}

\begin{ruledtabular}
\begin{tabular}{ccccccccccc}
$i$,$\bar\alpha$,$\alpha$ & 1 $(F_c=1)$ & 2 $(F_c=1)$& 3& 4& 5 & 6 & 7 $(F_c=0)$ & 8 & 9 & 10\\
\colrule
$\ket{i}$ & $(6s_{1/2})(np_{3/2})$& $(6s_{1/2})(np_{1/2})$& $5d$ \textbf{a} & $5d$ \textbf{b}& $5d$ \textbf{c}& $5d$ \textbf{d}& $(6s_{1/2})(np_{3/2})$ & $5d$ \textbf{e} & $5d$ \textbf{f} & $5d$ \textbf{g}\\ 
$I_i$ ($\mathrm{cm}^{-1}$) & $50\,443.217463$ & $50\,443.217463$ & $83\,967.7$ & $83\,967.7$ & $83\,967.7$& $83\,967.7$ &$50\,442.795744$ & $83\,967.7$& $83\,967.7$& $83\,967.7$\\
$\ket{\bar\alpha}$ & $6snp\,^1P_1$ & $6snp\,^3P_1$ & $5d$ \textbf{a} & $5d$ \textbf{b}& $5d$ \textbf{c}& $5d$ \textbf{d} & $6snp\,^3P_2$ & $5d$ \textbf{e} & $5d$ \textbf{f} & $5d$ \textbf{g} \\
$\mu_\alpha^{(0)}$ & 0.921706585 & 0.979638580 & 0.228828720& 0.205484818&0.193528629 & 0.181385000 & 0.924825736 &   0.236866903& 0.221055883 & 0.185599376  \\
$\mu_\alpha^{(2)}$ & 2.56569459 & $ -5.239904224$ & 0.0 & 0.0 & 0.0 & 0.0 &$-3.542481644$ &0.0 & 0.0& 0.0  \\
$\mu_\alpha^{(4)}$ & 0.0 & 0.0 & 0.0 & 0.0 & 0.0 & 0.0 &$81.5334687$ &0.0 & 0.0& 0.0  \\

\colrule
 & $\theta_{12}^{(0)}$& $\theta_{12}^{(2)}$ & $\theta_{12}^{(4)}$ & $\theta_{13}$ &$\theta_{14}$ & $\theta_{15}$ & $\theta_{16}$ &  &  &\\ 
$V_{\bar\alpha\alpha}$  & $-0.087127227$ & $135.400009$& $-12985.0162$ &$-0.073904060$ & $-0.063632668$ & $-0.021924569$  & $-0.106678810$ &  &   \\
\colrule

 &$\theta_{23}$&$\theta_{24}$ &$\theta_{25}$ &$\theta_{26}$ & $\theta_{78}$&$\theta_{79}$ &$\theta_{7\,10}$ & & &  \\

   & $0.032556999$  & $0.054105142$ & $0.086127672$ & $0.053804487$ & $0.071426685$ &$0.027464110$ &$-0.029741862$ & & \\  
\colrule
$U_{i\bar\alpha}$   & $\sqrt{5/3}/2$&$\sqrt{5/6}/2$&0&0&0&0& $-\sqrt{3/2}/2$ & 0 & 0& 0\\
                    & $-\sqrt{1/3}$&$\sqrt{2/3}$&0&0&0&0& 0& 0 & 0& 0\\
                    & 0&0&1&0&0&0&0&0&0&0\\
                    & 0&0&0&1&0&0&0&0&0&0\\
                    & 0&0&0&0&1&0&0&0&0&0\\
                    & 0&0&0&0&0&1&0&0&0&0\\
                    & $1/2$&$1/\left(2\sqrt{2}\right)$&0&0&0&0& $\sqrt{5/2}/2$&0&0&0\\
                    & 0&0&0&0&0&0&0&1&0&0\\
                    & 0&0&0&0&0&0&0&0&1&0\\
                    & 0&0&0&0&0&0&0&0&0&1\\
\end{tabular}
\end{ruledtabular}
\end{table}

\begin{table*}
\caption{\label{tab:171MQDT_D_F32} Six-channel MQDT model parameters for the $\ket{\nu,L=2,F=3/2}$ series of $^{171}$Yb obtained from fit to spectroscopic data presented in Tab.~\cref{tab:171Spec_D_F32,tab:171Spec_D_F52}. The electronic configuration of the perturbing Rydberg channels $4f^{13}5d6snp$ is abbreviated as $5d$. The rotations $\mathcal{R}(\theta_{ij})$ are applied in the order $\mathcal{R}(\theta_{12})\mathcal{R}(\theta_{13})\mathcal{R}(\theta_{14})\mathcal{R}(\theta_{24})\mathcal{R}(\theta_{15})\mathcal{R}(\theta_{25})$.
}
\begin{ruledtabular}
\begin{tabular}{ccccccc}
$i$,$\bar\alpha$,$\alpha$ & 1 $(F_c=1)$ & 2 $(F_c=1)$& 3& 4& 5 & 6 $(F_c=0)$\\
\colrule
$\ket{i}$ & $(6s_{1/2})(nd_{5/2})$&  $(6s_{1/2})(nd_{3/2})$ & $5d$ \textbf{a} & $5d$ \textbf{b} & $(6p)(np)$ &$(6s_{1/2})(nd_{3/2})$\\ 
$I_i$ ($\mathrm{cm}^{-1}$) & $50\,443.217463$ & $50\,443.217463$ & $83\,967.7$ &  $83\,967.7$ & $79\,725.35$ & $50\,442.795744$\\
$\ket{\bar\alpha}$ & $6snd\,^1D_2$ &  $6snd\,^3D_2$ & $5d$ \textbf{a} &  $5d$ \textbf{b} & $6pnp\,^1D_2$ & $6snd\,^3D_1$  \\ 
$\mu_\alpha^{(0)}$ & 0.730537124 & 0.751591782 & 0.196120394& 0.233742396&0.152905343 & $\mu_{^3D_1}^{rr}$  \\
$\mu_\alpha^{(2)}$ & $-0.000186828866$ & $-0.00114049637$ & 0.0 & 0.0 & 0.0 &\\
\colrule
$\mu_{^3D_1}^{rr}$ &$\mu_{^3D_1}^{(0)}=0.75258093$  &$\mu_{^3D_1}^{(2)}=0.382628525$  &$\mu_{^3D_1}^{(4)}=-483.120633$  &  & &  \\
\colrule
$V_{\bar\alpha\alpha}$  & $\theta_{12}=0.205496654$ & $\theta_{13}=0.00522401624$ & $\theta_{14}=0.0409502343$ & $\theta_{24}=-0.00378075773$  & $\theta_{15}=0.108563952$ & $\theta_{25}=0.0665700438$  \\
\colrule
$U_{i\bar\alpha}$   & $-\sqrt{3/5}$&$-\sqrt{2/5}$&0&0&0&0\\
                    & $\sqrt{3/5}/2$&$-3/\left(2\sqrt{10}\right)$&0&0&0&$\sqrt{5/2}/2$\\
                    & 0&0&1&0&0&0 \\
                    & 0&0&0&1&0&0\\
                    & 0&0&0&0&1&0\\
                    & $-1/2$&$\sqrt{3/2}/2$&0&0&0&$\sqrt{3/2}/2$\\

\end{tabular}
\end{ruledtabular}
\end{table*}

\begin{table*}
\caption{\label{tab:171MQDT_D_F52} Six-channel MQDT model parameters for the $\ket{\nu,L=2,F=5/2}$ series of $^{171}$Yb obtained from fit to spectroscopic data presented in Tab.~\cref{tab:171Spec_D_F32,tab:171Spec_D_F52}. The electronic configuration of the perturbing Rydberg channels $4f^{13}5d6snp$ is abbreviated as $5d$. The rotations $\mathcal{R}(\theta_{ij})$ are applied in the order $\mathcal{R}(\theta_{12})\mathcal{R}(\theta_{13})\mathcal{R}(\theta_{14})\mathcal{R}(\theta_{24})\mathcal{R}(\theta_{15})\mathcal{R}(\theta_{25})$.
}
\begin{ruledtabular}
\begin{tabular}{ccccccc}
$i$,$\bar\alpha$,$\alpha$ & 1 $(F_c=1)$ & 2 $(F_c=1)$& 3& 4& 5 & 6 $(F_c=0)$\\
\colrule
$\ket{i}$ & $(6s_{1/2})(nd_{5/2})$&  $(6s_{1/2})(nd_{3/2})$ & $5d$ \textbf{a} & $5d$ \textbf{b} & $(6p)(np)$ &$(6s_{1/2})(nd_{5/2})$\\ 
$I_i$ ($\mathrm{cm}^{-1}$) & $50\,443.217463$ & $50\,443.217463$ & $83\,967.7$ &  $83\,967.7$ & $79\,725.35$ & $50\,442.795744$\\
$\ket{\bar\alpha}$ & $6snd\,^1D_2$ &  $6snd\,^3D_2$ & $5d$ \textbf{a} &  $5d$ \textbf{b} & $6pnp\,^1D_2$ & $6snd\,^3D_3$  \\ 
$\mu_\alpha^{(0)}$ & 0.730537124 & 0.751591782 & 0.196120394& 0.233742396&0.152905343 & $\mu_{^3D_3}^{rr}$  \\
$\mu_\alpha^{(2)}$ & $-0.000186828866$ & $-0.00114049637$ & 0.0 & 0.0 & 0.0 &\\
\colrule
$\mu_{^3D_3}^{rr}$ &$\mu_{^3D_3}^{(0)}=0.72895315$  &$\mu_{^3D_3}^{(2)}=-0.20653489$  &$\mu_{^3D_3}^{(4)}=220.484722$  &  & &  \\
\colrule
$V_{\bar\alpha\alpha}$  & $\theta_{12}=0.205496654$ & $\theta_{13}=0.00522401624$ & $\theta_{14}=0.0409502343$ & $\theta_{24}=-0.00378075773$  & $\theta_{15}=0.108563952$ & $\theta_{25}=0.0665700438$  \\
\colrule
$U_{i\bar\alpha}$   & $\sqrt{7/5}/2$&$\sqrt{7/30}$&0&0&0&$-\sqrt{5/3}/2$\\
                    & $-\sqrt{2/5}$&$\sqrt{3/5}$&0&0&0&0\\
                    & 0&0&1&0&0&0 \\
                    & 0&0&0&1&0&0\\
                    & 0&0&0&0&1&0\\
                    & $1/2$&$\sqrt{1/6}$&0&0&0&$\sqrt{7/3}/2$\\

\end{tabular}
\end{ruledtabular}
\end{table*}

\end{turnpage}

\clearpage
\onecolumngrid
\section{Table of spectroscopic results}

This is a compilation of the previous \cite{SAymar1984three,SMeggers1970First,SCamus1980Highly,SAymar1980Highly,SLehec2018Laser,SMaeda1992Optical,SLehec2017PhD,SCamus1969spectre,SMartin1978Atomic,SBiRu1991The,SAli1999Two,SWyart1979Extended,SMajewski1985diploma} and new spectroscopic data used for the development of the MQDT models presented in this article. The dataset is available for download at the Harvard Dataverse \cite{SDVN/WOGVSN_2024}.

\label{sec:SpectroscopyResults}
\subsection{$^{174}$Yb $^1S_0$}

\begin{longtable*}{@{\extracolsep{\fill}}c  S[table-format=5.6]S[table-format=5.6]S[table-format=2.2]c}
\caption{Laser spectroscopic data of the $6sns\,^1S_0$ Rydberg series included in the determination of the MQDT models. The state energies and the respective uncertainties are taken from Refs.~\cite{SMeggers1970First,SCamus1980Highly} as compiled in \cite{SAymar1980Highly}, and from Ref.~\cite{SLehec2018Laser}. As in Ref.~\cite{SLehec2018Laser}, the observed wavenumbers from Ref.~\cite{SAymar1980Highly} are reduced by \SI{0.1}{\per\cm}.}\label{tab:174Spec_1S0} \\

\hline\hline\\
 [-6pt]{\multirow{2}{*}{State label}} & {\multirow{2}{*}{$\tilde\nu_\mathrm{exp.}$ ($\mathrm{cm}^{-1}$)}} & {\multirow{2}{*}{$\tilde\nu_\mathrm{th.}$ ($\mathrm{cm}^{-1}$ )}} & {$E_\mathrm{exp.}-E_\mathrm{th.}$}&{\multirow{2}{*}{Ref.}}\\
 & & & {($h\cdot\mathrm{MHz}$)}  &\\[3pt]
\hline\\ [-5pt] 
 \endfirsthead

 {{\tablename\ \thetable{} (\textit{continued})}} \\
\hline\hline\\
 [-6pt]{\multirow{2}{*}{State label}} & {\multirow{2}{*}{$\tilde\nu_\mathrm{exp.}$ ($\mathrm{cm}^{-1}$)}} & {\multirow{2}{*}{$\tilde\nu_\mathrm{th.}$ ($\mathrm{cm}^{-1}$ )}} & {$E_\mathrm{exp.}-E_\mathrm{th.}$}&{\multirow{2}{*}{Ref.}}\\
 & & & {($h\cdot\mathrm{MHz}$)}  &\\[3pt]
\hline\\ [-5pt] 
 \endhead

\hline
 \endfoot

 \hline
 \endlastfoot

$6s7s\,^1S_0$& 34350.55(10)& 34349.227809&39638&\cite{SAymar1980Highly}\\
$6s8s\,^1S_0$& 41939.78(10)& 41939.687015&2788&\cite{SAymar1980Highly}\\
$6p^2\,^3P_0$& 42436.84(10)& 42436.901198&-1835&\cite{SAymar1980Highly}\\
$4f^{13}5d6s6p\,B$& 46081.44(10)& 46081.618520&-5352&\cite{SAymar1980Highly}\\
$6s10s\,^1S_0$& 46893.14(10)& 46893.135082&147&\cite{SAymar1980Highly}\\
$6s11s\,^1S_0$& 47808.39(10)& 47808.418289&-848&\cite{SAymar1980Highly}\\
$6p^2\,^1S_0$& 48344.28(10)& 48344.184498&2863&\cite{SAymar1980Highly}\\
$6s12s\,^1S_0$& 48723.39(10)& 48723.404543&-436&\cite{SAymar1980Highly}\\
\hline\hline
\end{longtable*}

\begin{longtable*}{@{\extracolsep{\fill}}ccS[table-format=6.4]S[table-format=6.2]S[table-format=1.2]c}
\caption{\label{tab:174Spec_Microwave_SS} Spectroscopic data of $n\,^{1}S_0\leftrightarrow n'\,^{1}S_0$ Rydberg-Rydberg transitions obtained by microwave spectroscopy, as presented in Ref.~\cite{SMaeda1992Optical} and Ref.~\cite{SLehec2018Laser}. }\\

\hline\hline\\\multirow{2}{*}{$f$} & \multirow{2}{*}{$i$}& {\multirow{2}{*}{$\Delta E_{f\leftarrow i}^\mathrm{exp.}$ ($h\cdot\mathrm{MHz}$)}} & {\multirow{2}{*}{$\Delta E_{f\leftarrow i}^\mathrm{theo.}$ ($h\cdot\mathrm{MHz}$ )}} &  {$E_\mathrm{exp.}-E_\mathrm{th.}$}   &\multirow{2}{*}{Ref.}\\ 
 & & & & {($h\cdot\mathrm{MHz}$)} & \\\hline \\ 
 \endfirsthead

 {{\tablename\ \thetable{} (\textit{continued})}} \\
\hline\hline\\\multirow{2}{*}{$f$} & \multirow{2}{*}{$i$}& {\multirow{2}{*}{$\Delta E_{f\leftarrow i}^\mathrm{exp.}$ ($h\cdot\mathrm{MHz}$)}} & {\multirow{2}{*}{$\Delta E_{f\leftarrow i}^\mathrm{theo.}$ ($h\cdot\mathrm{MHz}$ )}} &  {$E_\mathrm{exp.}-E_\mathrm{th.}$}   &\multirow{2}{*}{Ref.}\\ 
 & & & & {($h\cdot\mathrm{MHz}$)} & \\\hline \\ 
 \endhead

\hline
 \endfoot
 
 \hline
 \endlastfoot
 
$6s35s\,^1S_0$& $6s34s\,^1S_0$& 238239.32(12)& 238239.57&-0.25&\cite{SLehec2018Laser}\\
$6s36s\,^1S_0$& $6s35s\,^1S_0$& 216088.70(14)& 216088.80&-0.10&\cite{SLehec2018Laser}\\
$6s37s\,^1S_0$& $6s36s\,^1S_0$& 196599.98(12)& 196599.91&0.07&\cite{SLehec2018Laser}\\
$6s38s\,^1S_0$& $6s37s\,^1S_0$& 179384.72(12)& 179384.94&-0.22&\cite{SLehec2018Laser}\\
$6s39s\,^1S_0$& $6s38s\,^1S_0$& 164121.48(12)& 164121.90&-0.42&\cite{SLehec2018Laser}\\
$6s40s\,^1S_0$& $6s39s\,^1S_0$& 150540.64(10)& 150541.89&-1.25&\cite{SLehec2018Laser}\\
$6s41s\,^1S_0$& $6s40s\,^1S_0$& 138420.36(80)& 138419.28&1.08&\cite{SMaeda1992Optical}\\
$6s42s\,^1S_0$& $6s41s\,^1S_0$& 127563.40(80)& 127563.77&-0.37&\cite{SMaeda1992Optical}\\
\hline\hline
\end{longtable*}

\begin{longtable*}{@{\extracolsep{\fill}}ccS[table-format=6.4]S[table-format=6.2]S[table-format=1.2]c}
\caption{\label{tab:174Spec_Microwave_SD} Spectroscopic data of $n\,^{1}S_0\leftrightarrow n'\,^{1}S_0$ Rydberg-Rydberg transitions obtained by microwave spectroscopy, as presented in Ref.~\cite{SMaeda1992Optical} and Ref.~\cite{SLehec2018Laser}. }\\

\hline\hline\\\multirow{2}{*}{$f$} & \multirow{2}{*}{$i$}& {\multirow{2}{*}{$\Delta E_{f\leftarrow i}^\mathrm{exp.}$ ($h\cdot\mathrm{MHz}$)}} & {\multirow{2}{*}{$\Delta E_{f\leftarrow i}^\mathrm{theo.}$ ($h\cdot\mathrm{MHz}$ )}} &  {$E_\mathrm{exp.}-E_\mathrm{th.}$}   &\multirow{2}{*}{Ref.}\\ 
 & & & & {($h\cdot\mathrm{MHz}$)} & \\\hline \\ 
 \endfirsthead

 {{\tablename\ \thetable{} (\textit{continued})}} \\
\hline\hline\\\multirow{2}{*}{$f$} & \multirow{2}{*}{$i$}& {\multirow{2}{*}{$\Delta E_{f\leftarrow i}^\mathrm{exp.}$ ($h\cdot\mathrm{MHz}$)}} & {\multirow{2}{*}{$\Delta E_{f\leftarrow i}^\mathrm{theo.}$ ($h\cdot\mathrm{MHz}$ )}} &  {$E_\mathrm{exp.}-E_\mathrm{th.}$}   &\multirow{2}{*}{Ref.}\\ 
 & & & & {($h\cdot\mathrm{MHz}$)} & \\\hline \\ \hline \\ 
 \endhead

\hline
 \endfoot

 \hline
 \endlastfoot
$6s29s\,^1S_0$& $6s28d\,^1D_2$& -235548.18(22)& -235549.07&0.89&\cite{SLehec2018Laser}\\
$6s30s\,^1S_0$& $6s29d\,^1D_2$& -209275.16(22)& -209275.15&-0.01&\cite{SLehec2018Laser}\\
$6s31s\,^1S_0$& $6s30d\,^1D_2$& -186918.48(12)& -186918.43&-0.05&\cite{SLehec2018Laser}\\
$6s32s\,^1S_0$& $6s31d\,^1D_2$& -167680.52(16)& -167679.89&-0.63&\cite{SLehec2018Laser}\\
$6s33s\,^1S_0$& $6s32d\,^1D_2$& -151009.20(14)& -151008.75&-0.45&\cite{SLehec2018Laser}\\
$6s34s\,^1S_0$& $6s33d\,^1D_2$& -136482.22(10)& -136481.90&-0.32&\cite{SLehec2018Laser}\\
$6s35s\,^1S_0$& $6s34d\,^1D_2$& -123762.40(16)& -123761.92&-0.48&\cite{SLehec2018Laser}\\
$6s36s\,^1S_0$& $6s35d\,^1D_2$& -112576.06(18)& -112574.97&-1.09&\cite{SLehec2018Laser}\\
\hline\hline
\end{longtable*}

\begin{longtable*}{@{\extracolsep{\fill}}c  S[table-format=2.5]S[table-format=2.5]}
\caption{Experimental ($\alpha_{0,\mathrm{exp.}}$) and predicted ($\alpha_{0,\mathrm{theo.}}$) static dipole polarizabilities of $6sns\,^1S_0$ Rydberg states of $^{174}$Yb. The experimental values for $\alpha_{0,\mathrm{exp.}}$ are obtained by measurement of quadratic Stark shifts as described in App.~\ref{sec:atomicbeam} of the main text. Prediction for the static dipole polarizability $\alpha_{0,\mathrm{theo.}}$ are obtained from a non-perturbative diagonalization of the atomic Hamiltonian including electric fields  \cite{Szimmerman1979stark}.}\label{tab:174Spec_Pol_1S0} \\

\hline\hline\\
 [-6pt]{State label} & {$\alpha_{0,\mathrm{exp.}}$ $\left(h\cdot\mathrm{MHz}/\left(\mathrm{V}/\mathrm{cm}\right)^2\right)$} & {$\alpha_{0,\mathrm{theo.}}$ $\left(h\cdot\mathrm{MHz}/\left(\mathrm{V}/\mathrm{cm}\right)^2\right)$} \\
\hline\\ [-5pt] 
 \endfirsthead

 {{\tablename\ \thetable{} (\textit{continued})}} \\
\hline\hline\\
 [-6pt]{State label} & {$\alpha_{0,\mathrm{exp.}}$ $\left(h\cdot\mathrm{MHz}/\left(\mathrm{V}/\mathrm{cm}\right)^2\right)$} & {$\alpha_{0,\mathrm{theo.}}$ $\left(h\cdot\mathrm{MHz}/\left(\mathrm{V}/\mathrm{cm}\right)^2\right)$} \\
\hline\\ [-5pt] 
 \endhead

\hline
 \endfoot

 \hline
 \endlastfoot
$6s35s\,^{1}S_0$& 22.6(7)&24.2\\
$6s36s\,^{1}S_0$& 32.3(7)&31.5\\
$6s37s\,^{1}S_0$& 42.7(11)&41.6\\
$6s38s\,^{1}S_0$& 57.1(6)&57.0\\
$6s39s\,^{1}S_0$& 87.1(9)&87.2\\
$6s40s\,^{1}S_0$& 231.4(11)&241.1\\
$6s41s\,^{1}S_0$& -62.5(5)&-62.8\\
$6s42s\,^{1}S_0$& 21.2(9)&21.3\\
$6s43s\,^{1}S_0$& 54.3(7)&53.4\\
$6s44s\,^{1}S_0$& 78.1(8)&78.8\\
$6s45s\,^{1}S_0$& 103.4(5)&103.8\\
$6s46s\,^{1}S_0$& 130.3(9)&130.6\\
$6s47s\,^{1}S_0$& 160.1(10)&160.4\\
$6s48s\,^{1}S_0$& 192.7(5)&194.0\\
$6s49s\,^{1}S_0$& 231.3(11)&232.0\\
$6s50s\,^{1}S_0$& 276.7(8)&275.3\\
\hline\hline
\end{longtable*}

\clearpage

\subsection{$^{174}$Yb $^{3}S_1$}

\begin{longtable*}{@{\extracolsep{\fill}}c  S[table-format=2.5]S[table-format=2.5]S[table-format=2.5]S[table-format=2.5]}
\caption{Experimental ($\alpha_{0,\mathrm{exp.}}$) and predicted ($\alpha_{0,\mathrm{theo.}}$) static dipole polarizabilities of $6sns\,^3S_1(m_J=0,\pm1)$ Rydberg states of $^{174}$Yb. The experimental values for $\alpha_{0,\mathrm{exp.}}$ are obtained by measurement of quadratic Stark shifts as described in App.~\ref{sec:atomicbeam} of the main text. Prediction for the static dipole polarizability $\alpha_{0,\mathrm{theo.}}$ are obtained from a non-perturbative diagonalization of the atomic Hamiltonian including electric fields \cite{Szimmerman1979stark}.}\label{tab:174Spec_Pol_3S1} \\

\hline\hline\\
 [-6pt]{\multirow{2}{*}{State label}} & {$\alpha_{0,\mathrm{exp.}}^{(m_J=0)}$ } & {$\alpha_{0,\mathrm{theo.}}^{(m_J=0)}$} & {$\alpha_{0,\mathrm{exp.}}^{(m_J=\pm1)}$} & {$\alpha_{0,\mathrm{theo.}}^{(m_J=\pm1)}$}  \\
  & {$\left(h\cdot\mathrm{MHz}/\left(\mathrm{V}/\mathrm{cm}\right)^2\right)$}&{$\left(h\cdot\mathrm{MHz}/\left(\mathrm{V}/\mathrm{cm}\right)^2\right)$} & {$\left(h\cdot\mathrm{MHz}/\left(\mathrm{V}/\mathrm{cm}\right)^2\right)$} & {$\left(h\cdot\mathrm{MHz}/\left(\mathrm{V}/\mathrm{cm}\right)^2\right)$}\\[3pt]
\hline\\ [-5pt] 
 \endfirsthead

 {{\tablename\ \thetable{} (\textit{continued})}} \\
\hline\hline\\
 [-6pt]{\multirow{2}{*}{State label}} & {$\alpha_{0,\mathrm{exp.}}^{(m_J=0)}$ } & {$\alpha_{0,\mathrm{theo.}}^{(m_J=0)}$} & {$\alpha_{0,\mathrm{exp.}}^{(m_J=\pm1)}$} & {$\alpha_{0,\mathrm{theo.}}^{(m_J=\pm1)}$}  \\
  & {$\left(h\cdot\mathrm{MHz}/\left(\mathrm{V}/\mathrm{cm}\right)^2\right)$}&{$\left(h\cdot\mathrm{MHz}/\left(\mathrm{V}/\mathrm{cm}\right)^2\right)$} & {$\left(h\cdot\mathrm{MHz}/\left(\mathrm{V}/\mathrm{cm}\right)^2\right)$} & {$\left(h\cdot\mathrm{MHz}/\left(\mathrm{V}/\mathrm{cm}\right)^2\right)$}\\[3pt]
\hline\\ [-5pt] 
 \endhead

\hline
 \endfoot

 \hline
 \endlastfoot
$6s31s\,^{3}S_1$& 29.9(8)&30.0&4.7(1)&3.1\\
$6s32s\,^{3}S_1$& 83.4(2)&87.6&3.8(1)&4.3\\
$6s33s\,^{3}S_1$& -751(3)&-697.5&6.9(1)&6.0\\
$6s34s\,^{3}S_1$& -113.4(3)&-116.0&9.2(3)&8.4\\
$6s35s\,^{3}S_1$& -79.2(2)&-81.4&11.3(2)&11.7\\
$6s36s\,^{3}S_1$& -69.3(10)&-72.4&15.5(2)&16.3\\
$6s39s\,^{3}S_1$& -78.5(14)&-78.7&38.9(3)&45.5\\
$6s40s\,^{3}S_1$& -85.8(13)&-85.8&62.8(3)&65.2\\
\hline\hline
\end{longtable*}

\clearpage

\subsection{$^{174}$Yb $^{1,3}D_2$}

\begin{longtable*}{@{\extracolsep{\fill}}c  S[table-format=5.6]S[table-format=5.6]S[table-format=2.2]c}
\caption{\label{tab:174Spec_13D2} Laser spectroscopic data of the $6snd\,^{1,3}D_2$ Rydberg series included in the determination of the MQDT models. The state energies and the respective uncertainties are taken from Refs.~\cite{SMeggers1970First,SCamus1980Highly} as compiled in \cite{SAymar1980Highly}, and from Ref.~\cite{SLehec2018Laser}.}\\

\hline\hline\\
 [-6pt]{\multirow{2}{*}{State label}} & {\multirow{2}{*}{$\tilde\nu_\mathrm{exp.}$ ($\mathrm{cm}^{-1}$)}} & {\multirow{2}{*}{$\tilde\nu_\mathrm{th.}$ ($\mathrm{cm}^{-1}$ )}} & {$E_\mathrm{exp.}-E_\mathrm{th.}$}&{\multirow{2}{*}{Ref.}}\\
 & & & {($h\cdot\mathrm{MHz}$)}  &\\[3pt]
\hline\\ [-5pt] 
 \endfirsthead

 {{\tablename\ \thetable{} (\textit{continued})}} \\
\hline\hline\\
 [-6pt]{\multirow{2}{*}{State label}} & {\multirow{2}{*}{$\tilde\nu_\mathrm{exp.}$ ($\mathrm{cm}^{-1}$)}} & {\multirow{2}{*}{$\tilde\nu_\mathrm{th.}$ ($\mathrm{cm}^{-1}$ )}} & {$E_\mathrm{exp.}-E_\mathrm{th.}$}&{\multirow{2}{*}{Ref.}}\\
 & & & {($h\cdot\mathrm{MHz}$)}  &\\[3pt]
\hline\\ [-5pt] 
 \endhead

\hline
 \endfoot

 \hline
 \endlastfoot
$6s8d\,^1D_2$& 46405.62(10)& 46403.272104&70388&\cite{SAymar1980Highly}\\
$6s8d\,^3D_2$& 46467.69(10)& 46468.043298&-10592&\cite{SAymar1980Highly}\\
$6p^2\,^1D_2$& 47420.97(10)& 47421.132756&-4879&\cite{SAymar1980Highly}\\
$6s9d\,^3D_2$& 47634.40(10)& 47634.131975&8035&\cite{SAymar1980Highly}\\
$6s9d\,^1D_2$& 47821.74(10)& 47822.162808&-12675&\cite{SAymar1980Highly}\\
$6s10d\,^3D_2$& 48357.63(10)& 48353.002109&138741&\cite{SAymar1980Highly}\\
$6s10d\,^1D_2$& 48403.49(10)& 48394.683988&263998&\cite{SAymar1980Highly}\\
$4f^{13}5d6s6p\,B$& 48762.52(10)& 48760.869520&49480&\cite{SAymar1980Highly}\\
\hline\hline
\end{longtable*}

\begin{longtable*}{@{\extracolsep{\fill}}ccS[table-format=6.4]S[table-format=6.2]S[table-format=1.2]c}
\caption{\label{tab:174Spec_Microwave_DD} Spectroscopic data of $n\,^{1}D_2\leftrightarrow n'\,^{1}D_2$ Rydberg-Rydberg transitions obtained by microwave spectroscopy, as presented in Ref.~\cite{SMaeda1992Optical} and Ref.~\cite{SLehec2018Laser}. }\\

\hline\hline\\\multirow{2}{*}{$f$} & \multirow{2}{*}{$i$}& {\multirow{2}{*}{$\Delta E_{f\leftarrow i}^\mathrm{exp.}$ ($h\cdot\mathrm{MHz}$)}} & {\multirow{2}{*}{$\Delta E_{f\leftarrow i}^\mathrm{theo.}$ ($h\cdot\mathrm{MHz}$ )}} &  {$E_\mathrm{exp.}-E_\mathrm{th.}$}   &\multirow{2}{*}{Ref.}\\ 
 & & & & {($h\cdot\mathrm{MHz}$)} & \\\hline \\ 
 \endfirsthead

 {{\tablename\ \thetable{} (\textit{continued})}} \\
\hline\hline\\\multirow{2}{*}{$f$} & \multirow{2}{*}{$i$}& {\multirow{2}{*}{$\Delta E_{f\leftarrow i}^\mathrm{exp.}$ ($h\cdot\mathrm{MHz}$)}} & {\multirow{2}{*}{$\Delta E_{f\leftarrow i}^\mathrm{theo.}$ ($h\cdot\mathrm{MHz}$ )}} &  {$E_\mathrm{exp.}-E_\mathrm{th.}$}   &\multirow{2}{*}{Ref.}\\ 
 & & & & {($h\cdot\mathrm{MHz}$)} & \\\hline \\  \hline \\ 
 \endhead

\hline
 \endfoot

 \hline
 \endlastfoot
$6s33d\,^1D_2$& $6s32d\,^1D_2$& 248996.64(28)& 248995.32&1.32&\cite{SLehec2018Laser}\\
$6s34d\,^1D_2$& $6s33d\,^1D_2$& 225519.66(12)& 225519.59&0.07&\cite{SLehec2018Laser}\\
$6s35d\,^1D_2$& $6s34d\,^1D_2$& 204902.10(12)& 204901.85&0.25&\cite{SLehec2018Laser}\\
$6s36d\,^1D_2$& $6s35d\,^1D_2$& 186721.48(12)& 186721.21&0.27&\cite{SLehec2018Laser}\\
$6s37d\,^1D_2$& $6s36d\,^1D_2$& 170628.56(14)& 170628.47&0.09&\cite{SLehec2018Laser}\\
$6s38d\,^1D_2$& $6s37d\,^1D_2$& 156333.20(14)& 156332.64&0.56&\cite{SLehec2018Laser}\\
$6s39d\,^1D_2$& $6s38d\,^1D_2$& 143589.46(8)& 143589.88&-0.42&\cite{SLehec2018Laser}\\
$6s40d\,^1D_2$& $6s39d\,^1D_2$& 132194.90(80)& 132195.01&-0.11&\cite{SMaeda1992Optical}\\
\hline\hline
\end{longtable*}

\clearpage

\subsection{$^{174}$Yb $^{1,3}P_1$}

\begin{longtable*}{@{\extracolsep{\fill}}ccS[table-format=6.4]S[table-format=5.6]S[table-format=5.6]S[table-format=1.2]c}
\caption{\label{tab:174Spec_Microwave_13p1_S} Spectroscopic data of $n\,^{1,3}P_1\leftrightarrow n'\,^1S_0$ Rydberg-Rydberg transitions obtained by microwave spectroscopy in an atomic beam setup as described in App.~\ref{sec:atomicbeam} of the main text and from Ref.~\cite{SLehec2017PhD}. }\\

\hline\hline\\\multirow{2}{*}{$f$} & \multirow{2}{*}{$i$}& {\multirow{2}{*}{$\nu_{f\leftarrow i}$ (MHz)}} & {\multirow{2}{*}{$\tilde\nu_\mathrm{exp.}$ ($\mathrm{cm}^{-1}$ )}}& {\multirow{2}{*}{$\tilde\nu_\mathrm{th.}$ ($\mathrm{cm}^{-1}$ )}} &  {$E_\mathrm{exp.}-E_\mathrm{th.}$ }  &\multirow{2}{*}{Ref.}\\ 
    & & & & & {($h\cdot\mathrm{MHz}$)} & \\\hline \\ 
 \endfirsthead

 {{\tablename\ \thetable{} (\textit{continued})}} \\
\hline\hline\\\multirow{2}{*}{$f$} & \multirow{2}{*}{$i$}& {\multirow{2}{*}{$\nu_{f\leftarrow i}$ (MHz)}} & {\multirow{2}{*}{$\tilde\nu_\mathrm{exp.}$ ($\mathrm{cm}^{-1}$ )}}& {\multirow{2}{*}{$\tilde\nu_\mathrm{th.}$ ($\mathrm{cm}^{-1}$ )}} &  {$E_\mathrm{exp.}-E_\mathrm{th.}$ }  &\multirow{2}{*}{Ref.}\\ 
    & & & & & {($h\cdot\mathrm{MHz}$)} & \\\hline \\ 
 \endhead

\hline
 \endfoot

 \hline
 \endlastfoot
$6s30p\,^1P_1$& $6s30s\,^1S_0$& 125083.25(10)& 50281.494629& 50281.494630&-0.02&This work\\
$6s31p\,^1P_1$& $6s31s\,^1S_0$& 111552.79(10)& 50293.204338& 50293.204334&0.14&This work\\
$6s32p\,^1P_1$& $6s32s\,^1S_0$& 99908.78(10)& 50303.686967& 50303.686965&0.07&This work\\
$6s33p\,^1P_1$& $6s33s\,^1S_0$& 89832.20(10)& 50313.107919& 50313.107921&-0.05&This work\\
$6s37p\,^1P_1$& $6s38s\,^1S_0$& -118631.36(10)& 50342.640744& 50342.640748&-0.12&This work\\
$6s38p\,^1P_1$& $6s39s\,^1S_0$& -108614.16(10)& 50348.449399& 50348.449403&-0.11&This work\\
$6s39p\,^1P_1$& $6s40s\,^1S_0$& -99693.21(10)& 50353.768507& 50353.768511&-0.13&This work\\
$6s42p\,^3P_1$& $6s41s\,^1S_0$& 122420.24(10)& 50365.794584& 50365.794585&-0.01&This work\\
\hline\hline
\end{longtable*}

\begin{longtable*}{@{\extracolsep{\fill}}ccS[table-format=6.4]S[table-format=5.6]S[table-format=5.6]S[table-format=1.2]c}
\caption{\label{tab:174Spec_Microwave_13p1_D} Spectroscopic data of $n\,^{1,3}P_1\leftrightarrow n'\,^{1,3}D_2$ Rydberg-Rydberg transitions obtained by microwave spectroscopy in an atomic beam setup as described in App.~\ref{sec:atomicbeam} of the main text and from Ref.~\cite{SLehec2017PhD}. }\\

\hline\hline\\\multirow{2}{*}{$f$} & \multirow{2}{*}{$i$}&{ \multirow{2}{*}{$\nu_{f\leftarrow i}$ (MHz)}} & {\multirow{2}{*}{$\tilde\nu_\mathrm{exp.}$ ($\mathrm{cm}^{-1}$ )}}& {\multirow{2}{*}{$\tilde\nu_\mathrm{th.}$ ($\mathrm{cm}^{-1}$ )}} &  {$E_\mathrm{exp.}-E_\mathrm{th.}$}   &\multirow{2}{*}{Ref.}\\ 
    & & & & & {($h\cdot\mathrm{MHz}$)} & \\\hline \\ 
 \endfirsthead

 {{\tablename\ \thetable{} (\textit{continued})}} \\
\hline\hline\\\multirow{2}{*}{$f$} & \multirow{2}{*}{$i$}& {\multirow{2}{*}{$\nu_{f\leftarrow i}$ (MHz)}} & {\multirow{2}{*}{$\tilde\nu_\mathrm{exp.}$ ($\mathrm{cm}^{-1}$ )}}& {\multirow{2}{*}{$\tilde\nu_\mathrm{th.}$ ($\mathrm{cm}^{-1}$ )}} &  {$E_\mathrm{exp.}-E_\mathrm{th.}$}   &\multirow{2}{*}{Ref.}\\ 
    & & & & & {($h\cdot\mathrm{MHz}$)} & \\\hline \\ 
 \endhead

\hline
 \endfoot

 \hline
 \endlastfoot
$6s31p\,^1P_1$& $6s30d\,^3D_2$& -114223.63(10)& 50291.495583& 50291.495586&-0.09&This work\\
$6s32p\,^1P_1$& $6s31d\,^3D_2$& -106695.56(10)& 50302.022020& 50302.022017&0.08&This work\\
$6s34p\,^1P_1$& $6s33d\,^3D_2$& -94535.43(10)& 50320.005993& 50320.005989&0.11&This work\\
$6s35p\,^1P_1$& $6s34d\,^3D_2$& -89631.26(10)& 50327.720675& 50327.720672&0.09&This work\\
$6s36p\,^1P_1$& $6s35d\,^3D_2$& -85360.87(10)& 50334.722784& 50334.722784&0.01&This work\\
$6s37p\,^1P_1$& $6s36d\,^3D_2$& -81628.49(10)& 50341.097425& 50341.097427&-0.06&This work\\
$6s38p\,^1P_1$& $6s37d\,^3D_2$& -78347.73(10)& 50346.917607& 50346.917609&-0.06&This work\\
$6s39p\,^1P_1$& $6s38d\,^3D_2$& -75439.74(10)& 50352.246287& 50352.246291&-0.12&This work\\
\hline\hline
\end{longtable*}

\begin{longtable*}{@{\extracolsep{\fill}}c  S[table-format=5.6]S[table-format=5.6]S[table-format=2.2]c}
\caption{\label{tab:174YbSpec_3P1} Laser spectroscopic data of the $6snp\,^{1,3}P_1$ Rydberg series included in the determination of the MQDT models. The state energies and the respective uncertainties are taken from Refs.~\cite{SCamus1969spectre,SAymar1984three}.}\\

\hline\hline\\
 [-6pt]{\multirow{2}{*}{State label}} & {\multirow{2}{*}{$\tilde\nu_\mathrm{exp.}$ ($\mathrm{cm}^{-1}$)}} & {\multirow{2}{*}{$\tilde\nu_\mathrm{th.}$ ($\mathrm{cm}^{-1}$ )}} & {$E_\mathrm{exp.}-E_\mathrm{th.}$}&{\multirow{2}{*}{Ref.}}\\
 & & & {($h\cdot\mathrm{MHz}$)}  &\\[3pt]
\hline\\ [-5pt] 
 \endfirsthead

 {{\tablename\ \thetable{} (\textit{continued})}} \\
\hline\hline\\
 [-6pt]{\multirow{2}{*}{State label}} & {\multirow{2}{*}{$\tilde\nu_\mathrm{exp.}$ ($\mathrm{cm}^{-1}$)}} & {\multirow{2}{*}{$\tilde\nu_\mathrm{th.}$ ($\mathrm{cm}^{-1}$ )}} & {$E_\mathrm{exp.}-E_\mathrm{th.}$}&{\multirow{2}{*}{Ref.}}\\
 & & & {($h\cdot\mathrm{MHz}$)}  &\\[3pt]
\hline\\ [-5pt] 
 \endhead

\hline
 \endfoot

 \hline
 \endlastfoot

$6s12p\,^1P_1$& 48719.18(10)& 48719.416230&-6954&\cite{SCamus1969spectre}\\
$6s12p\,^3P_1$& 48761.81(10)& 48763.202662&-41756&\cite{SCamus1969spectre}\\
$4f^{13}5d^26s\,A$& 49005.89(10)& 49006.059508&-5082&\cite{SCamus1969spectre}\\
$6s13p\,^3P_1$& 49110.07(10)& 49107.770988&68918&\cite{SCamus1969spectre}\\
$6s13p\,^1P_1$& 49127.40(10)& 49127.675624&-8263&\cite{SCamus1969spectre}\\
$6s14p\,^1P_1$& 49352.44(10)& 49352.942891&-15206&\cite{SCamus1969spectre}\\
$6s14p\,^3P_1$& 49360.60(10)& 49359.516364&32481&\cite{SCamus1969spectre}\\
$6s15p\,^1P_1$& 49536.42(10)& 49537.093278&-20296&\cite{SCamus1969spectre}\\
\hline\hline
\end{longtable*}

\clearpage

\subsection{$^{174}$Yb $^{3}P_2$}

\begin{longtable*}{@{\extracolsep{\fill}}ccS[table-format=6.4]S[table-format=5.6]S[table-format=5.6]S[table-format=1.2]c}
\caption{\label{tab:174Spec_Microwave_3p2} Spectroscopic data of the  $n\,^{3}P_2$ Rydberg series obtained by microwave spectroscopy in an atomic beam setup as described in App.~\ref{sec:atomicbeam} of the main text and from Ref.~\cite{SLehec2017PhD}. }\\

\hline\hline\\\multirow{2}{*}{$f$} & \multirow{2}{*}{$i$}& {\multirow{2}{*}{$\nu_{f\leftarrow i}$ (MHz)}} & {\multirow{2}{*}{$\tilde\nu_\mathrm{exp.}$ ($\mathrm{cm}^{-1}$ )}}& {\multirow{2}{*}{$\tilde\nu_\mathrm{th.}$ ($\mathrm{cm}^{-1}$ )}} &  {$E_\mathrm{exp.}-E_\mathrm{th.}$ }  &\multirow{2}{*}{Ref.}\\ 
    & & & & & {($h\cdot\mathrm{MHz}$)} & \\\hline \\ 
 \endfirsthead

 {{\tablename\ \thetable{} (\textit{continued})}} \\
\hline\hline\\\multirow{2}{*}{$f$} & \multirow{2}{*}{$i$}& {\multirow{2}{*}{$\nu_{f\leftarrow i}$ (MHz)}} & {\multirow{2}{*}{$\tilde\nu_\mathrm{exp.}$ ($\mathrm{cm}^{-1}$ )}}& {\multirow{2}{*}{$\tilde\nu_\mathrm{th.}$ ($\mathrm{cm}^{-1}$ )}} &  {$E_\mathrm{exp.}-E_\mathrm{th.}$ }  &\multirow{2}{*}{Ref.}\\ 
    & & & & & {($h\cdot\mathrm{MHz}$)} & \\\hline \\ 
 \endhead

\hline
 \endfoot

 \hline
 \endlastfoot
$6s31p\,^3P_2$& $6s30d\,^3D_2$& -54957.40(10)& 50293.472491& 50293.472497&-0.16&This work\\
$6s36p\,^3P_2$& $6s34d\,^3D_2$& 172078.74(10)& 50336.450381& 50336.450366&0.46&This work\\
$6s37p\,^3P_2$& $6s35d\,^3D_2$& 156685.04(10)& 50342.796567& 50342.796557&0.30&This work\\
$6s38p\,^3P_2$& $6s36d\,^3D_2$& 143074.41(10)& 50348.592707& 50348.592706&0.03&This work\\
$6s39p\,^3P_2$& $6s37d\,^3D_2$& 130996.14(10)& 50353.900567& 50353.900572&-0.15&This work\\
$6s40p\,^3P_2$& $6s38d\,^3D_2$& 120240.54(10)& 50358.773479& 50358.773486&-0.21&This work\\
$6s43p\,^3P_2$& $6s41d\,^3D_2$& 94279.05(10)& 50371.215928& 50371.215944&-0.49&This work\\
$6s43p\,^3P_2$& $6s41d\,^3D_2$& 94279.0(5)& 50371.215926& 50371.215944&-0.53&\cite{SLehec2017PhD}\\
\hline\hline
\end{longtable*}

\newpage

\begin{longtable*}{@{\extracolsep{\fill}}c  S[table-format=5.6]S[table-format=5.6]S[table-format=2.2]c}
\caption{\label{tab:174YbSpec_3P2} Laser spectroscopic data of the $6snp\,^3P_2$ Rydberg series included in the determination of the MQDT model. The state energies and the respective uncertainties are taken from Ref.~\cite{SAymar1984three}.}\\

\hline\hline\\
 [-6pt]{\multirow{2}{*}{State label}} & {\multirow{2}{*}{$\tilde\nu_\mathrm{exp.}$ ($\mathrm{cm}^{-1}$)}} & {\multirow{2}{*}{$\tilde\nu_\mathrm{th.}$ ($\mathrm{cm}^{-1}$ )}} & {$E_\mathrm{exp.}-E_\mathrm{th.}$}&{\multirow{2}{*}{Ref.}}\\
 & & & {($h\cdot\mathrm{MHz}$)}  &\\[3pt]
\hline\\ [-5pt] 
 \endfirsthead

 {{\tablename\ \thetable{} (\textit{continued})}} \\
\hline\hline\\
 [-6pt]{\multirow{2}{*}{State label}} & {\multirow{2}{*}{$\tilde\nu_\mathrm{exp.}$ ($\mathrm{cm}^{-1}$)}} & {\multirow{2}{*}{$\tilde\nu_\mathrm{th.}$ ($\mathrm{cm}^{-1}$ )}} & {$E_\mathrm{exp.}-E_\mathrm{th.}$}&{\multirow{2}{*}{Ref.}}\\
 & & & {($h\cdot\mathrm{MHz}$)}  &\\[3pt]
\hline\\ [-5pt] 
 \endhead

\hline
 \endfoot

 \hline
 \endlastfoot
$6s9p\,^3P_2$& 46184.15& 46159.155515&749316&\cite{SMartin1978Atomic}\\
$6s10p\,^3P_2$& 47471.10& 47476.475100&-161141&\cite{SMartin1978Atomic}\\
$6s11p\,^3P_2$& 48233.7(5)& 48251.376961&-529942&\cite{SBiRu1991The}\\
$6s12p\,^3P_2$& 48745.6(5)& 48741.398200&125967&\cite{SAli1999Two}\\
$4f^{13}6s5d^2\,^3P_2$& 48965.15& 48961.554736&112280&\cite{SWyart1979Extended}\\
$6s13p\,^3P_2$& 49136.4(5)& 49138.766072&-70933&\cite{SAli1999Two}\\
$6s14p\,^3P_2$& 49370.5(5)& 49372.087230&-47584&\cite{SAli1999Two}\\
$6s15p\,^3P_2$& 49553.0(5)& 49553.219876&-6592&\cite{SAli1999Two}\\
\hline\hline
\end{longtable*}

\clearpage

\subsection{$^{174}$Yb $^{3}P_0$}

\begin{longtable*}{@{\extracolsep{\fill}}ccS[table-format=6.4]S[table-format=5.6]S[table-format=5.6]S[table-format=1.2]c}
\caption{\label{tab:174Spec_Microwave_3p0} Spectroscopic data of the  $n\,^{3}P_0$ Rydberg series obtained by microwave spectroscopy in an atomic beam setup as described in App.~\ref{sec:atomicbeam} of the main text and from Ref.~\cite{SLehec2017PhD}. }\\

\hline\hline\\\multirow{2}{*}{$f$} & \multirow{2}{*}{$i$}& {\multirow{2}{*}{$\Delta E_{f\leftarrow i}^\mathrm{exp}$ ($h\cdot\mathrm{MHz}$)}} & {\multirow{2}{*}{$\tilde\nu_\mathrm{exp.}$ ($\mathrm{cm}^{-1}$ )}}& {\multirow{2}{*}{$\tilde\nu_\mathrm{th.}$ ($\mathrm{cm}^{-1}$ )}} &  {$E_\mathrm{exp.}-E_\mathrm{th.}$ }  &\multirow{2}{*}{Ref.}\\ 
    & & & & & {($h\cdot\mathrm{MHz}$)} & \\\hline \\ 
 \endfirsthead

 {{\tablename\ \thetable{} (\textit{continued})}} \\
\hline\hline\\\multirow{2}{*}{$f$} & \multirow{2}{*}{$i$}& {\multirow{2}{*}{$\Delta E_{f\leftarrow i}^\mathrm{exp}$ ($h\cdot\mathrm{MHz}$)}} & {\multirow{2}{*}{$\tilde\nu_\mathrm{exp.}$ ($\mathrm{cm}^{-1}$ )}}& {\multirow{2}{*}{$\tilde\nu_\mathrm{th.}$ ($\mathrm{cm}^{-1}$ )}} &  {$E_\mathrm{exp.}-E_\mathrm{th.}$ }  &\multirow{2}{*}{Ref.}\\ 
    & & & & & {($h\cdot\mathrm{MHz}$)} & \\\hline \\ 
 \endhead

\hline
 \endfoot

 \hline
 \endlastfoot
$6s31p\,^3\mathrm{P}_0$& $6s30d\,^3\mathrm{D}_2$& -218431.0(10)& 50288.019600& 50288.019606&-0.18&This work\\
$6s32p\,^3\mathrm{P}_0$& $6s31d\,^3\mathrm{D}_2$& -202698.6(10)& 50298.819704& 50298.819689&0.43&This work\\
$6s34p\,^3\mathrm{P}_0$& $6s33d\,^3\mathrm{D}_2$& -174533.8(10)& 50317.337535& 50317.337507&0.83&This work\\
$6s35p\,^3\mathrm{P}_0$& $6s34d\,^3\mathrm{D}_2$& -161992.9(10)& 50325.306952& 50325.307034&-2.48&This work\\
$6s36p\,^3\mathrm{P}_0$& $6s35d\,^3\mathrm{D}_2$& -150399.2(10)& 50332.553339& 50332.553286&1.56&This work\\
$6s50p\,^3\mathrm{P}_0$& $6s47d\,^3\mathrm{D}_2$& 83958.5(10)& 50389.830942& 50389.830943&-0.01&This work\\
\hline\hline
\end{longtable*}

\begin{longtable*}{@{\extracolsep{\fill}}c  S[table-format=5.6]S[table-format=5.6]S[table-format=2.2]c}
\caption{\label{tab:174YbSpec_3P0} Laser spectroscopic data of the $6snp\,^3P_0$ Rydberg series included in the determination of the MQDT model. The state energies and the respective uncertainties are taken from Ref.~\cite{SAymar1984three}.}\\

\hline\hline\\
 [-6pt]{\multirow{2}{*}{State label}} & {\multirow{2}{*}{$\tilde\nu_\mathrm{exp.}$ ($\mathrm{cm}^{-1}$)}} & {\multirow{2}{*}{$\tilde\nu_\mathrm{th.}$ ($\mathrm{cm}^{-1}$ )}} & {$E_\mathrm{exp.}-E_\mathrm{th.}$}&{\multirow{2}{*}{Ref.}}\\
 & & & {($h\cdot\mathrm{MHz}$)}  &\\[3pt]
\hline\\ [-5pt] 
 \endfirsthead

 {{\tablename\ \thetable{} (\textit{continued})}} \\
\hline\hline\\
 [-6pt]{\multirow{2}{*}{State label}} & {\multirow{2}{*}{$\tilde\nu_\mathrm{exp.}$ ($\mathrm{cm}^{-1}$)}} & {\multirow{2}{*}{$\tilde\nu_\mathrm{th.}$ ($\mathrm{cm}^{-1}$ )}} & {$E_\mathrm{exp.}-E_\mathrm{th.}$}&{\multirow{2}{*}{Ref.}}\\
 & & & {($h\cdot\mathrm{MHz}$)}  &\\[3pt]
\hline\\ [-5pt] 
 \endhead

\hline
 \endfoot

 \hline
 \endlastfoot
$6s19p\,^3P_0$& 49950.00(10)& 49949.916956&2490&\cite{SAymar1984three}\\
$6s20p\,^3P_0$& 50008.89(10)& 50008.797212&2782&\cite{SAymar1984three}\\
$6s21p\,^3P_0$& 50057.80(10)& 50057.639704&4806&\cite{SAymar1984three}\\
$6s22p\,^3P_0$& 50098.66(10)& 50098.610920&1471&\cite{SAymar1984three}\\
$6s23p\,^3P_0$& 50133.38(10)& 50133.325703&1628&\cite{SAymar1984three}\\
$6s24p\,^3P_0$& 50163.01(10)& 50163.007489&75&\cite{SAymar1984three}\\
$6s25p\,^3P_0$& 50188.69(10)& 50188.596367&2807&\cite{SAymar1984three}\\
$6s26p\,^3P_0$& 50210.87(10)& 50210.823496&1394&\cite{SAymar1984three}\\
$6s27p\,^3P_0$& 50230.30(10)& 50230.263498&1094&\cite{SAymar1984three}\\
$6s28p\,^3P_0$& 50247.59(10)& 50247.372132&6532&\cite{SAymar1984three}\\
$6s29p\,^3P_0$& 50262.63(10)& 50262.514002&3478&\cite{SAymar1984three}\\
$6s30p\,^3P_0$& 50276.24(10)& 50275.983317&7695&\cite{SAymar1984three}\\
\hline\hline
\end{longtable*}

\clearpage

\subsection{$^{171}$Yb S $F=1/2$}

\begin{longtable*}{@{\extracolsep{\fill}}S[table-format=3.7]  S[table-format=5.6]S[table-format=5.6]S[table-format=2.2]}
\caption{\label{tab:171Spec_S_F12} Laser spectroscopic data of the $\ket{\nu,L=0,F=1/2}$ Rydberg series obtained from laser spectroscopy in an atomic beam setup as described in App.~\ref{sec:atomicbeam} of the main text.}\\

\hline\hline\\
 [-6pt]{\multirow{2}{*}{$\nu_{F_c=1}$}} & {\multirow{2}{*}{$\tilde\nu_\mathrm{exp.}$ ($\mathrm{cm}^{-1}$)}} & {\multirow{2}{*}{$\tilde\nu_\mathrm{th.}$ ($\mathrm{cm}^{-1}$ )}} & {$E_\mathrm{exp.}-E_\mathrm{th.}$}  \\ 
&&& {($h\cdot\mathrm{MHz}$)}\\[3pt]
\hline\\ [-5pt] 
 \endfirsthead

 {{\tablename\ \thetable{} (\textit{continued})}} \\
\hline\hline \\
 [-6pt]{\multirow{2}{*}{$\nu_{F_c=1}$}} & {\multirow{2}{*}{$\tilde\nu_\mathrm{exp.}$ ($\mathrm{cm}^{-1}$)}} & {\multirow{2}{*}{$\tilde\nu_\mathrm{th.}$ ($\mathrm{cm}^{-1}$ )}} & {$E_\mathrm{exp.}-E_\mathrm{th.}$}  \\ 
&&& {($h\cdot\mathrm{MHz}$)}\\[3pt]
\hline\\ [-5pt] 
 \endhead

\hline
 \endfoot

 \hline
 \endlastfoot
26.7220724& 50289.53900(10)& 50289.539148&-4.31\\
27.5326737& 50298.45485(10)& 50298.454960&-3.24\\
27.7206406& 50300.41147(10)& 50300.411503&-0.92\\
28.5284580& 50308.38457(10)& 50308.384445&3.70\\
28.7192557& 50310.17000(10)& 50310.170033&-0.98\\
29.5240542& 50317.32469(10)& 50317.324673&0.50\\
29.7179082& 50318.96181(10)& 50318.961749&1.92\\
30.5194120& 50325.40248(10)& 50325.402483&-0.13\\
\hline\hline
\end{longtable*}

\begin{longtable*}{@{\extracolsep{\fill}}S[table-format=3.7]  S[table-format=1.3] S[table-format=1.3]}
	\caption{\label{tab:171Spec_S_F12_gfactors} Predicted $g_\mathrm{th}$ and experimental $g_\mathrm{exp}$ factors of $\ket{\nu,L=0,F=1/2}$ Rydberg states obtained from laser spectroscopy in an atomic beam setup as described in App.~\ref{sec:atomicbeam} of the main text.}\\
	
	\hline\hline\\
	[-6pt]{$\nu_{F_c=1}$} & {$g_\mathrm{exp}$}   & {$g_\mathrm{th}$}    \\[3pt]
	\hline\\ [-5pt] 
	\endfirsthead
	
	{{\tablename\ \thetable{} (\textit{continued})}} \\
	\hline\hline \\
	[-6pt]{$\nu_{F_c=1}$} & {$g_\mathrm{exp}$}  & {$g_\mathrm{th}$}   \\[3pt]
	\hline\\ [-5pt] 
	\endhead
	
	\hline
	\endfoot
	
	\hline
	\endlastfoot
	30.5194120& 2.70(16)&2.63\\
	30.7165912& 0.03(2)&0.04\\
	35.7103372& 0.06(2)&0.07\\
	40.4541871& 2.54(14)&2.54\\
	40.7045350& 0.13(2)&0.13\\
	50.3436170& 2.44(13)&2.42\\
	50.6942249& 0.25(5)&0.25\\
	60.1761918& 2.42(13)&2.29\\
\hline\hline
\end{longtable*}

\begin{longtable*}{@{\extracolsep{\fill}}S[table-format=3.7]  S[table-format=1.3] S[table-format=1.3]}
	\caption{\label{tab:171Spec_S_F12_pol} Predicted $\alpha_{0,\mathrm{th}}$ and experimental $\alpha_{0,\mathrm{exp}}$ static dipole polarizabilities of $\ket{\nu,L=0,F=1/2}$ Rydberg states obtained from laser spectroscopy in an atomic beam setup as described in App.~\ref{sec:atomicbeam} of the main text.}\\
	
	\hline\hline\\
	[-6pt]{$\nu_{F_c=1}$} & {$\alpha_{0,\mathrm{exp}}$ $\left(\mathrm{MHz}/(\mathrm{V}/\mathrm{cm})^2\right)$}   & {$\alpha_{0,\mathrm{th}}$ $\left(\mathrm{MHz}/(\mathrm{V}/\mathrm{cm})^2\right)$}     \\[3pt]
	\hline\\ [-5pt] 
	\endfirsthead
	
	{{\tablename\ \thetable{} (\textit{continued})}} \\
	\hline\hline \\
	[-6pt]{$\nu_{F_c=1}$} & {$\alpha_{0,\mathrm{exp}}$ $\left(\mathrm{MHz}/(\mathrm{V}/\mathrm{cm})^2\right)$}   & {$\alpha_{0,\mathrm{th}}$ $\left(\mathrm{MHz}/(\mathrm{V}/\mathrm{cm})^2\right)$}   \\[3pt]
	\hline\\ [-5pt] 
	\endhead
	
	\hline
	\endfoot
	
	\hline
	\endlastfoot
27.5326737& 18.1(14)&16.8\\
28.5284580& 37.1(9)&37.1\\
29.5240542& 161.5(11)&167.9\\
30.5194120& -128.2(11)&-135.1\\
31.5144948& 25.6(14)&-63.5\\
31.7152997& -63.4(11)&31.8\\
32.5092738& -49.1(9)&-48.0\\
32.7140304& 43.6(19)&42.4\\
\hline\hline
\end{longtable*}

\begin{longtable*}{@{\extracolsep{\fill}}S[table-format=3.7]  S[table-format=1.4] S[table-format=1.1]}
	\caption{\label{tab:171Spec_S_F12_stark} Predicted $\Delta E_{\mathrm{Stark},\mathrm{th}}$ and experimental $\Delta E_{\mathrm{Stark},\mathrm{exp}}$ Stark shifts of $\ket{\nu,L=0,F=1/2}$ Rydberg states obtained from laser spectroscopy in an atomic beam setup as described in App.~\ref{sec:atomicbeam} of the main text at static electric fields of \SI{23.8}{mV\per cm} ($^a$) or \SI{41.6}{mV\per cm} ($^b$).}\\
	
	\hline\hline\\
	[-6pt]{$\nu_{F_c=1}$} & {$\Delta E_{\mathrm{Stark},\mathrm{exp}}$ $\left(\mathrm{MHz}\right)$}   & {$\Delta E_{\mathrm{Stark},\mathrm{th}}$ $\left(\mathrm{MHz}\right)$}     \\[3pt]
	\hline\\ [-5pt] 
	\endfirsthead
	
	{{\tablename\ \thetable{} (\textit{continued})}} \\
	\hline\hline \\
	[-6pt]{$\nu_{F_c=1}$} & {$\Delta E_{\mathrm{Stark},\mathrm{exp}}$ $\left(\mathrm{MHz}\right)$}   & {$\Delta E_{\mathrm{Stark},\mathrm{th}}$ $\left(\mathrm{MHz}\right)$}   \\[3pt]
	\hline\\ [-5pt] 
	\endhead
	
	\hline
	\endfoot
	
	\hline
	\endlastfoot
64.0916080& 12.9(6)$^a$&11.8$^a$\\
65.6785389& -4.7(6)$^b$&-4.8$^b$\\
66.6770495& -5.0(6)$^b$&-5.3$^b$\\
67.6753964& 17.3(6)$^b$&18.3$^b$\\
	\hline\hline
\end{longtable*}

\clearpage

\subsection{$^{171}$Yb S $F=3/2$}

\begin{longtable*}{@{\extracolsep{\fill}}S[table-format=2.7] S[table-format=2.7] S[table-format=7.4] S[table-format=5.6]S[table-format=5.6]S[table-format=2.2]}
\caption{\label{tab:171Spec_S_F32} Spectroscopic data of the  $\ket{\nu,L=0,F=3/2}$ Rydberg series obtained from microwave spectroscopy in an atomic beam setup as described in App.~\ref{sec:atomicbeam} of the main text between $f=\ket{\nu,L=0,F=3/2}$ and $i=\ket{\nu',L=0,F=1/2}$ (a) or $^\mathrm{b}$ $i=\ket{\nu',L=2,F=5/2}$ (b) states. }\\

\hline\hline\\
[-6pt]{\multirow{2}{*}{$\nu_{F_c=1}^f$} }& {\multirow{2}{*}{$\nu_{F_c=1}^i$}}& {\multirow{2}{*}{$\nu_{f\leftarrow i}$ (MHz)}} & {\multirow{2}{*}{$\tilde\nu_\mathrm{exp.}$ ($\mathrm{cm}^{-1}$ )}}& {\multirow{2}{*}{$\tilde\nu_\mathrm{th.}$ ($\mathrm{cm}^{-1}$ )}} & {$E_\mathrm{exp.}-E_\mathrm{th.}$}\\ 
&&&&& {($h\cdot\mathrm{MHz}$)}\\[3pt]
\hline\\ [-5pt]
 \endfirsthead

 {{\tablename\ \thetable{} (\textit{continued})}} \\
 \hline\hline\\
[-6pt]{\multirow{2}{*}{$\nu_{F_c=1}^f$} }& {\multirow{2}{*}{$\nu_{F_c=1}^i$}}& {\multirow{2}{*}{$\nu_{f\leftarrow i}$ (MHz)}} & {\multirow{2}{*}{$\tilde\nu_\mathrm{exp.}$ ($\mathrm{cm}^{-1}$ )}}& {\multirow{2}{*}{$\tilde\nu_\mathrm{th.}$ ($\mathrm{cm}^{-1}$ )}} & {$E_\mathrm{exp.}-E_\mathrm{th.}$} \\ 
&&&&& {($h\cdot\mathrm{MHz}$)}\\[3pt]
\hline\\ [-5pt]
 \endhead

\hline
 \endfoot

 \hline
 \endlastfoot
42.5615357 & 41.7034283$^\mathrm{a}$ & 75510.49(10) & 50382.639145 &50382.639012 & 4.01\\
43.5614817 & 42.2815828$^\mathrm{b}$ & 106550.02(10)&50385.388283 & 50385.388228& 1.66 \\
44.5614339 & 43.2815256$^\mathrm{b}$ & 99434.19(10)& 50387.954467& 50387.954469 & -0.06\\

\hline\hline
\end{longtable*}

\clearpage

\subsection{$^{171}$Yb P $F=1/2$}

\begin{longtable*}{@{\extracolsep{\fill}}S[table-format=2.7] S[table-format=2.7] S[table-format=7.4] S[table-format=5.6]S[table-format=5.6]S[table-format=2.2]}
\caption{\label{tab:171Spec_P_F12} Spectroscopic data of the  $\ket{\nu,L=1,F=1/2}$ Rydberg series obtained from microwave spectroscopy in an atomic beam setup as described in App.~\ref{sec:atomicbeam} of the main text between $i=\ket{\nu,L=0,F=1/2}$ and $f=\ket{\nu',L=1,F=1/2}$. }\\

\hline\hline\\
[-6pt]{\multirow{2}{*}{$\nu_{F_c=1}^f$} }& {\multirow{2}{*}{$\nu_{F_c=1}^i$}}& {\multirow{2}{*}{$\nu_{f\leftarrow i}$ (MHz)}} & {\multirow{2}{*}{$\tilde\nu_\mathrm{exp.}$ ($\mathrm{cm}^{-1}$ )}}& {\multirow{2}{*}{$\tilde\nu_\mathrm{th.}$ ($\mathrm{cm}^{-1}$ )}} & {$E_\mathrm{exp.}-E_\mathrm{th.}$}\\ 
&&&&& {($h\cdot\mathrm{MHz}$)}\\[3pt]
\hline\\ [-5pt]
 \endfirsthead

 {{\tablename\ \thetable{} (\textit{continued})}} \\
 \hline\hline\\
[-6pt]{\multirow{2}{*}{$\nu_{F_c=1}^f$} }& {\multirow{2}{*}{$\nu_{F_c=1}^i$}}& {\multirow{2}{*}{$\nu_{f\leftarrow i}$ (MHz)}} & {\multirow{2}{*}{$\tilde\nu_\mathrm{exp.}$ ($\mathrm{cm}^{-1}$ )}}& {\multirow{2}{*}{$\tilde\nu_\mathrm{th.}$ ($\mathrm{cm}^{-1}$ )}} & {$E_\mathrm{exp.}-E_\mathrm{th.}$} \\ 
&&&&& {($h\cdot\mathrm{MHz}$)}\\[3pt]
\hline\\ [-5pt]
 \endhead

\hline
 \endfoot

 \hline
 \endlastfoot
28.0560523& 27.7206406& 101759.03(10)& 50303.805819& 50303.805600&6.57\\
29.5295597& 29.7179082& -47673.09(10)& 50317.371545& 50317.371612&-2.00\\
29.8278618& 29.7179082& 27406.71(10)& 50319.875938& 50319.876141&-6.09\\
30.5113566& 30.7165912& -47064.40(10)& 50325.340310& 50325.340265&1.34\\
31.0525836& 30.7165912& 75043.07(10)& 50329.413376& 50329.413508&-3.94\\
32.0514332& 31.7152997& 68236.65(10)& 50336.395994& 50336.396148&-4.61\\
33.0502594& 32.7140304& 62222.71(10)& 50342.755019& 50342.755172&-4.61\\
33.7359503& 34.7115502& -160206.00(10)& 50346.797389& 50346.797509&-3.57\\
\hline\hline
\end{longtable*}

\clearpage

\begin{longtable*}{@{\extracolsep{\fill}}S[table-format=2.6]  S[table-format=5.6]S[table-format=5.6]S[table-format=2.2]c}
\caption{\label{tab:171Spec_P_F12_Laser} Laser spectroscopic data of the $\ket{\nu,L=1,F=1/2}$ Rydberg series from three-photon laser spectroscopy as presented in Ref.~\cite{SMajewski1985diploma}.}\\

\hline\hline\\
 [-6pt]{\multirow{2}{*}{$\nu_{F_c=1}$}} & {\multirow{2}{*}{$\tilde\nu_\mathrm{exp.}$ ($\mathrm{cm}^{-1}$)}} & {\multirow{2}{*}{$\tilde\nu_\mathrm{th.}$ ($\mathrm{cm}^{-1}$ )}} & {$E_\mathrm{exp.}-E_\mathrm{th.}$} & Ref. \\ 
&&& {($h\cdot\mathrm{MHz}$)}& \\[3pt]
\hline\\ [-5pt] 
 \endfirsthead

 {{\tablename\ \thetable{} (\textit{continued})}} \\
\hline\hline \\
 [-6pt]{\multirow{2}{*}{$\nu_{F_c=1}$}} & {\multirow{2}{*}{$\tilde\nu_\mathrm{exp.}$ ($\mathrm{cm}^{-1}$)}} & {\multirow{2}{*}{$\tilde\nu_\mathrm{th.}$ ($\mathrm{cm}^{-1}$ )}} & {$E_\mathrm{exp.}-E_\mathrm{th.}$} & Ref.  \\ 
&&& {($h\cdot\mathrm{MHz}$)}& \\[3pt]
\hline\\ [-5pt] 
 \endhead

\hline
 \endfoot

 \hline
 \endlastfoot
9.0666242& 49110.29823& 49108.277609&60576.73& \cite{SMajewski1985diploma}\\
9.1344553& 49127.33009& 49128.030119&-20986.44& \cite{SMajewski1985diploma}\\
10.0393585& 49352.40117& 49354.435282&-60981.24& \cite{SMajewski1985diploma}\\
10.0621527& 49360.61396& 49359.362629&37513.80& \cite{SMajewski1985diploma}\\
11.0117215& 49536.19710& 49538.230163&-60949.58& \cite{SMajewski1985diploma}\\
11.0565325& 49546.66039& 49545.550934&33260.53& \cite{SMajewski1985diploma}\\
11.9663443& 49677.50027& 49676.862568&19117.72& \cite{SMajewski1985diploma}\\
12.0524108& 49688.69245& 49687.768596&27696.37& \cite{SMajewski1985diploma}\\
\hline\hline
\end{longtable*}

\begin{longtable*}{@{\extracolsep{\fill}}S[table-format=3.7]  S[table-format=1.3] S[table-format=1.3]}
	\caption{\label{tab:171Spec_P_F12_gfactors} Predicted $g_\mathrm{th}$ and experimental $g_\mathrm{exp}$ factors of $\ket{\nu,L=1,F=1/2}$ Rydberg states obtained from microwave spectroscopy in an atomic beam setup as described in App.~\ref{sec:atomicbeam} of the main text.}\\
	
	\hline\hline\\
	[-6pt]{$\nu_{F_c=1}$} & {$g_\mathrm{exp}$}   & {$g_\mathrm{th}$}    \\[3pt]
	\hline\\ [-5pt] 
	\endfirsthead
	
	{{\tablename\ \thetable{} (\textit{continued})}} \\
	\hline\hline \\
	[-6pt]{$\nu_{F_c=1}$} & {$g_\mathrm{exp}$}  & {$g_\mathrm{th}$}   \\[3pt]
	\hline\\ [-5pt] 
	\endhead
	
	\hline
	\endfoot
	
	\hline
	\endlastfoot
28.0560523& 1.418(20)&1.404\\
29.8278618& 2.033(20)&1.971\\
31.0525836& 1.414(20)&1.402\\
33.7359502& 2.064(20)&2.014\\
34.7093704& 2.088(20)&2.033\\
35.0477819& 1.395(20)&1.396\\
35.4386363& -0.167(20)&-0.165\\
35.6815855& 2.112(20)&2.057\\
37.6235720& 2.100(20)&2.120\\
38.0434931& 1.387(20)&1.389\\
40.3786900& -0.326(20)&-0.325\\
41.5113915& 2.217(20)&2.245\\
42.0357614& 1.396(20)&1.374\\
42.4877437& 2.203(20)&2.232\\
45.0271345& 1.353(20)&1.352\\
46.2790376& 0.107(20)&0.103\\
46.4219443& 1.865(20)&1.855\\
	\hline\hline
\end{longtable*}

\clearpage

\subsection{$^{171}$Yb P $F=3/2$}

\begin{longtable*}{@{\extracolsep{\fill}}S[table-format=2.7] S[table-format=2.7] S[table-format=7.4] S[table-format=5.6]S[table-format=5.6]S[table-format=2.2]}
\caption{\label{tab:171Spec_P_F32} Spectroscopic data of the  $\ket{\nu,L=1,F=3/2}$ Rydberg series obtained from microwave spectroscopy in an atomic beam setup as described in App.~\ref{sec:atomicbeam} of the main text between $i=\ket{\nu,L=0,F=1/2}$ and $f=\ket{\nu',L=1,F=3/2}$. }\\

\hline\hline\\
[-6pt]{\multirow{2}{*}{$\nu_{F_c=1}^f$} }& {\multirow{2}{*}{$\nu_{F_c=1}^i$}}& {\multirow{2}{*}{$\nu_{f\leftarrow i}$ (MHz)}} & {\multirow{2}{*}{$\tilde\nu_\mathrm{exp.}$ ($\mathrm{cm}^{-1}$ )}}& {\multirow{2}{*}{$\tilde\nu_\mathrm{th.}$ ($\mathrm{cm}^{-1}$ )}} & {$E_\mathrm{exp.}-E_\mathrm{th.}$} \\ 
&&&&& {($h\cdot\mathrm{MHz}$)}\\[3pt]
\hline\\ [-5pt]
 \endfirsthead

 {{\tablename\ \thetable{} (\textit{continued})}} \\
 \hline\hline\\
[-6pt]{\multirow{2}{*}{$\nu_{F_c=1}^f$} }& {\multirow{2}{*}{$\nu_{F_c=1}^i$}}& {\multirow{2}{*}{$\nu_{f\leftarrow i}$ (MHz)}} & {\multirow{2}{*}{$\tilde\nu_\mathrm{exp.}$ ($\mathrm{cm}^{-1}$ )}}& {\multirow{2}{*}{$\tilde\nu_\mathrm{th.}$ ($\mathrm{cm}^{-1}$ )}} & {$E_\mathrm{exp.}-E_\mathrm{th.}$} \\ 
&&&&& {($h\cdot\mathrm{MHz}$)}\\[3pt]
\hline\\ [-5pt]
 \endhead

\hline
 \endfoot

\hline\\
 \endlastfoot
28.0307908& 27.7206406& 94221.66(10)& 50303.554399& 50303.554209&5.69\\
28.0711961& 27.7206406& 106260.55(10)& 50303.955973& 50303.955979&-0.17\\
29.8585747& 29.7179082& 35019.09(10)& 50320.129860& 50320.129752&3.24\\
30.8406204& 30.7165912& 27991.62(10)& 50327.843909& 50327.843815&2.81\\
31.0158452& 30.7165912& 66958.10(10)& 50329.143691& 50329.143745&-1.64\\
31.0684215& 30.7165912& 78522.04(10)& 50329.529422& 50329.529507&-2.55\\
32.8005012& 33.7127811& -163250.19(10)& 50341.219493& 50341.219414&2.38\\
33.0043868& 32.7140304& 53847.84(10)& 50342.475663& 50342.475714&-1.53\\
\hline\hline
\end{longtable*}

\clearpage

\begin{longtable*}{@{\extracolsep{\fill}}S[table-format=2.6]  S[table-format=5.6]S[table-format=5.6]S[table-format=2.2]c}
\caption{\label{tab:171Spec_P_F32_Laser} Laser spectroscopic data of the $\ket{\nu,L=1,F=3/2}$ Rydberg series from three-photon laser spectroscopy as presented in Ref.~\cite{SMajewski1985diploma}.}\\

\hline\hline\\
 [-6pt]{\multirow{2}{*}{$\nu_{F_c=1}$}} & {\multirow{2}{*}{$\tilde\nu_\mathrm{exp.}$ ($\mathrm{cm}^{-1}$)}} & {\multirow{2}{*}{$\tilde\nu_\mathrm{th.}$ ($\mathrm{cm}^{-1}$ )}} & {$E_\mathrm{exp.}-E_\mathrm{th.}$} & Ref. \\ 
&&& {($h\cdot\mathrm{MHz}$)}& \\[3pt]
\hline\\ [-5pt] 
 \endfirsthead

 {{\tablename\ \thetable{} (\textit{continued})}} \\
\hline\hline \\
 [-6pt]{\multirow{2}{*}{$\nu_{F_c=1}$}} & {\multirow{2}{*}{$\tilde\nu_\mathrm{exp.}$ ($\mathrm{cm}^{-1}$)}} & {\multirow{2}{*}{$\tilde\nu_\mathrm{th.}$ ($\mathrm{cm}^{-1}$ )}} & {$E_\mathrm{exp.}-E_\mathrm{th.}$} & Ref.  \\ 
&&& {($h\cdot\mathrm{MHz}$)}& \\[3pt]
\hline\\ [-5pt] 
 \endhead

\hline
 \endfoot

 \hline
 \endlastfoot
9.0661376& 49110.14299& 49108.134316&60218.56& \cite{SMajewski1985diploma}\\
9.1352231& 49127.55521& 49128.251203&-20865.36& \cite{SMajewski1985diploma}\\
10.0407478& 49352.67672& 49354.736575&-61752.79& \cite{SMajewski1985diploma}\\
10.0614184& 49360.47786& 49359.204427&38176.60& \cite{SMajewski1985diploma}\\
11.0134422& 49536.46659& 49538.512920&-61347.41& \cite{SMajewski1985diploma}\\
11.0556682& 49546.52743& 49545.410580&33482.26& \cite{SMajewski1985diploma}\\
11.9870812& 49677.72159& 49679.511771&-53668.37& \cite{SMajewski1985diploma}\\
12.0514733& 49688.56786& 49687.651058&27485.07& \cite{SMajewski1985diploma}\\
\hline\hline
\end{longtable*}

\begin{longtable*}{@{\extracolsep{\fill}}S[table-format=3.7]  S[table-format=1.3] S[table-format=1.3]}
	\caption{\label{tab:171Spec_P_F32_gfactors} Predicted $g_\mathrm{th}$ and experimental $g_\mathrm{exp}$ factors of $\ket{\nu,L=1,F=3/2}$ Rydberg states obtained from microwave spectroscopy in an atomic beam setup as described in App.~\ref{sec:atomicbeam} of the main text.}\\
	
	\hline\hline\\
	[-6pt]{$\nu_{F_c=1}$} & {$g_\mathrm{exp}$}   & {$g_\mathrm{th}$}    \\[3pt]
	\hline\\ [-5pt] 
	\endfirsthead
	
	{{\tablename\ \thetable{} (\textit{continued})}} \\
	\hline\hline \\
	[-6pt]{$\nu_{F_c=1}$} & {$g_\mathrm{exp}$}  & {$g_\mathrm{th}$}   \\[3pt]
	\hline\\ [-5pt] 
	\endhead
	
	\hline
	\endfoot
	
	\hline
	\endlastfoot
	28.0307908& 1.047(20)&1.037\\
	28.0711961& 1.477(20)&1.461\\
	29.8585748& 0.964(20)&0.954\\
	30.8406204& 0.967(20)&0.953\\
	31.0158452& 1.125(20)&1.116\\
	31.0684215& 1.393(20)&1.380\\
	33.7782917& 0.974(20)&0.951\\
	34.7547035& 0.980(20)&0.951\\
	34.9915709& 1.196(20)&1.197\\
	35.0656773& 1.295(20)&1.296\\
	35.7298575& 0.974(20)&0.951\\
	37.6772350& 0.981(20)&0.952\\
	37.9695264& 1.239(20)&1.240\\
	38.0641210& 1.252(20)&1.249\\
	41.5692557& 0.950(20)&0.958\\
	42.0625036& 1.219(20)&1.203\\
	42.5437375& 0.952(20)&0.961\\
	45.8913416& 1.304(20)&1.308\\
	46.0612539& 1.165(20)&1.169\\
	46.4547077& 0.989(20)&0.974\\
	50.0602610& 1.157(20)&1.143\\
	52.3628809& 1.021(20)&0.996\\
	57.7095044& 1.362(20)&1.338\\
	58.0587732& 1.122(20)&1.108\\
	58.3068555& 1.031(20)&1.016\\
	58.6904017& 1.414(20)&1.339\\
	59.0586194& 1.129(20)&1.105\\
	59.2997598& 1.044(20)&1.019\\
	59.6706572& 1.353(20)&1.339\\
	60.0584708& 1.108(20)&1.101\\
	60.2931358& 1.046(20)&1.022\\
	60.6502661& 1.361(20)&1.340\\
	61.0583267& 1.119(20)&1.098\\
	61.2869307& 1.050(20)&1.025\\
	62.6075323& 1.351(20)&1.340\\
	63.0580502& 1.104(20)&1.092\\
	63.2755815& 1.057(20)&1.031\\
	\hline\hline
\end{longtable*}

\clearpage

\subsection{Hyperfine splitting of $^{171}$Yb $P$ $F=1/2$ and $F=3/2$ Rydberg states}

\begin{longtable*}{@{\extracolsep{\fill}}c  S[table-format=4.0] S[table-format=4.0] S[table-format=4.0]}
	\caption{\label{tab:171Spec_P_1p1_majewski} Hyperfine splitting of $^{171}$Yb $P$ $F=1/2$ and $F=3/2$ Rydberg states referenced to the $6s^2\,^1S_0\rightarrow6snp\,^1P_1$ transition of $^{176}$Yb (see Tab.~\ref{tab:176Spec_P_majewski}). The data is reproduced from Tab.~3.3a or from Fig.~3.9 (marked by$^\dag$) of Ref.~\cite{SMajewski1985diploma}.}\\
	
	\hline\hline\\
	[-6pt]\multirow{2}{*}{$n$} & \multicolumn{3}{c}{$\Delta E^{(171-176)}$ $(h\cdot\mathrm{MHz})$} \\[3pt]
    \cline{2-4}\rule{0pt}{3ex} 
              & {$F=3/2$} &  {$F=1/2$} & {$F=1/2^\dag$}\\[3pt]
	\hline\\ [-5pt] 
	\endfirsthead
	
	{{\tablename\ \thetable{} (\textit{continued})}} \\
	\hline\hline \\
	[-6pt]\multirow{2}{*}{$n$} & \multicolumn{3}{c}{$\Delta E^{(171-176)}$ $(h\cdot\mathrm{MHz})$} \\[3pt]
    \cline{2-4}\rule{0pt}{3ex} 
              & {$F=3/2$} &  {$F=1/2$} & {$F=1/2^\dag$}\\[3pt]
	\hline\\ [-5pt] 
	\endhead
	
	\hline
	\endfoot
	
	\hline
	\endlastfoot
13& 4134& -2615& {\textemdash}\\
14& 4779& -3482& {\textemdash}\\
15& 4775& -3304& {\textemdash}\\
16& 4625& -2010& {\textemdash}\\
17& 4298& -3080& {\textemdash}\\
18& 3478& -1599& {\textemdash}\\
19& 2980& -91& {\textemdash}\\
20& 2824& 1241& {\textemdash}\\
21& 3851& 2546& {\textemdash}\\
22& -2236& 3368& {\textemdash}\\
23& -1425& 3819& {\textemdash}\\
24& -1401& 4061& {\textemdash}\\
25& -1568& 4210& {\textemdash}\\
28& -2245& 4368& {\textemdash}\\
30& -2727& 4378& 9833\\
35& -3814& 4278& 7655\\
40& -4615& 4092& 6624\\
41& -4749& 4046& 6486\\
42& -4870& 3995& 6349\\
43& -4986& 3928& 6235\\
44& -5086& 3873& 6154\\
45& -5202& 3802& 6063\\
51& -5731& 3154& 5697\\
55& -6004& {\textemdash}& 5526\\
60& -6258& 736& 5413\\
	\hline\hline
\end{longtable*}

\begin{longtable*}{@{\extracolsep{\fill}}c  S[table-format=4.0] S[table-format=4.0] S[table-format=4.0]}
	\caption{\label{tab:171Spec_P_3p1_majewski}  Hyperfine splitting of $^{171}$Yb $P$ $F=1/2$ and $F=3/2$ Rydberg states referenced to the $6s^2\,^1S_0\rightarrow6snp\,^3P_1$ transition of $^{176}$Yb (see Tab.~\ref{tab:176Spec_P_majewski}). The data is reproduced from Tab.~3.3b or from Fig.~3.11 (marked by$^\dag$) of Ref.~\cite{SMajewski1985diploma}.}\\
	
	\hline\hline\\
	[-6pt]\multirow{2}{*}{$n$} & \multicolumn{3}{c}{$\Delta \nu^{(171-176)}$ $(h\cdot\mathrm{MHz})$} \\[3pt]
    \cline{2-4}\rule{0pt}{3ex} 
              & {$F=3/2$} &  {$F=1/2$} &  {$F=3/2^\dag$}\\[3pt]
	\hline\\ [-5pt] 
	\endfirsthead
	
	{{\tablename\ \thetable{} (\textit{continued})}} \\
	\hline\hline \\
	[-6pt]\multirow{2}{*}{$n$} & \multicolumn{3}{c}{$\Delta \nu^{(171-176)}$ $(h\cdot\mathrm{MHz})$} \\[3pt]
    \cline{2-4}\rule{0pt}{3ex} 
              & {$F=3/2$} &  {$F=1/2$} &  {$F=3/2^\dag$}\\[3pt]
	\hline\\ [-5pt] 
	\endhead
	
	\hline
	\endfoot
	
	\hline
	\endlastfoot
13& 470& 5124& {\textemdash}\\
14& 616& 4696& {\textemdash}\\
15& 603& 4589& {\textemdash}\\
16& 616& 4351& {\textemdash}\\
17& 968& 3930& {\textemdash}\\
18& 1271& 3143& {\textemdash}\\
19& 1924& 1767& {\textemdash}\\
20& 2795& 30& {\textemdash}\\
21& 3537& -1390& {\textemdash}\\
22& 4002& -2249& {\textemdash}\\
23& 4270& -2707& {\textemdash}\\
24& 4415& -2963& {\textemdash}\\
25& 4502& -3108& {\textemdash}\\
28& 4573& -3240& {\textemdash}\\
30& 4567& -3245& {\textemdash}\\
35& 4474& -3068& {\textemdash}\\
40& 4394& -2598& {\textemdash}\\
41& 4391& -2431& {\textemdash}\\
42& 4373& -2240& {\textemdash}\\
43& 4371& -2002& {\textemdash}\\
44& 4370& -1714& {\textemdash}\\
45& 4372& -1382& {\textemdash}\\
51& 4451& {\textemdash}& {\textemdash}\\
55& 4521& 4521& -5060\\
59& 4596& 5717& -3447\\
61& 4630& 6026& -2633\\
	\hline\hline
\end{longtable*}

\begin{longtable*}{@{\extracolsep{\fill}}c  S[table-format=5.5] S[table-format=5.5]}
	\caption{\label{tab:176Spec_P_majewski}  Experimental term values of $6snp\,^{1,3}P_1$ Rydberg states of $^{176}$Yb. The data is reproduced from Tab.~3.1 of Ref.~\cite{SMajewski1985diploma}.}\\
	
	\hline\hline\\
	[-6pt]$n$ & \multicolumn{2}{c}{$\tilde{\nu}$ (cm$^{-1}$)} \\[3pt]
     \cline{2-3}\rule{0pt}{3ex} 
              & {$6snp\,^1P_1$} &  {$6snp\,^3P_1$} \\[3pt]
	\hline\\ [-5pt] 
	\endfirsthead
	
	{{\tablename\ \thetable{} (\textit{continued})}} \\
	\hline\hline \\
	[-6pt]$n$ & \multicolumn{2}{c}{$\tilde{\nu}$ (cm$^{-1}$)} \\[3pt]
    \cline{2-3}\rule{0pt}{3ex} 
              & {$6snp\,^1P_1$} &  {$6snp\,^3P_1$} \\[3pt]
	\hline\\ [-5pt] 
	\endhead
	
	\hline
	\endfoot
	
	\hline
	\endlastfoot
13& 49127.37(14)& 49110.08(14)\\
14& 49352.47(14)& 49360.41(14)\\
15& 49536.26(14)& 49546.46(14)\\
16& 49677.52(14)& 49688.50(14)\\
17& 49786.39(14)& 49799.03(14)\\
18& 49867.13(14)& 49886.80(14)\\
19& 49969.11(14)& 49957.77(14)\\
20& 50021.24(14)& 50015.82(14)\\
\hline\hline
\end{longtable*}

\clearpage

\subsection{$^{171}$Yb D $F=3/2$}

%

\clearpage

\subsection{$^{171}$Yb D $F=5/2$}

%

\end{document}